%% file: Elices_Fouque_2010_Perturbed_Copula-Introducing_the_skew_effect_in_the_co-dependence.tex
\documentclass[12pt]{article}
\usepackage{graphicx}
\usepackage{lscape}
\bibstyle{plain}

\newcommand{\be}{\begin{equation}}
\newcommand{\en}{\end{equation}}

\newcommand{\PP}{{\mathord{I\kern -.33em P}}}
\newcommand{\EE}{{\mathord{I\kern -.33em E}}}
\newcommand{\RR}{{\mathord{I\kern -.33em R}}}

\newcommand{\ea}{\end{eqnarray}}
\newcommand{\ba}{\begin{eqnarray}}
\newcommand{\ean}{\end{eqnarray*}}
\newcommand{\ban}{\begin{eqnarray*}}

\begin{document}

\title{Perturbed Copula: Introducing the skew effect in the co-dependence.}

\author{Alberto Elices\thanks{Model Validation group, Methodology Department, Risk Division, Grupo Santander, Ciudad Grupo Santander, Ed. Encinar Pl. 2, Avda. Cantabria s/n, 28660 Boadilla del Monte, Spain, {\em aelices@gruposantander.com}.} \and Jean-Pierre Fouque\thanks{Department of Statistics \& Applied Probability, University of California, Santa Barbara, CA 93106-3110, {\em fouque@pstat.ucsb.edu}.}}
\date{\today}
\maketitle

\bigskip

\begin{abstract}
Gaussian copulas are widely used in the industry to correlate two random variables when there is no prior knowledge about the co-dependence between them. The perturbed Gaussian copula approach allows introducing the skew information of both random variables into the co-dependence structure. The analytical expression of this copula is derived through an asymptotic expansion under the assumption of a common fast mean reverting stochastic volatility factor. This paper applies this new perturbed copula to the valuation of derivative products; in particular FX quanto options to a third currency. A calibration procedure to fit the skew of both underlying securities is presented. The action of the perturbed copula is interpreted compared to the Gaussian copula. A real worked example is carried out comparing both copulas and a local volatility model with constant correlation for varying maturities, correlations and skew configurations.
\end{abstract}

\section{Introduction}
\label{sec:Introduction}

Copula models arise in the market place when only quoted information about the behaviour of single assets is available but nothing or very little is known about their joint relations. Products that depend on several assets allow to imply from the market some information about their joint behaviour. However, there is a lot of market information embedded in just a single price and therefore, it is necessary to make some assumptions about the joint relations in order to imply some meaningful parameters out of market prices. On the other hand, when products which depend on several assets are not available in the market or they are not liquid enough, it is necessary to make some assumptions about the joint relations of their underlying assets in order to describe them in a simple and intuitive way through some parameters whose values might be given as input. Then, the dependence and sensitivity of a given product to these ``unobserved" parameters allow to know the risk associated with them and allows to take a conservative position for trading and managing them. Most of the assumptions of co-dependence among assets can be highly simplified through a copula model. One of the most popular contexts in which copula models have been used is credit (e.g. the valuation of Colateralized Debt Obligations). Another popular application appears in hybrid models which combine two different asset classes for which co-dependence information is not available. The particular application addressed in this research work is FX (foreign exchange) quanto options to a third currency different from the two currencies of the underlying FX pair of the option.

A copula model allows to obtain a joint probability distribution of two random variables when only their marginal distributions are known. As it is clearly explained in \cite{Meucci2005}, the joint distribution of two random variables can be cleanly decomposed into two separate contributions: the distribution of the co-dependence (the copula function) and the marginal distribution of each random variable. This means that the copula function embeds completely the co-dependence information. The gaussian copula is the most popular, well-known and widely used copula model in the industry. It has become a reference proxy in the industry for it is analytical, easy to use and provides intuition. The main assumption of the gaussian copula is to consider that the random variables are normally distributed and their joint distribution is multi-variate normal. The copula function would only depend on the correlation matrix of the multivariate-normal distribution. This allows to embed the whole information of the co-dependence in just the correlation parameter. This hypothesis might not be reasonable when the distribution of the underlyings is skewed (the slope of the volatility surface with respect to strike is different from zero) or in the credit context, the probability of extreme events (the tail of the gaussian distribution) is too low.

The perturbed gaussian copula comes into play to improve the gaussian copula under the assumption that the two random variables have a common fast mean reverting stochastic volatility factor. This hypothesis allows to introduce the skew effect in the co-dependence of the two random variables. The joint distribution of the random variables is approximated through an asymptotic expansion to first order calculated using perturbation theory under the asumption that the common stochastic volatility factor is fast mean-reverting. This allows to obtain an analytical expression for the joint and marginal distributions of both random variables and therefore the copula funtion (the joint density divided by the product of the two marginal densities). These ``pertubed" marginal distributions will in general be different to the empirical ones obtained from the market. However, the closer these ``pertubed" marginal distributions are to the empirical ones, the better their co-dependence will be modelled by the copula.

The general formulation of the perturbed copula has been simplified so that the whole information of the co-dependence of two random variables is condensed in just five parameters with a very intuitive interpretation. The skew information of each random variable is introduced through two parameters: one controls the volatility level and the other the slope of the volatility with respect to strike. Finally, the remaining parameter related to the co-dependence is the traditional correlation used for the gaussian copula. The current formulation of the perturbed copula only reflects properly the skew effect (the slope of the volatility with respect to the strike). In order to incorporate the smile effect (slope and convexity) it would be necessary to continue the asymptotic expansion up to order two. That's why only two parameters (level and slope) are added to each of the random variables in addition to the tradicional correlation parameter.

The contribution of this work is the generalization of the original perturbed copula \cite{Fouque2006} to cope with lognornal-inspired underlyings, the simplification of the ``perturbed copula" general formulation in terms of only five easy-to-interpret parameters, their calibration to market data, the application to a concrete case (FX quanto options to a third currency), the interpretation of the action of the perturbed copula compared to the gaussian copula and the comparison of both models with some market standard such the local volatility model for a set of scenarios varying market data, moneyness, maturity and correlation.

The organization of this paper starts with the generalization of the original perturbed copula approach to cope with lognormal-inspired underlyings (section \ref{sec:Formulation}). Section \ref{sec:Reduction2IntuitiveParameters} reduces the general formulation through some additional hypothesis to only depend on five easy-to-interpret parameters. Section \ref{sec:Calibration} shows the calibration procedure of the perturbed copula parameters to market data based on a regular Newton-Raphson. Initial parameters based on analytical formulas based on approximations calculated through asymptotic expansions are derived in appendix \ref{app:InitialParams}. The interpretation of the action of the perturbed copula compared to the gaussian copula is presented in section \ref{sec:Interpretation}. To justify the interpretation, both copulas and a Monte Carlo method with local volatility and constant correlation are compared among each other through a collection of scenarios pricing FX quanto options to a third currency varying skew and correlation. Section \ref{sec:CaseStudy} applies the same tests to a real market scenario. Section \ref{sec:Conclusions} concludes.

\section{Perturbed copula formulation}
\label{sec:Formulation}

Consider the process $( {S_t^{(1)} ,S_t^{(2)} ,Y_t } )$ which follows the dynamics of equation (\ref{eq:PerCopLogModel}), where $W_t^{(1)}$, $W_t^{(2)}$ and $W_t^{(Y)}$ are standard Brownian motions correlated with correlations given by equation (\ref{eq:PerCopCorr}), $\alpha_t^{(i)}$ (i=1,2) are the drifts of $S_t^{(i)}$, $m$ is the long term value of $Y_t$, $\nu$ is a parameter which controls the volatility of the process $Y_t$ and $\varepsilon$ is a small constant ($\varepsilon << 1$) which is the inverse of the mean reversion speed (the smaller $\varepsilon$, the faster mean reversion).

\begin{equation}
\begin{array}{l}
 dS_t^{(1)}  = \alpha _t^{(1)} S_t^{(1)} dt + S_t^{(1)} f_1 \left( {Y_t } \right)dW_t^{(1)}  \\ 
 dS_t^{(2)}  = \alpha _t^{(2)} S_t^{(2)} dt + S_t^{(2)} f_2 \left( {Y_t } \right)dW_t^{(2)}  \\ 
 dY_t  = \frac{1}{\varepsilon }\left( {m - Y_t } \right)dt + \frac{{\nu \sqrt 2 }}{{\sqrt \varepsilon  }}dW_t^{(Y)}  \\ 
 \end{array}
  \label{eq:PerCopLogModel}
\end{equation}

\begin{equation}
\begin{array}{*{20}c}
   {d({W_t^{(1)} ,W_t^{(2)} }) = \rho dt} & {d( {W_t^{(1)} ,W_t^{(Y)} }) = \rho _{1Y} dt} & {d( {W_t^{(2)} ,W_t^{(Y)} }) = \rho _{2Y} dt} \\
\end{array}
   \label{eq:PerCopCorr}
\end{equation}

The processes $S_t^{(1)}$ and $S_t^{(2)}$ represent two correlated lognormal-inspired underyings. The functions $f_i(Y_t)$ are the volatilities of the underlyings. They depend on a common stochastic volatility factor $Y_t$. The correlations $\rho_{iY}$ control the slope of the skew of each underlying and the parameter $\nu$ controls the convexity of the smile. Doing the change of variables $X_t^{(i)} = \ln(S_t^{(i)})$ gives the usual normal-inspired processes given by equation (\ref{eq:PerCopModel}).

\begin{equation}
\begin{array}{l}
 dX_t^{(1)}  = \left( {\alpha _t^{(1)}  - \frac{1}{2}f_1^2 \left( {Y_t } \right)} \right)dt + f_1 \left( {Y_t } \right)dW_t^{(1)}  \\ 
 dX_t^{(2)}  = \left( {\alpha _t^{(2)}  - \frac{1}{2}f_2^2 \left( {Y_t } \right)} \right)dt + f_2 \left( {Y_t } \right)dW_t^{(2)}  \\ 
 dY_t  = \frac{1}{\varepsilon }\left( {m - Y_t } \right)dt + \frac{{\nu \sqrt 2 }}{{\sqrt \varepsilon  }}dW_t^{(Y)}  \\ 
 \end{array}
  \label{eq:PerCopModel}
\end{equation}

The process $Y_t$ is an Ornstein-Uhlenbeck process which is ergodic and therefore it has a stationary probability distribution (see section 3.2.3 of \cite{Fouque2000} for more information). This means that there exists a probability distribution so that the expectation of an arbitrary function $g$ of $Y_t$ does not depend on time ($\frac{d}{{dt}}\left( {{\bf E}\left[ {g(Y_t )} \right]} \right) = 0$). This stationary distribution is reached after some amount of time depending on the parameters of the process. For the Ornstein-Uhlenbeck process $Y_t$ of equation (\ref{eq:PerCopModel}), this invariant distribution is $N(m,\nu^2)$ (normal with mean $m$ and standard deviation $\nu$). See that changing the parameter $\varepsilon$ is equivalent to do a time scale change (the drift term is multiplied by $1/\varepsilon$ and the diffusion term by the square root of time). Increasing $\varepsilon$ is equivalent to compress the time scale and thus the result is a faster mean reversion.

The objective is to derive the joint and marginal transition probability distributions $u^{\varepsilon}$, $u_1^{\varepsilon}$ and $u_2^{\varepsilon}$ from an initial point ${\bf x} = (x_1,x_2)$ to the end point $(\xi_1,\xi_2)$, considered constant in this development. These transition probability distributions are given by equation (\ref{eq:JointMargProb}), where ${\bf X}_t = (X_t^{(1)}X_t^{(2)})$. These probability distributions are given by the dynamics of equation (\ref{eq:PerCopModel}).

\begin{equation}
\begin{array}{l}
 u^\varepsilon   = P\left( {X_T^{(1)}  \in d\xi _1 ,X_T^{(2)}  \in d\xi _2 |{\bf X}_t  = {\bf x},Y_t  = y} \right) \\ 
 v_1^\varepsilon   = P\left( {X_T^{(1)}  \in d\xi _1 |{X}_t^{(1)}  = {x_1},Y_t  = y} \right) \\ 
 v_2^\varepsilon   = P\left( {X_T^{(2)}  \in d\xi _2 |{X}_t^{(2)}  = {x_2},Y_t  = y} \right) \\ 
 \end{array}
   \label{eq:JointMargProb}
\end{equation}

Applying Ito's Lemma to a function which depends on $t$, $X_t^{(1)}$, $X_t^{(2)}$ and $Y_t$ and setting the drift term to zero (to impose that the evolution of the function is a martingale) yields the Kolmogorov backward equation given by (\ref{eq:PDEprob}), where the operator $\mathcal{L}_{\varepsilon}$ is given by equation (\ref{eq:Lepsilon}) in terms of $\mathcal{L}_0$ (the infinitesimal generator of the $Y_t$ process), $\mathcal{L}_1$ (the crossed correlation terms of processes $X_t^{(1)}$ and $X_t^{(2)}$ with $Y_t$) and $\mathcal{L}_2$ (the infinitesimal generator of the joint evolution of $X_t^{(1)}$ and $X_t^{(2)}$). These operators are given by equations (\ref{eq:L0}) to (\ref{eq:L2}). The terminal condition for the joint probatility density function $u^{\varepsilon}$ is a Dirac delta function $\delta(\xi_i;x_i)$ with the spike at $\xi _i = x_i$.

\begin{equation}
\begin{array}{l}
 \mathcal{L}^\varepsilon  u^\varepsilon  (t,x_1 ,x_2 ,y) = 0 \\ 
 u^\varepsilon  (T,x_1 ,x_2 ,y) = \delta (\xi _1 ;x_1 )\delta (\xi _2 ;x_2 ) \\ 
 \end{array}
  \label{eq:PDEprob}
\end{equation}

\begin{equation}
\mathcal{L}^\varepsilon   = \frac{1}{\varepsilon }\mathcal{L}_0  + \frac{1}{{\sqrt \varepsilon  }}\mathcal{L}_1  + \mathcal{L}_2 
  \label{eq:Lepsilon}
\end{equation}

\begin{equation}
\mathcal{L}_0  = (m - y)\frac{\partial }{{\partial y}} + \nu ^2 \frac{{\partial ^2 }}{{\partial y^2 }}
  \label{eq:L0}
\end{equation}

\begin{equation}
\mathcal{L}_1  = \nu \sqrt 2 \rho _{1Y} f_1 (y)\frac{{\partial ^2 }}{{\partial x_1 \partial y}} + \nu \sqrt 2 \rho _{2Y} f_2 (y)\frac{{\partial ^2 }}{{\partial x_2 \partial y}}
  \label{eq:L1}
\end{equation}

\begin{equation}
\begin{array}{l}
\mathcal{L}_2  = \frac{\partial }{{\partial t}} + \frac{1}{2}f_1^2 \left( y \right)\frac{{\partial ^2 }}{{\partial x_1^2 }} + \frac{1}{2}f_2^2 \left( y \right)\frac{{\partial ^2 }}{{\partial x_2^2 }} + \rho f_1 \left( y \right)f_2 \left( y \right)\frac{{\partial ^2 }}{{\partial x_1 \partial x_2 }} \\ 
 \begin{array}{*{20}c}
   {} & {} & {}  \\
\end{array} + \left( {\alpha _t^{(1)}  - \frac{1}{2}f_1^2 \left( y \right)} \right)\frac{\partial }{{\partial x_1 }} + \left( {\alpha _t^{(2)}  - \frac{1}{2}f_2^2 \left( y \right)} \right)\frac{\partial }{{\partial x_2 }} \\ 
 \end{array}
  \label{eq:L2}
\end{equation}

The solution of equation (\ref{eq:PDEprob}) is now expanded in powers of $\sqrt{\varepsilon}$ according to equation (\ref{eq:Expansion}). The approximation of the solution will only consider up to order 1 ($u^{\varepsilon}=u_0 + \sqrt{\varepsilon} u_1$). If the solution (\ref{eq:Expansion}) is substituted in equation (\ref{eq:PDEprob}) and the terms of the same order are grouped (terms multiplying $\frac{1}{\varepsilon}$, $\frac{1}{\sqrt{\varepsilon}}$, 1, $\sqrt{\varepsilon}$ and so on), equation (\ref{eq:ExpandedEq}) us obtained. If each term is set to zero independently, the system of partial differential equations of equation (\ref {eq:SystemEq}) is obtained. This system is easier to solve than equation (\ref{eq:PDEprob}).

\begin{equation}
u^\varepsilon   = u_0  + \sqrt \varepsilon  u_1  + \varepsilon u_2  + \varepsilon ^{3/2} u_3  +  \cdots 
  \label{eq:Expansion}
\end{equation}

\begin{equation}
\begin{array}{l}
 \frac{1}{\varepsilon }\mathcal{L}_0 u_0  + \frac{1}{{\sqrt \varepsilon  }}\left( {\mathcal{L}_0 u_1  + \mathcal{L}_1 u_0 } \right) + \left( {\mathcal{L}_0 u_2  + \mathcal{L}_1 u_1  + \mathcal{L}_2 u_0 } \right) +  \\ 
  \begin{array}{*{20}c}
   {} & {} & {}  \\
	\end{array}
+ \sqrt \varepsilon  \left( {\mathcal{L}_0 u_3  + \mathcal{L}_1 u_2  + \mathcal{L}_2 u_1 } \right) +  \cdots  = 0 \\ 
 \end{array}
  \label{eq:ExpandedEq}
\end{equation}

\begin{equation}
\left\{ \begin{array}{l}
 \mathcal{L}_0 u_0  = 0 \\ 
 \mathcal{L}_0 u_1  + \mathcal{L}_1 u_0  = 0 \\ 
 \mathcal{L}_0 u_2  + \mathcal{L}_1 u_1  + \mathcal{L}_2 u_0  = 0 \\ 
 \mathcal{L}_0 u_3  + \mathcal{L}_1 u_2  + \mathcal{L}_2 u_1  = 0 \\ 
 \end{array} \right.
  \label{eq:SystemEq}
\end{equation}

As the operator $\mathcal{L}_0$ of the first equation of the system ($\mathcal{L}_0 u_0 = 0$) is ergodic (it is the operator associated with the Ornstein-Uhlenbeck process) and acts on $y$, the solution is constant with respect to $y$, $u_0 = u_0(t,x_1,x_2)$ (see section 3.2.4 of \cite{Fouque2000}). From the second equation of the system (\ref{eq:SystemEq}) ($\mathcal{L}_0 u_1  + \mathcal{L}_1 u_0  = 0$), as $\mathcal{L}_1$ takes derivatives with respect to $y$ and $u_0$ does not depend on $y$, it implies that $\mathcal{L}_1 u_0 = 0$. Therefore, $\mathcal{L}_0 u_1 = 0$. As again $\mathcal{L}_0$ is ergodic and acts on $y$, $u_1$ must be constant with respect to $y$ ($u_1 = u_1(t,x_1,x_2)$). This implies that the combination of the first two terms ($u_0 + \sqrt{\epsilon} u_1$) is independent of the volatility factor $y$.

The third equation of the system (\ref{eq:SystemEq}), reduces to equation (\ref{eq:OrderZeroPDE}) as $\mathcal{L}_1 u_1 = 0$, because $\mathcal{L}_1$ takes derivatives on $y$ and $u_1$ does not depend on $y$. This last equation is a Poisson equation with respect to the operator $\mathcal{L}_0$ in the variable $y$. The general form of this equation is given by (\ref{eq:PoissonEqDef}). For this equation to have a solution, the operator must be ergodic (with an invariant distribution) and the function $g(y)$ must be centered with respect to the invariant distribution of the operator. The invariant distribution of the operator $\mathcal{L}_0$ is a normal distribution $N(m,\nu^2)$ and the centering condition reduces to $\left\langle g \right\rangle  = 0$, where the notation $\left\langle \cdot \right\rangle$ means the expectation with respect to the invariant distribution as given by equation (\ref{eq:ExpInvariant}).

\begin{equation}
\mathcal{L}_0 u_2 + \mathcal{L}_2 u_0  = 0
  \label{eq:OrderZeroPDE}
\end{equation}

\begin{equation}
\mathcal{L}_0 \chi + g(y) = 0
  \label{eq:PoissonEqDef}
\end{equation}

\begin{equation}
\left\langle g^2 \right\rangle  = \int_{ - \infty }^\infty  {g(y)^2\frac{1}{{\nu \sqrt {2\pi } }}\exp \left( { - \frac{{(y - m)^2 }}{{2\nu ^2 }}} \right)dy} 
  \label{eq:ExpInvariant}
\end{equation}

The centering condition for equation (\ref{eq:OrderZeroPDE}), $\left\langle \mathcal{L}_2 u_0 \right\rangle = 0$, gives the zero order term of the solution $u_0$. As it does not depend on $y$, $u_0$ can be taken out of the expectation. Therefore, $u_0$ must satisfy the partial differential equation (\ref{eq:PDEu0}).

\begin{equation}
\begin{array}{l}
 \left\langle {\mathcal{L}_2 } \right\rangle u_0 (t,x_1 ,x_2 ) = 0 \\ 
 u_0 (T,x_1 ,x_2 ) = \delta (\xi _1 ;x_1 )\delta (\xi _2 ;x_2 ) \\ 
 \end{array}
  \label{eq:PDEu0}
\end{equation}

Taking the expectation $\left\langle \cdot \right\rangle$ over the operator $\mathcal{L}_2$ means calculating the expectation of each of the coefficients of its partial derivative components as they do not take derivatives with respect to $y$, the variable over which the expectation is computed. Equation (\ref{eq:SigDash}) defines the effective volatilities $\bar \sigma_1$ and $\bar \sigma_2$ and the effective correlation $\bar \rho$ which appear when this expectation is calculated. If the effective parameters are substituted in equation (\ref{eq:PDEu0}), equation (\ref{eq:PDEu02}) is obtained. This last equation depends only on the constant effective parameters and it is well known. The solution is the transition joint density function of two correlated brownian motions, scaled by the effective volatilities and with an instantaneous correlation equal to $\bar \rho$. The solution is given by equation (\ref{eq:u0}) where $\tilde x_i$ is given by equation (\ref{eq:xtilde_i}). It gives the transition density of starting at $x_1$ and $x_2$ at $t$ and ending at $\xi _1$ and $\xi _2$ at $T$. The volatilities of the two underlying brownian motions are now constant and equal to the effective volatilities and the instantaneous correlation is constant and equal to the effective correlation. See that the zero order term does not depend at all on the stochastic factor $Y_t$. It only provides the contribution of the average volatility of each random variable (the effective volatilities) and the average correlation.

\begin{equation}
\bar \sigma _1  = \sqrt {\left\langle {f_1^2 } \right\rangle } \begin{array}{*{20}c}
   {} & {}  \\
\end{array}\bar \sigma _2  = \sqrt {\left\langle {f_2^2 } \right\rangle } \begin{array}{*{20}c}
   {} & {}  \\
\end{array}\bar \rho  = \frac{{\rho \left\langle {f_1 f_2 } \right\rangle }}{{\bar \sigma _1 \bar \sigma _2 }}
  \label{eq:SigDash}
\end{equation}

\begin{equation}
\begin{array}{l}
 \frac{{\partial u_0 }}{{\partial t}} + \bar \rho \bar \sigma _1 \bar \sigma _2 \frac{{\partial ^2 u_0 }}{{\partial x_1 \partial x_2 }} + \sum\limits_{i = 1}^2 {\left[ {\left( {\alpha _t^{(i)}  - \frac{1}{2}\bar \sigma _i^2 } \right)\frac{{\partial u_0 }}{{\partial x_i }} + \frac{1}{2}\bar \sigma _i^2 \frac{{\partial ^2 u_0 }}{{\partial x_i^2 }}} \right]}  = 0 \\ 
 u_0 \left( {T,x_1 ,x_2 } \right) = \delta \left( {\xi _1 ;x_1 } \right)\delta \left( {\xi _2 ;x_2 } \right) \\ 
 \end{array}
  \label{eq:PDEu02}
\end{equation}

\begin{equation}
u_0  = \frac{1}{{2\pi \bar \sigma _1 \bar \sigma _2 (T - t)\sqrt {1 - \bar \rho ^2 } }}e^{ - \frac{1}{{2(1 - \bar \rho ^2 )}}\left( {\frac{{(\xi _1  - \tilde x_1 )^2 }}{{\bar \sigma _1^2 (T - t)}} - 2\bar \rho \frac{{(\xi _1  - \tilde x_1 )(\xi _2  - \tilde x_2 )}}{{\bar \sigma _1 \bar \sigma _2 (T - t)}} + \frac{{(\xi _2  - \tilde x_2 )^2 }}{{\bar \sigma _2^2 (T - t)}}} \right)} 
  \label{eq:u0}
\end{equation}

\begin{equation}
\tilde x_i  = x_i  + \int_t^T {\alpha _s^{(i)} ds - \frac{1}{2}} \bar \sigma _i^2 (T - t)
  \label{eq:xtilde_i}
\end{equation}

As the centering condition $\left\langle \mathcal{L}_2 u_0 \right\rangle = 0$ is satisfied, $\mathcal{L}_2 u_0$ is given by equation (\ref{eq:CenteredL2}) using the definition of the operator (\ref{eq:L2}) and taking into account that $u_0$ does not depend on $y$ (the variable over which the expectation is calculated).

\begin{equation}
\begin{array}{l}
 \mathcal{L}_2 u_0  = \mathcal{L}_2 u_0  - \left\langle {\mathcal{L}_2 u_0 } \right\rangle  = \frac{1}{2}\left( {f_1^2  - \left\langle {f_1^2 } \right\rangle } \right)\frac{{\partial ^2 u_0}}{{\partial x_1^2 }} + \frac{1}{2}\left( {f_2^2  - \left\langle {f_2^2 } \right\rangle } \right)\frac{{\partial ^2 u_0}}{{\partial x_2^2 }} \\ 
 \begin{array}{*{20}c}
   {}  \\
\end{array} + \rho \left( {f_1 f_2  - \left\langle {f_1 f_2 } \right\rangle } \right)\frac{{\partial ^2 u_0}}{{\partial x_1 \partial x_2 }} - \frac{1}{2}\left( {f_1^2  - \left\langle {f_1^2 } \right\rangle } \right)\frac{\partial u_0}{{\partial x_1 }} - \frac{1}{2}\left( {f_2^2  - \left\langle {f_2^2 } \right\rangle } \right)\frac{\partial u_0}{{\partial x_2 }} \\ 
 \end{array}
  \label{eq:CenteredL2}
\end{equation}

From equation (\ref{eq:OrderZeroPDE}) the second order order term $u_2$ is given by equation (\ref{eq:Solution_u2}), where $\mathcal{L}_0^{-1}$ is the inverse operator ($\phi  = \mathcal{L}^{ - 1} \left( H \right) \Leftrightarrow \mathcal{L}\phi  = H$) taken on the centered quantity $\mathcal{L}_2 - \left\langle {\mathcal{L} _2} \right\rangle$ given by equation (\ref{eq:CenteredL2}).

\begin{equation}
u_2 (t,x_1 ,x_2 ) = - \mathcal{L}_0^{ - 1} \left( {\mathcal{L}_2  - \left\langle {\mathcal{L}_2 } \right\rangle } \right)u_0 
  \label{eq:Solution_u2}
\end{equation}

The last equation of the system (\ref{eq:SystemEq}) is again a Poisson equation whose centering condition is $\left\langle {\mathcal{L}_2 u_1  + \mathcal{L}_1 u_2 } \right\rangle = 0$. Replacing $u_2$ by equation (\ref{eq:Solution_u2}) in the latter centering condition and taking into account that $u_0$ does not depend on $y$ yields the partial differential equation (\ref{eq:PDEu1}) which gives the first order term of the solution. The initial condition imposed on this term is zero because the initial condition of the solution was already fitted by the zero order term.

\begin{equation}
\begin{array}{l}
 \left\langle {\mathcal{L}_2 } \right\rangle u_1 (t,x_1 ,x_2 ) = \left\langle {\mathcal{L}_1 \mathcal{L}_0^{ - 1} \left( {\mathcal{L}_2  - \left\langle {\mathcal{L}_2 } \right\rangle } \right)} \right\rangle u_0  \\ 
 u_1 (T,x_1 ,x_2 ) = 0 \\ 
 \end{array}
  \label{eq:PDEu1}
\end{equation}

Consider now the functions $\phi_1(y)$, $\phi_2(y)$ and $\phi_{12}(y)$ as the solutions of the Poisson equations (\ref{eq:PoissonEq}). See that they are already centered and therefore they have a solution. These functions are defined up to an additive constant $c(t,x_1,x_2)$ with respect to $y$. This constant will be eliminated after applying the operator $\mathcal{L}_1$ later on in equation (\ref{eq:L1invL0}) as this operator takes derivatives with respect to $y$.

\begin{equation}
\begin{array}{l}
 \mathcal{L}_0 \phi _1 (y) = f_1^2 (y) - \left\langle {f_1^2 } \right\rangle  \\ 
 \mathcal{L}_0 \phi _2 (y) = f_2^2 (y) - \left\langle {f_2^2 } \right\rangle  \\ 
 \mathcal{L}_0 \phi _{12} (y) = f_1 (y)f_2 (y) - \left\langle {f_1 f_2 } \right\rangle  \\ 
 \end{array}
  \label{eq:PoissonEq}
\end{equation}

Applying the inverse operator $\mathcal{L}_0^{-1}$ to equation (\ref{eq:CenteredL2}) and taking into account the solutions of the Poisson equations (\ref{eq:PoissonEq}), yields equation (\ref{eq:invL0}). If the operator $\mathcal{L}_1$ is now applied to equation (\ref{eq:invL0}), equation (\ref{eq:L1invL0}) is obtained, where $A=\mathcal{L}_0^{ - 1} \left( {\mathcal{L}_2  - \left\langle {\mathcal{L}_2 } \right\rangle } \right)$.

\begin{equation}
\begin{array}{l}
 \mathcal{L}_0^{ - 1} \left( {\mathcal{L}_2  - \left\langle {\mathcal{L}_2 } \right\rangle } \right)u_0  = \frac{1}{2}\phi _1 \left( y \right)\frac{{\partial ^2 u_0 }}{{\partial x_1^2 }} + \frac{1}{2}\phi _2 \left( y \right)\frac{{\partial ^2 u_0 }}{{\partial x_2^2 }} + \rho \phi _{12} \frac{{\partial ^2 u_0 }}{{\partial x_1 \partial x_2 }} \\ 
 \begin{array}{*{20}c}
   {} & {} & {}  \\
\end{array} - \frac{1}{2}\phi _1 \left( y \right)\frac{{\partial u_0 }}{{\partial x_1 }} - \frac{1}{2}\phi _2 \left( y \right)\frac{{\partial u_0 }}{{\partial x_2 }} \\ 
 \end{array}
  \label{eq:invL0}
\end{equation}

\begin{equation}
\begin{array}{l}
 \mathcal{L}_1 Au_0  =  \\ 
  + \nu \frac{{\sqrt 2 }}{2}\rho _{1Y} f_1 \left( {\phi _1 '\frac{{\partial ^3 u_0 }}{{\partial x_1^3 }} + \phi _2 '\frac{{\partial ^3 u_0 }}{{\partial x_1 \partial x_2^2 }} + 2\rho \phi _{12} '\frac{{\partial ^3 u_0 }}{{\partial x_1^2 \partial x_2 }} - \phi _1 '\frac{{\partial ^2 u_0 }}{{\partial x_1^2 }} - \phi _2 '\frac{{\partial ^2 u_0 }}{{\partial x_1 \partial x_2 }}} \right) \\ 
  + \nu \frac{{\sqrt 2 }}{2}\rho _{2Y} f_2 \left( {\phi _1 '\frac{{\partial ^3 u_0 }}{{\partial x_1^2 \partial x_2 }} + \phi _2 '\frac{{\partial ^3 u_0 }}{{\partial x_2^3 }} + 2\rho \phi _{12} '\frac{{\partial ^3 u_0 }}{{\partial x_1 \partial x_2^2 }} - \phi _1 '\frac{{\partial ^2 u_0 }}{{\partial x_1 \partial x_2 }} - \phi _2 '\frac{{\partial ^2 u_0 }}{{\partial x_2^2 }}} \right) \\ 
 \end{array}
  \label{eq:L1invL0}
\end{equation}

Taking expectations of $\mathcal{L}_1 A$ with respect to $y$ to obtain the right hand side of equation (\ref{eq:PDEu1}) and multiplying by $\sqrt{\varepsilon}$ yields equation (\ref{eq:sqrtEpsilonA}), where the constants $R_1$, $R_2$, $R_{12}$, $R_{21}$, $Q_{12}$ and $Q_{21}$ are given by equation (\ref{eq:Rij}) (see that they all are of the order of $\sqrt{\varepsilon}$). This constants are small and will be calibrated from market data (explicit expresions of $f_i$, $\phi '_i$ will not be necessary).

\begin{equation}
\begin{array}{l}
 \sqrt \varepsilon  \left\langle {\mathcal{L}_1 A} \right\rangle  u_0 = R_1 \left( {\frac{{\partial ^3 u_0 }}{{\partial x_1^3 }} - \frac{{\partial ^2 u_0 }}{{\partial x_1^2 }}} \right) + R_2 \left( {\frac{{\partial ^3 u_0 }}{{\partial x_2^3 }} - \frac{{\partial ^2 u_0 }}{{\partial x_2^2 }}} \right) \\ 
 \begin{array}{*{20}c}
   {} & {} & {}  \\
\end{array} + R_{12} \frac{{\partial ^3 u_0 }}{{\partial x_1 \partial x_2^2 }} + R_{21} \frac{{\partial ^3 u_0 }}{{\partial x_1^2 \partial x_2 }} - \left( {Q_{12}  + Q_{21} } \right)\frac{{\partial ^2 u_0 }}{{\partial x_1 \partial x_2 }} \\ 
 \end{array}
  \label{eq:sqrtEpsilonA}
\end{equation}

\begin{equation}
\begin{array}{l}
 R_1  = \frac{{\nu \rho _{1Y} \sqrt \varepsilon  }}{{\sqrt 2 }}\left\langle {f_1 \phi '_1 } \right\rangle  \\ 
 R_2  = \frac{{\nu \rho _{2Y} \sqrt \varepsilon  }}{{\sqrt 2 }}\left\langle {f_2 \phi '_2 } \right\rangle  \\ 
 R_{12}  = \frac{{\nu \rho _{1Y} \sqrt \varepsilon  }}{{\sqrt 2 }}\left\langle {f_1 \phi '_2 } \right\rangle  + \nu \sqrt {2\varepsilon } \rho \rho _{2Y} \left\langle {f_2 \phi '_{12} } \right\rangle  \\ 
 R_{21}  = \frac{{\nu \rho _{2Y} \sqrt \varepsilon  }}{{\sqrt 2 }}\left\langle {f_2 \phi '_1 } \right\rangle  + \nu \sqrt {2\varepsilon } \rho \rho _{1Y} \left\langle {f_1 \phi '_{12} } \right\rangle  \\ 
 Q_{12}  = \frac{{\nu \rho _{1Y} \sqrt \varepsilon  }}{{\sqrt 2 }}\left\langle {f_1 \phi _2 '} \right\rangle  \\ 
 Q_{21}  = \frac{{\nu \rho _{2Y} \sqrt \varepsilon  }}{{\sqrt 2 }}\left\langle {f_2 \phi _1 '} \right\rangle  \\ 
 \end{array}
  \label{eq:Rij}
\end{equation}

Taking into account the definition of $\left\langle {\mathcal{L}_2 } \right\rangle$ and that equation (\ref{eq:L2Derivatives}) is satisfied ($j+k = n$), it can be checked directly that the solution of equation (\ref{eq:PDEu1}) is given by equation (\ref{eq:FirstOrderSol}) and therefore, the first order correction is given by equation (\ref{eq:FirstOrderSolexpanded}).

\begin{equation}
\left\langle {\mathcal{L}_2 } \right\rangle \frac{{\partial ^n u_0 }}{{\partial x_1^j \partial x_1^k }} = \frac{{\partial ^n }}{{\partial x_1^j \partial x_1^k }}\left\langle {\mathcal{L}_2 } \right\rangle u_0  = 0
  \label{eq:L2Derivatives}
\end{equation}

\begin{equation}
u_1  =  - \left( {T - t} \right)\left\langle \mathcal{L}_1 A \right\rangle u_0 
  \label{eq:FirstOrderSol}
\end{equation}

\begin{equation}
\begin{array}{l}
 \sqrt \varepsilon  u_1  =  - \left( {T - t} \right)\left\{ {R_1 \left( {\frac{{\partial ^3 u_0 }}{{\partial x_1^3 }} - \frac{{\partial ^2 u_0 }}{{\partial x_1^2 }}} \right) + R_2 \left( {\frac{{\partial ^3 u_0 }}{{\partial x_2^3 }} - \frac{{\partial ^2 u_0 }}{{\partial x_2^2 }}} \right)} \right. \\ 
 \begin{array}{*{20}c}
   {} & {} & {}  \\
\end{array}\left. { + R_{12} \frac{{\partial ^3 u_0 }}{{\partial x_1 \partial x_2^2 }} + R_{21} \frac{{\partial ^3 u_0 }}{{\partial x_1^2 \partial x_2 }} - \left( {Q_{12}  + Q_{21} } \right)\frac{{\partial ^2 u_0 }}{{\partial x_1 \partial x_2 }}} \right\} \\ 
 \end{array}
  \label{eq:FirstOrderSolexpanded}
\end{equation}
	
It is not guaranteed that the asymptotic approximation for the joint density function will neither be positive nor integrate 1 in the whole domain. To get positive densities the multiplicative expression $\hat u_0 \left( {1 + \tanh (\sqrt \varepsilon  \hat u_1 )} \right)$ is used instead of $u_0 + \sqrt{\epsilon} u_1$. The first order Taylor expansions of both expressions are matched leading to $\hat u_0 = u_0$ and $\hat u_1 = u_1/u_0$. This procedure is explained in \cite{Fouque2006} and it is also shown that the multiplicative approximation has the same accuracy than the additive approximation. Therefore, the final solution for the joint probability density function is given by equation (\ref{eq:FirstOrderSolfinal}), where $u_0$ and $\sqrt{\varepsilon}u_1$ are respectively given by equations (\ref{eq:u0}) and (\ref{eq:FirstOrderSolexpanded}) and $W$ is the normalizing constant so that the density integrates 1 in the whole domain. In the context of \cite{Fouque2006}, the multiplicative approximation also guarantees that the density integrates 1 in the whole domain. Unfortunately, the presence of the second partial derivatives (even functions) in (\ref{eq:FirstOrderSolexpanded}) makes the density (\ref{eq:FirstOrderSolfinal}) not integrate 1 and the normalizing constant $W$ has to be introduced.  Explicit expressions for the third partial derivatives of $u_0$ can be found in appendix A of \cite{Fouque2006}.

\begin{equation}
u^\varepsilon(t,x_1 ,x_2 ;T,\xi _1 ,\xi _2 )   = \frac{1}{W} u_0 \left\{ {1 + \tanh \left( {\frac{{\sqrt \varepsilon  u_1 }}{{u_0 }}} \right)} \right\}
  \label{eq:FirstOrderSolfinal}
\end{equation}

For the marginal transition density functions $v_1^{\varepsilon}$ and $v_2^{\varepsilon}$, the same argument used to obtain the solution for the joint distribution can be applied. The solution is given by equation (\ref{eq:vi}) where the zero order approximation is a regular normal density function given by equation (\ref{eq:pi}), where $\tilde x_i$ is again given by (\ref{eq:xtilde_i}) and $W_i$ is the normalizing constant so that the marginal density integrates 1 in the whole domain.

\begin{equation}
v_i^\varepsilon(t,x_1;T,\xi _i)   = \frac{1}{W_i} p_i \left[ {1 + \tanh \left( { - (T - t)\frac{{R_i }}{{p_i }} \left\{ \frac{{\partial ^3 p_i }}{{\partial x_i^3 }} - \frac{{\partial ^2 p_i }}{{\partial x_i^2 }} \right\} } \right)} \right]
  \label{eq:vi}
\end{equation}

\begin{equation}
p_i (t,x_i ,T,\xi _i ) = \frac{1}{{\bar \sigma _i \sqrt {2\pi (T - t)} }}\exp \left( { - \frac{{\left( {\xi _i  - \tilde x_i } \right)^2 }}{{2\bar \sigma _i^2 \left( {T - t} \right)}}} \right)
  \label{eq:pi}
\end{equation}

The perturbed copula function will be given by the ratio of the joint density and the product of the two marginal densities as shown by equation (\ref{eq:fcop}), where $z_1$ and $z_2$ are related to $\xi _1$ and $\xi _2$ by equation (\ref{eq:zi}). The variables $z_i$ (they are in the interval $[0,1]$) are the marginal cummulative probabilities of $\xi_i$. The copula function is expressed in terms of $z_i$ and for each of those values, the values of $\xi_i$ should be calculated through the quantile function (the inverse of the cummulative density function). It is proved in \cite{Fouque2006} that $f_{cop}$ is indeed a copula function.

\begin{equation}
f_{cop} \left( {z_1 ,z_2 } \right) = \frac{{u^\varepsilon  (t,x_1 ,x_2 ;T,\xi _1 ,\xi _2 )}}{{v_1^\varepsilon  (t,x_1 ,T,\xi _1 )v_2^\varepsilon  (t,x_2 ,T,\xi _2 )}}
  \label{eq:fcop}
\end{equation}

\begin{equation}
z_i  = P\left( {\left. {X_T^{(i)}  \le \xi _i } \right|{\bf X}_t  = {\bf x},Y_t  = y} \right)
  \label{eq:zi}
\end{equation}

See that for the purpose of calculating the copula density, it can be assumed that $\tilde x_i = 0$, because the quantile function provides a value $\xi_i = \tilde \xi_i + \tilde x_i$, where $\tilde \xi_i$ is the increment of $\xi_i$ with respect to the mean $\tilde x_i$. When this value is replaced in the terms $(\xi_i - \tilde x_i)$ of equations (\ref{eq:u0}) and (\ref{eq:pi}), the result is $(\xi_i - \tilde x_i) = \tilde \xi_i + \tilde x_i - \tilde x_i = \tilde \xi_i$. Therefore, if we had started with $\tilde \xi_i$ instead of $\xi_i$ the result would have been the same (for numerical reasons it is better to assume $\tilde x_i = 0$ for the purpose of calculating the copula function).
 
\section{Reduction to five intuitive parameters}
\label{sec:Reduction2IntuitiveParameters}

This section incorporates some hypothesis in order to express the perturbed copula in terms of just five easy-to-calibrate parameters with a clear interpretation. Equation (\ref{eq:fiRed}) shows the main hypothesis for this parameter reduction: the volatility dynamics of both random variables have the same dependence with respect to the common volatility factor $Y_t$; only the level of volatility might be different. This common dependence with respect to $Y_t$ is expressed through the function $g(y)$.

\begin{equation}
f_1 (y) = \sigma _1 g(y)\begin{array}{*{20}c}
   {} & {} & {}  \\
\end{array}f_2 (y) = \sigma _2 g(y)
  \label{eq:fiRed}
\end{equation}

If the functions $f_i$ from equation (\ref{eq:fiRed}) are replaced in equation (\ref{eq:SigDash}) and the new hypothesis $\left\langle {g^2 } \right\rangle  = 1$ is added, the expression for $\bar \sigma_i$ and $\bar \rho$ turns very simple into equations (\ref{eq:SigDashRed}) and (\ref{eq:rhoRed}).

\begin{equation}
\bar \sigma _1  = \sqrt {\left\langle {f_1^2 } \right\rangle }  = \sigma _1 \sqrt {\left\langle {g^2 } \right\rangle }  = \sigma _1 \begin{array}{*{20}c}
   {} & {}  \\
\end{array}\bar \sigma _2  = \sqrt {\left\langle {f_2^2 } \right\rangle }  = \sigma _2 \sqrt {\left\langle {g^2 } \right\rangle }  = \sigma _2 
  \label{eq:SigDashRed}
\end{equation}

\begin{equation}
\bar \rho  = \frac{{\rho \left\langle {f_1 f_2 } \right\rangle }}{{\bar \sigma _1 \bar \sigma _2 }} = \rho \frac{{\sigma _1 \sigma _2 \left\langle {g^2 } \right\rangle }}{{\sigma _1 \sqrt {\left\langle {g^2 } \right\rangle } \sigma _2 \sqrt {\left\langle {g^2 } \right\rangle } }} = \rho 
  \label{eq:rhoRed}
\end{equation}

Using the linearity property of the operators, the solutions of the three Poisson equations (\ref{eq:PoissonEq}), $\phi_1$, $\phi_2$ and $\phi_{12}$ can be expressed in terms of a single solution $\phi$ according to equation (\ref{eq:Phi_iRed}) which satisfies the Poisson equation (\ref{eq:PoissonEqRed}).

\begin{equation}
\phi _1 (y) = \sigma _1^2 \phi (y)\begin{array}{*{20}c}
   {} & {}  \\
\end{array}\phi _2 (y) = \sigma _2^2 \phi (y)\begin{array}{*{20}c}
   {} & {}  \\
\end{array}\phi _{12} (y) = \sigma _1 \sigma _2 \phi (y)
  \label{eq:Phi_iRed}
\end{equation}

\begin{equation}
\mathcal{L}_0 \phi (y) = g^2 (y) - \left\langle {g^2 } \right\rangle 
  \label{eq:PoissonEqRed}
\end{equation}

Then the parameters $R_1$, $R_2$, $R_{12}$, $R_{21}$, $Q_{12}$ and $Q_{21}$ from equation (\ref{eq:Rij}) turn into equation (\ref{eq:RijRed}), where $R_1$, $R_2$, $\sigma_1$ and $\sigma_2$ will be calibrated to market (see section \ref{sec:Calibration}) and $R_{12}$, $R_{21}$, $Q_{12}$, $Q_{21}$ will be calculated according to equation (\ref{eq:RijRed}).

\begin{equation}
\begin{array}{l}
 R_1  = \frac{{\nu \rho _{1Y} \sqrt \varepsilon  }}{{\sqrt 2 }}\sigma _1^3 \left\langle {g\phi '} \right\rangle  \\ 
 R_2  = \frac{{\nu \rho _{2Y} \sqrt \varepsilon  }}{{\sqrt 2 }}\sigma _2^3 \left\langle {g\phi '} \right\rangle  \\ 
 R_{12}  = \left( {\frac{{\nu \rho _{1Y} \sqrt \varepsilon  }}{{\sqrt 2 }}\sigma _1 \sigma _2^2  + \nu \sqrt {2\varepsilon } \rho \rho _{2Y} \sigma _2^2 \sigma _1 } \right)\left\langle {g\phi '} \right\rangle  = \left( {\frac{{\sigma _2 }}{{\sigma _1 }}} \right)^2 R_1  + 2\left( {\frac{{\sigma _1 }}{{\sigma _2 }}} \right)R_2 \rho  \\ 
 R_{21}  = \left( {\frac{{\nu \rho _{2Y} \sqrt \varepsilon  }}{{\sqrt 2 }}\sigma _2 \sigma _1^2  + \nu \sqrt {2\varepsilon } \rho \rho _{1Y} \sigma _1^2 \sigma _2 } \right)\left\langle {g\phi '} \right\rangle  = \left( {\frac{{\sigma _1 }}{{\sigma _2 }}} \right)^2 R_2  + 2\left( {\frac{{\sigma _2 }}{{\sigma _1 }}} \right)R_1 \rho  \\ 
 Q_{12}  = \frac{{\nu \rho _{1Y} \sqrt \varepsilon  }}{{\sqrt 2 }}\sigma _1 \sigma _2^2 \left\langle {g\phi '} \right\rangle  = \left( {\frac{{\sigma _2 }}{{\sigma _1 }}} \right)^2 R_1  \\ 
 Q_{21}  = \frac{{\nu \rho _{2Y} \sqrt \varepsilon  }}{{\sqrt 2 }}\sigma _2 \sigma _1^2 \left\langle {g\phi '} \right\rangle  = \left( {\frac{{\sigma _1 }}{{\sigma _2 }}} \right)^2 R_2  \\ 
 \end{array}
  \label{eq:RijRed}
\end{equation}

Section \ref{sec:Calibration} along with appendix \ref{app:InitialParams} will show that the marginal densities given by equation (\ref{eq:vi}) produce an implied volatility curve that is a straight line with respect to a given definition of moneyness. This line is defined by the slope and the intercept (two parameters). Therefore, the parameters $\sigma_i$ and $R_i$ ($i=1,2$) govern respectively the ATM (at-the-money) level and the skew slope for each of the marginal distributions. The parameter $\rho$ measures the correlation of the co-dependence (it has the same meaning for the gaussian copula). The parameters $\sigma_i = \bar \sigma_i$ govern the volatility level because they are the volatilities for both the joint and marginal density functions of equations (\ref{eq:u0}) and (\ref{eq:pi}) provided by the zero order term of the asymptotic approximation. The parameters $R_i$ govern the slope of the skew because for a fixed level of $\sigma_i$, they only depend on the correlation $\rho_{iY}$ and some common parameters ($\nu$, $\left\langle g\phi ' \right\rangle$ and $\sqrt{\varepsilon}$). The correlation $\rho_{iY}$ between the brownian motions driving the two random variables $W_t^{(i)}$ and the common factor $Y_t$ controls the slope of the skew. Finally, the parameter $\rho$ controls the correlation between both random variables because it is the correlation for the zero order term of the joint density function according to equation (\ref{eq:u0}). The hypothesis considered assume that the internal dynamics for both random variables are the same. However, the side and amount of skew is independent for each variable (it is controlled independently by each correlation $\rho_{iY}$).

\section{Calibration}
\label{sec:Calibration}

The calibration procedure involves finding the parameters $\sigma_i$ (implied volatility level) and $R_i$ (implied volatility slope) for both underlyings (2 degrees of freedom for each underlying). These values along with the correlation $\rho$ (which is not calibrated but input by the trader) will allow calculating the joint and marginal densities and therefore the copula function. The method proposed is an exact calibration to two vanilla options for each underlying using a simple Newton-Raphson algorithm. However, for the algorithm to converge, initial values close enough to the solution are needed. These initial values are calculated using asymptotic expansions for vanilla option prices similar to those described in section \ref{sec:Formulation} for the pertubed copula. These initial values are given by a calibration procedure described in \cite{Fouque2000}, but this calibration is carried out for processes formulated in the real market measure (rather than the risk free measure), assuming a risk premium different from zero. Therefore, in order to better relate the calibration procedure of \cite{Fouque2000} with the estimation of the initial parameters, this section formulates the process in the real measure.

Equation (\ref{eq:ProcProbReal}) presents the evolution of the underlying processes $S_t^{(i)}$ (for i = 1,2), where $\mu_t^{(i)}$ is the unknown market drift of the underlying, $q_t^{(i)}$ is the continuously compounded dividend yield and the rest of the parameters are the same as those defined in equation (\ref{eq:PerCopLogModel}). The asterisk in the brownian motions expresses that process is referred the real market measure rather than the risk neutral measure. The noise driving the stochastic volatility $dW_t^{*(Y)}$ has already been decomposed into a linear combination of independent brownian motions $dW_t^*$ and $dZ_t^*$ to fix the appropriate correlation $\rho_{iY}$ between the underlying and the stochastic volatility brownian motions.

\begin{equation}
\begin{array}{l}
 dS_t^{(i)}  = (\mu_t^{(i)} - q_t^{(i)}) S_t^{(i)} dt + f_i(Y_t )S_t^{(i)} dW_t^*  \\ 
 dY_t  = \frac{1}{\varepsilon }\left( {m - Y_t } \right)dt + \frac{{\nu \sqrt 2 }}{{\sqrt \varepsilon  }}\left( {\rho_{iY} dW_t^*  + \sqrt {1 - \rho_{iY} ^2 } dZ_t^* } \right) \\ 
 \end{array}
  \label{eq:ProcProbReal}
\end{equation}

\begin{equation}
dW_t  = dW_t^*  + \frac{{\mu_t^{(i)}  - r_t }}{{f\left( {Y_t } \right)}}\begin{array}{*{20}c}
   {} & {} & {}  \\
\end{array}dZ_t  = dZ_t^*  + \gamma _t dt
  \label{eq:Real2NeutralProb}
\end{equation}

Equation (\ref{eq:Real2NeutralProb}) shows the change of measure to turn the real market measure into the risk neutral measure. See that $\frac{{\mu_i(t)  - r_t }}{{f\left( {Y_t } \right)}}$ is the market risk premium and $\gamma_t$ is the risk premium for volatility risk. When $dW_t^*$ and $dZ_t^*$ are replaced in equation (\ref{eq:ProcProbReal}), equation (\ref{eq:ProcProbNeutral}) is obtained. It expresses the evolution of the underlying in the risk neutral measure. The function $\Lambda(Y_t)$ in equation (\ref{eq:MarketValueRisk}) is a combined market premium and volatility risk. See that the unknown drift $\mu_t^{(i)}$ of $S_t^{(i)}$ turns into the known risk free rate. Therefore, the process of the discounted underlying ($S_t^{(i)} \exp( -\int_0^t {r_s ds} )$) will be a martingale (zero drift) which is the condition required for the absence of arbitrage. When only pricing is considered, the risk premium is set to zero ($\Lambda(Y_t) = 0$).

\begin{equation}
\begin{array}{l}
 dS_t^{(i)}  = (r_t - q_t^{(i)}) S_t^{(i)} dt + f_i(Y_t )S_t^{(i)} dW_t  \\ 
 dY_t  = \left( {\frac{1}{\varepsilon }\left( {m - Y_t } \right)dt - \frac{{\nu \sqrt 2 }}{{\sqrt \varepsilon  }}\Lambda \left( {Y_t } \right)} \right) + \frac{{\nu \sqrt 2 }}{{\sqrt \varepsilon  }}\left( {\rho_{iY} dW_t  + \sqrt {1 - \rho_{iY} ^2 } dZ_t } \right) \\ 
 \end{array}
  \label{eq:ProcProbNeutral}
\end{equation}

\begin{equation}
\Lambda \left( y \right) = \rho _t \frac{{\mu_i(t)  - r_t }}{{f\left( {Y_t } \right)}} + \gamma _t \left( y \right)\sqrt {1 - \rho ^2 } 
  \label{eq:MarketValueRisk}
\end{equation}

%Doing the change of variables $X_t^{(i)} = \ln(S_t^{(i)})$ and applying It\^{o}'s lemma to equation \ref{eq:ProcProbNeutral} leads to equation \ref{eq:LogUnd}. This equation is similar to equation \ref{eq:PerCopModel} but only with one underlying. Therefore the calibration procedure for the lognormal-inspired process of equation \ref{eq:PerCopLogModel} described in 
%
%\begin{equation}
%\begin{array}{l}
% dX_t^{(i)}  = \left( r_t-q_i(t)  - \frac{1}{2}f_i \left( {Y_t } \right)^2 \right)dt + f_i \left( {Y_t } \right)dW_t^{(i)}  \\ 
% dY_t  = \left( {\frac{1}{\varepsilon }\left( {m - Y_t } \right)dt - \frac{{\nu \sqrt 2 }}{{\sqrt \varepsilon  }}\Lambda \left( {Y_t } \right)} \right) + \frac{{\nu \sqrt 2 }}{{\sqrt \varepsilon  }}\left( {\rho dW_t  + \sqrt {1 - \rho ^2 } dV_t } \right) \\ 
% \end{array}
%  \label{eq:LogUnd}
%\end{equation}

The calibration of the process of equation (\ref{eq:ProcProbReal}) is described in \cite{Fouque2000}. However, it was done considering that the process was not written in the risk free measure but the real market measure. Indeed, \cite{Fouque2000} estimates the effective volatilities $\bar \sigma _i$ of equation (\ref{eq:SigDash}) from historical returns whereas section \ref{sec:Reduction2IntuitiveParameters} sets them equal to $\sigma_i$ according to equation (\ref{eq:SigDashRed}) under the assumption that $\left\langle {g^2 } \right\rangle  = 1$. This can be done because in the context of this paper, the process will be only used in the risk neutral measure (the market risk premium will be set equal to zero) and therefore, no historical data will be needed to estimate the level and slope of the implied volatility.

\begin{equation}
\sigma _i^{imp} \left( K \right) = a\left( {\ln \left( {\frac{K}{{F_{iT} }}} \right)\frac{1}{{T - t}}} \right) + b =  - \frac{{R_i }}{{\bar \sigma _i^3 }}\left( {\ln \left( {\frac{K}{{F_{iT} }}} \right)\frac{1}{{T - t}}} \right) + \bar \sigma _i  - \frac{{R_i }}{{2\bar \sigma _i }}
  \label{eq:SigImpl}
\end{equation}

\begin{equation}
R_i  =  - a\bar \sigma _i^3 \begin{array}{*{20}c}
   {} & {}  \\
\end{array}\bar \sigma _i^2  + \frac{2}{a}\bar \sigma _i  - \frac{{2b}}{a} = 0\begin{array}{*{20}c}
   {} & {}  \\
\end{array}\bar \sigma _i  \approx b - \frac{{ab^2 }}{2}
  \label{eq:RiSigi}
\end{equation}

Equation (\ref{eq:SigImpl}) shows the approximation up to first order of the implied volatility $\sigma_i^{imp}$ of a vanilla option (see appendix \ref{app:InitialParams} to see where this expression comes from) when the underlying follows the process of equation (\ref{eq:ProcProbNeutral}) assuming that $\epsilon$ is small (fast mean reversion). $F_{iT} = S_{i0} \exp(\int_t^T {\alpha_S^{(i)}ds})$ is the forward value of the underlying $i$ at time $T$. For pricing purposes, the risk neutral meaure will be used and $\alpha_s^{(i)} = (r_s  - q_s )$ where $r_s$ is the domestic risk free rate and $q_s$ is the dividend yield (or the foreign risk free rate if an FX pair is considered). See that up to first order, the implied volatility behaves as a straight line where $a$ is the slope and $b$ is the intercept, when the independent variable is equal to the log-moneyness-to-maturity ratio (${\ln \left( {\frac{K}{{F_{iT} }}} \right)\frac{1}{{T - t}}}$). The parameters $a$ and $b$ are estimated through a linear regression of the volatility with respect to the log-moneyness-to-maturity ratio. Equation (\ref{eq:RiSigi}) gives the parameters of the model $R_i$ and $\bar \sigma_i$ in terms of $a$ and $b$. The second order equation for $\bar \sigma_i$ can be simplified assuming that $\bar \sigma_i \approx b$, yielding the right expression of equation (\ref{eq:RiSigi}).

\begin{equation}
c_i(K,T) = P(t,T)\int_{ - \infty }^\infty  {\left( {\exp \left( {\beta _i  + \xi _i } \right) - K} \right)} ^ +  v_i^\varepsilon  \left( {t,0,T,\xi _i } \right)d\xi _i 
  \label{eq:Callprice}
\end{equation}

\begin{equation}
\beta _{i0}  = \ln S_{i0}^{(i)}  + \int_t^T {(r_s^{(i)}  - q_s^{(i)} )ds}  - \frac{1}{2}\bar \sigma _i^2 \left( {T - t} \right)^2 
  \label{eq:InitialBeta}
\end{equation}

Equation (\ref{eq:Callprice}) shows how to calculate the price of a call option out of the perturbed marginal transition probability density $v_i^{\epsilon}$ given by equations (\ref{eq:pi}) and (\ref{eq:vi}), where $P(t,T)$ is the discount factor from the expiration date $T$ to present time $t$ and $\beta_i$ is a parameter to fix the mean of the distribution. See that the transition probability is defined from 0 to $\xi_i$ (it has no mean), because the mean of the distribution is taken into account by $\beta_i$. An initial value for this parameter $\beta_{i0}$ is given by equation (\ref{eq:InitialBeta}) which is the mean of the zero order term of the solution given by equation (\ref{eq:xtilde_i}), where $\alpha_t^{(i)} = r_t^{(i)} - q_t^{(i)}$ and $x_i = \ln(S_t^{(i)})$.

The parameter $\beta_{i0}$ is only an asymptotic approximation of the actual parameter $\beta_i$ which satisfies equation (\ref{eq:FwdiT_int}) (the expected value of the underlying must be equal to its forward value). The parameter $\beta_i$ is easily obtained calculating the forward $F_{iT}^{prev}$ from the initial $\beta_{i0}$ and integrating numerically equation (\ref{eq:FwdiT_int}). This value will not in general match the actual forward $F_{iT}$. However, if the next beta is corrected by $\beta_i^{next} = \beta_i^{prev} - ln(F_{iT}^{prev}) + ln(F_{iT})$ the new value of the forward $F_{iT}^{prev}$ obtained out of $\beta_i^{next}$ will be a lot closer to $F_{iT}$. After a few iterations, $\beta_i$ converges.

\begin{equation}
F_{iT}  = \int_{ - \infty }^\infty  {\exp \left( {\beta _i  + \xi _i } \right)v_i^\varepsilon  \left( {t,0,T,\xi _i } \right)d\xi _i } 
  \label{eq:FwdiT_int}
\end{equation}

Once the forward of the distribution $F_{iT}$ is matched, the price of two vanilla options can be matched using a simple Newton-Raphson method with step shortening starting from the initial $R_i$ and $\sigma_i$ given by equation (\ref{eq:RiSigi}). %The asymptotic approximation given by equation (5.43) of \cite{Fouque2000} for vanilla options is more exact than equation \ref{eq:Callprice} which uses the perturbed marginal density. The underlying model is the same but the normalization of the perturbed density to sum 1 (through the hyperbolic tangent of equation \ref{eq:vi}) may slightly distort the distribution (that's why the Newton-Raphson method is used to get an exact match).

\section{Interpretation of perturbed copula}
\label{sec:Interpretation}
 
This section interprets the calibration and the effect of the perturbed copula compared to the gaussian copula. In terms of pricing, the interpretation is carried out applying the perturbed gaussian copula approach to the valuation of FX quanto options to a third currency different to the currencies of the underlying pair. In particular, the underlying FX pair considered is the XAU/USD which is quoted in USD per ounce of gold (XAU). The option is quantoed to EUR and therefore the second underlying pair involved is the EUR/USD. Equation (\ref{eq:Payoff}) shows the payoff function of this option, where $S_T$ is the price of the XAU/USD (in USD per ounce of gold), $X_T$ is the price of the EUR/USD (in USD per unit of EUR), $DF_T^{USD}$ is the discount factor of the USD curve and $K$ is the strike price. If the USD money market account is chosen as numeraire, both $S_T$ and $X_T$ are denominated in the numeraire currency (USD) and their drifts are simply calculated as the difference between their domestic (USD) and foreign interest rates at maturity. However, as the option is quanto, the $(S_T-K)^+$ will be paid in EUR and therefore, as it should be discounted with the numeraire in USD, the quantity must be converted to USD first, multiplying by the EUR/USD exchange rate $X_T$. The spot price of the XAU/USD is $S_0=981.3$ and the spot price of the EUR/USD is $X_0=1.422$.

\begin{equation}
p = {\bf E}\left[ {\left( {S_T  - K} \right)^ +  X_T DF_T^{USD} } \right]
  \label{eq:Payoff}
\end{equation}

The expectation given by equation (\ref{eq:Payoff}) is computed through the double integral given by equation (\ref{eq:PayoffInt}), where $\xi_i$ are the logarithms of the underlyings at expiration ($\xi_1 = \ln(S_T)$ and $\xi_2 = \ln(X_T)$) and $f^{joint}$ is the joint probability density function of both underlyings given by equation (\ref{eq:jpdf}) (see chapter 2 of \cite{Meucci2005} to see where this equation comes from), where $f_{cop}$ is the copula function defined in equation (\ref{eq:fcop}), $z_i$ are given by equation (\ref{eq:zi}) and $f_i^{marg}$ are the empirical marginal distributions. The latter are obtained through equation (\ref{eq:mpdf}) (see \cite{Gatheral2006} to find out where this equation comes from), where $P_i(K,T,\sigma_i^{impl})$ is the Black Scholes price of a put option of underlying $i$ with strike $K$, maturity $T$ and $\sigma _i^{impl}(K,T)$ is the interpolated implied volatility for strike level $K$ of underlying $i$ at maturity $T$.

\begin{equation}
p = DF_T^{USD} \int\limits_{{\bf R}^2 } {\left( {e^{\xi _1 }  - K} \right)^ +  e^{\xi _2 } f^{{\rm joint}} \left( {\xi _1 ,\xi _2 } \right)d\xi _1 d\xi _2 } 
  \label{eq:PayoffInt}
\end{equation}

\begin{equation}
f^{{\rm joint}} \left( {\xi _1 ,\xi _2 } \right) = f_{cop} \left( {z_1 ,z_2 } \right)f_1^{{\rm marg}} \left( {\xi _1 } \right)f_2^{{\rm marg}} \left( {\xi _2 } \right)
  \label{eq:jpdf}
\end{equation}

\begin{equation}
f_i^{{\rm marg}} \left( {\xi _i } \right) = \frac{1}{{DF_T^{USD} }}\frac{{d^2 P_i \left( {e^{\xi _i } ,T,\sigma _i^{impl} (e^{\xi _i } ,T)} \right)}}{{dK^2 }}
  \label{eq:mpdf}
\end{equation}

The perturbed copula has been tested in a set of 25 scenarios varying the correlation and the skew of both underlyings. These scenarios have zero interest rates and have been created with volatility surfaces given out of a Heston model with constant parameters varying the correlation between the brownian motions of the underlying and the variance process (this parameter controls the skew). The time horizon considered is one year. A skew which favors out-of-the money (OTM) puts (lower strikes have higher volatility) is referred to in this paper as left skew. When OTM calls are favored (higher strikes have greater volatility), the resulting skew is called right skew. There are five groups of scenarios with different skews: ``LR" (left-right), ``RL" (right-left), ``SS" (smile-smile), ``RR" (right-right) and ``LL" (left-left). Each group has five differerent correlations between both underlyings (0.6, 0.3, 0, -0.3, -0.6). The perturbed copula has been calibrated to the 25 delta out-of-the-money call and put options (the most liquid products).

\begin{figure}[htbp]
	\centering
	\includegraphics[width=0.31\textwidth]{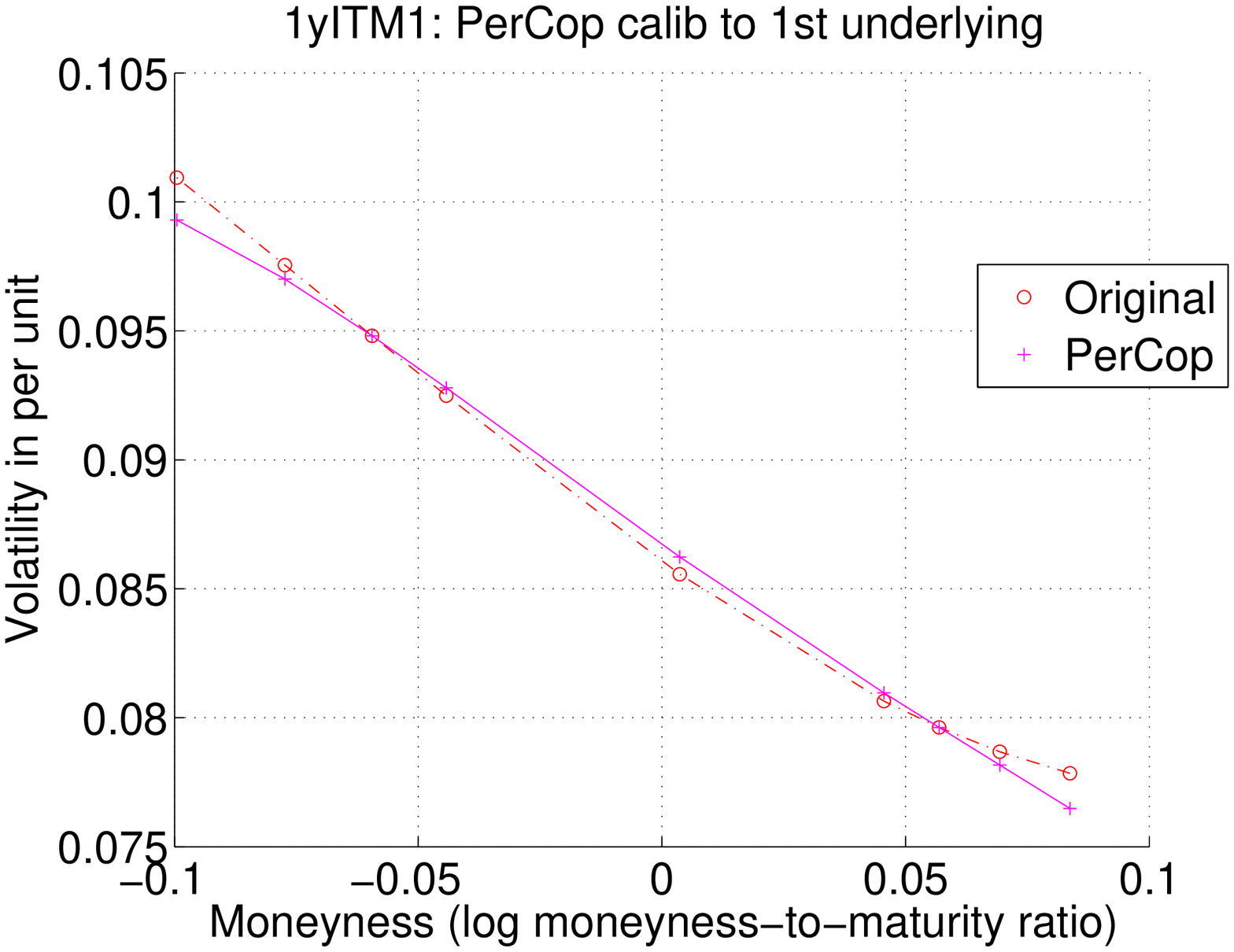}
	\includegraphics[width=0.31\textwidth]{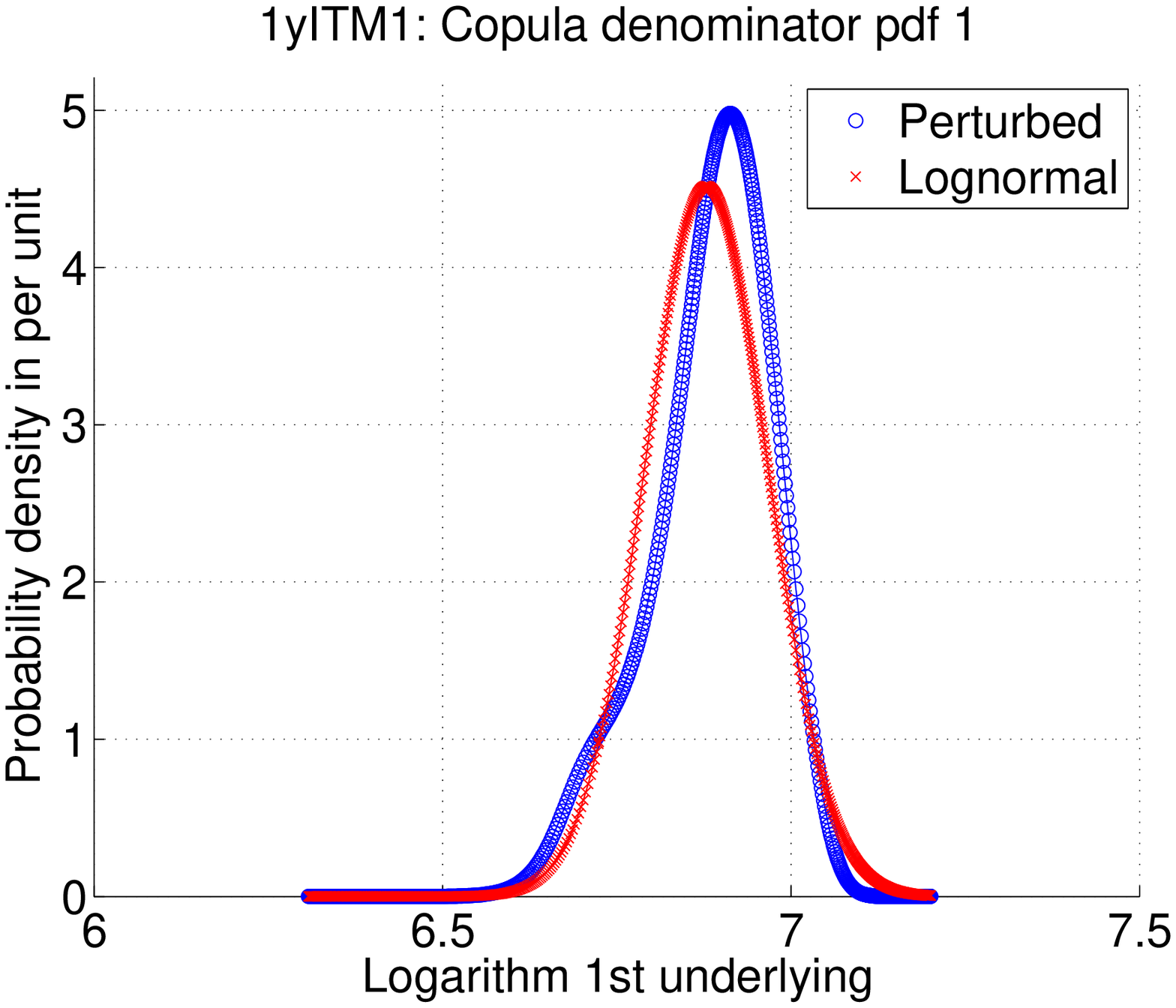}
	\includegraphics[width=0.31\textwidth]{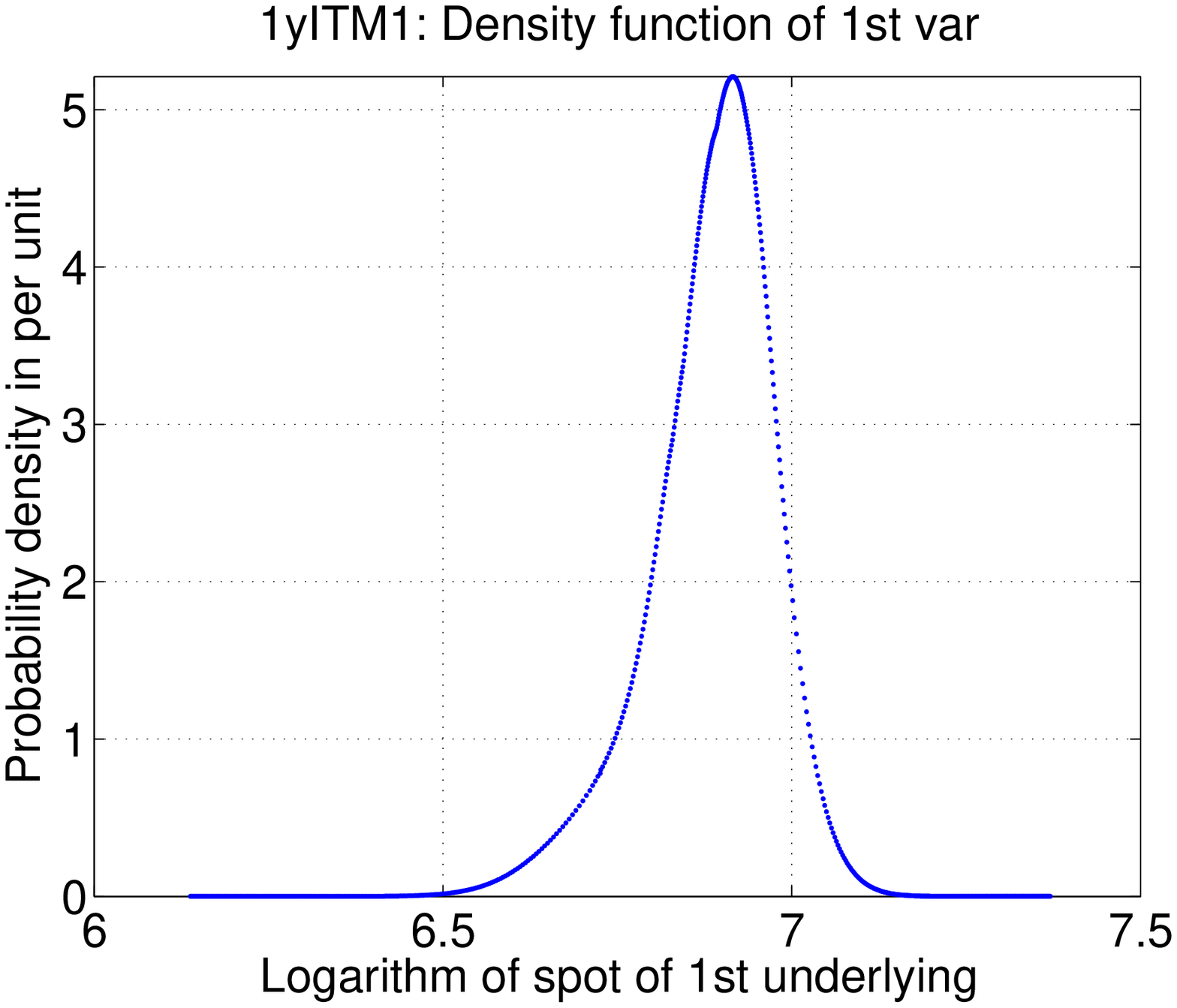}
	\includegraphics[width=0.31\textwidth]{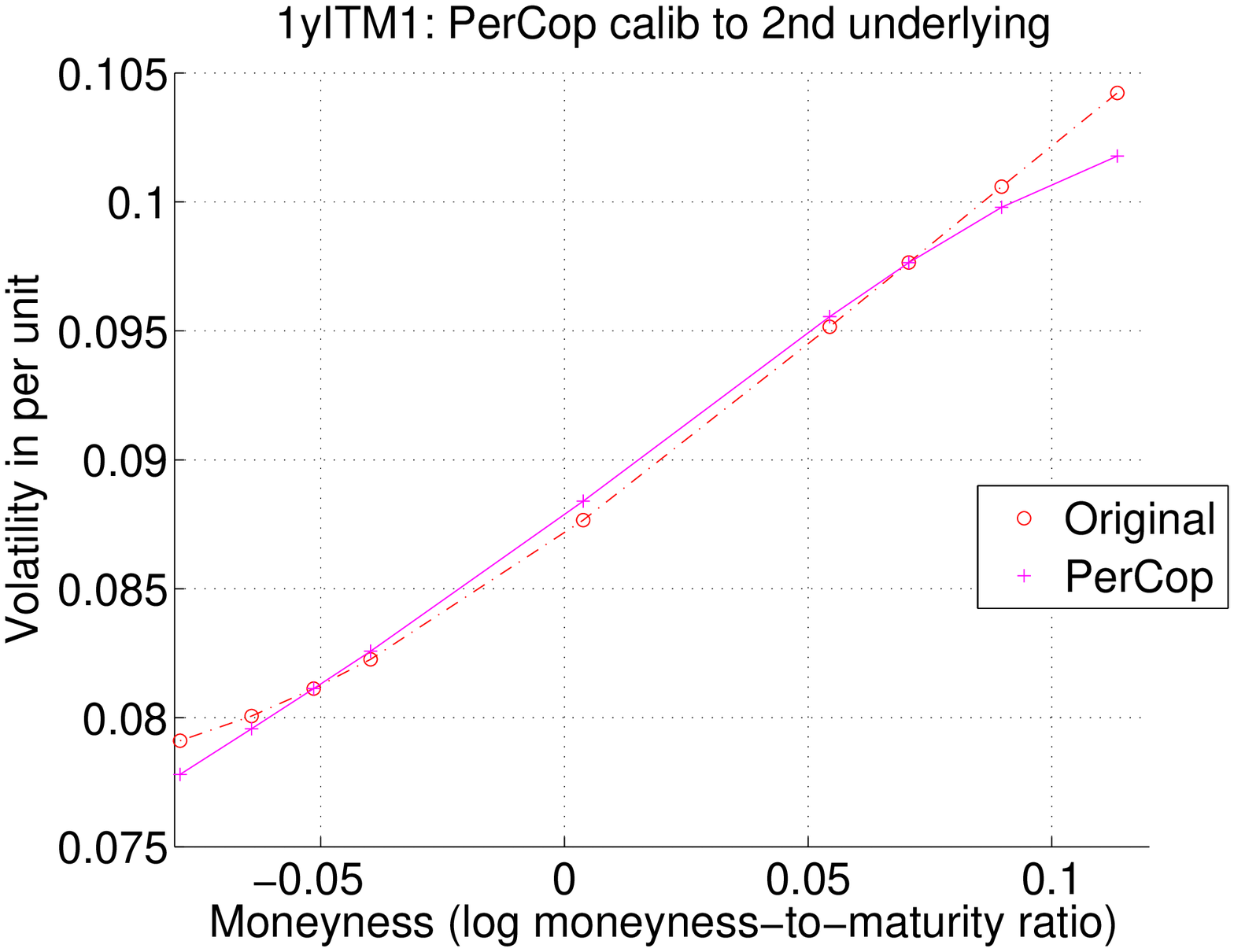}
	\includegraphics[width=0.31\textwidth]{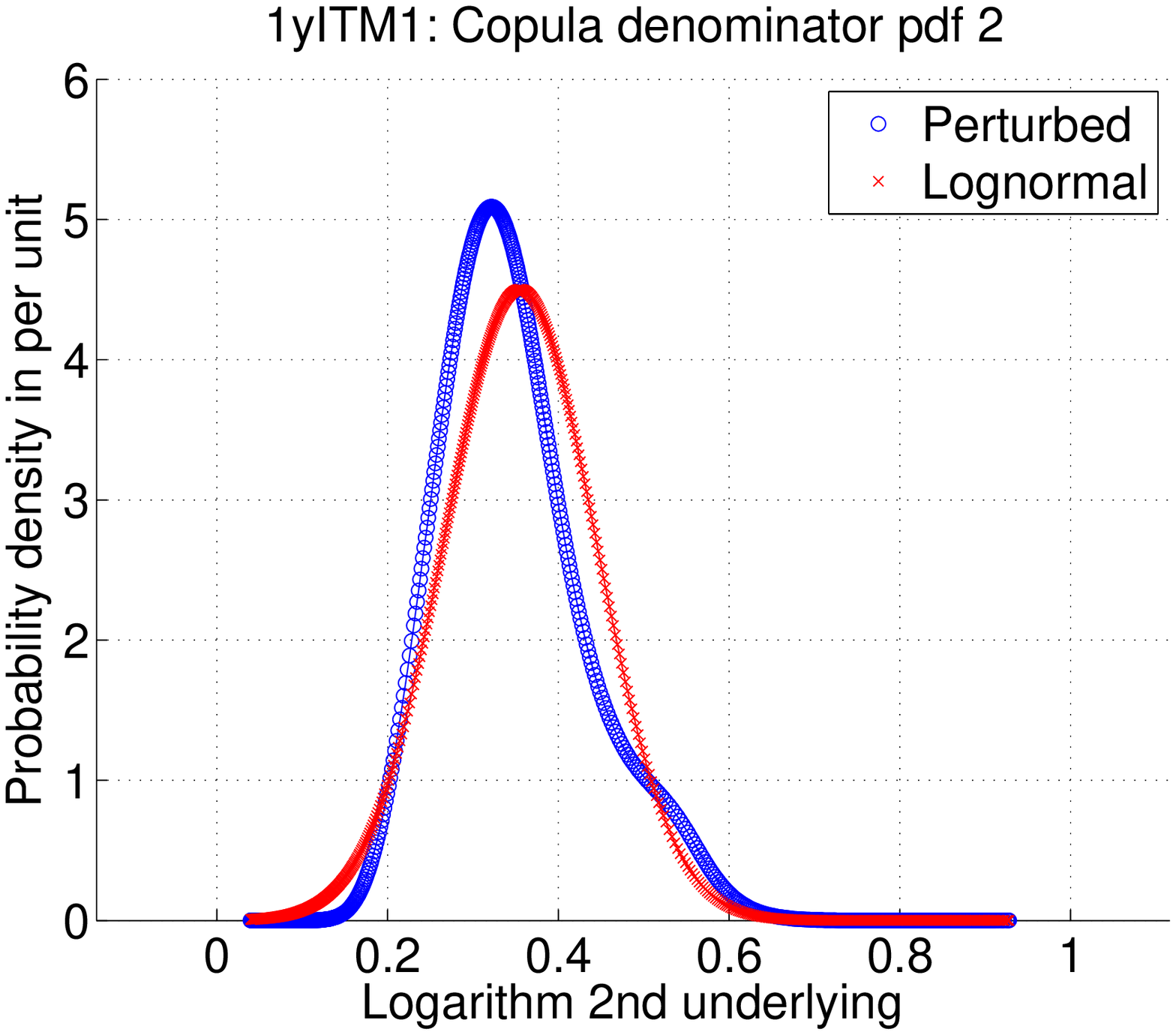}
	\includegraphics[width=0.31\textwidth]{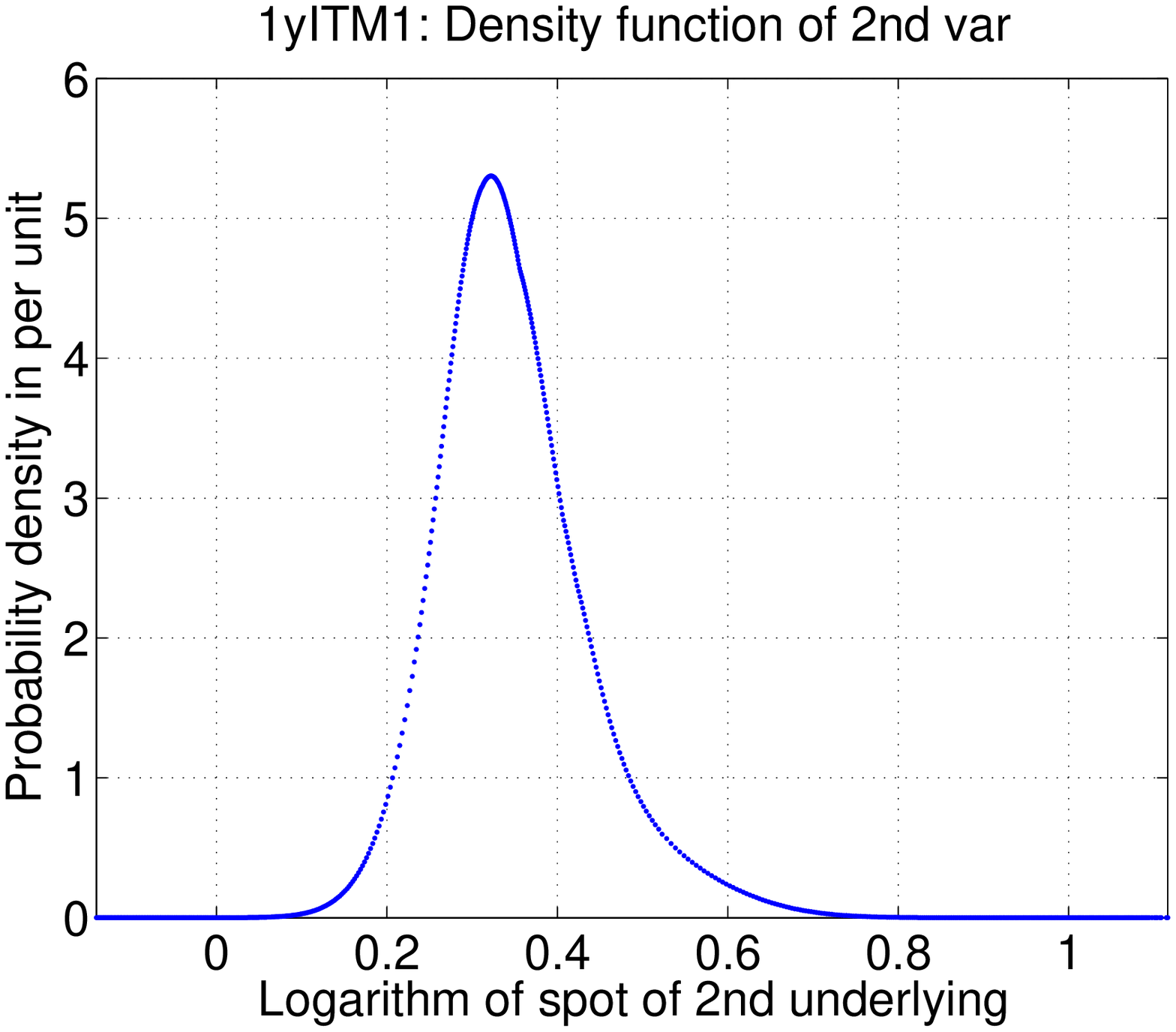}
	\caption{Implied volatility of left and right skew Heston scenarios compared to their calibrations (left plots), corresponding perturbed pdf (middle plots) and empirical Heston pdf (right plots). Calibrated parameters: $\beta_1 = 6.8783$, $\sigma_1 = 0.0892$, $R_1 = 1.31 \cdot 10^{-4}$ (up) and $\beta_2 = 0.3550$, $\sigma_2 = 0.0877$, $R_1 = -1.29 \cdot 10^{-4}$ (down).}
	\label{fig:LRcalib}
\end{figure}

Figure \ref{fig:LRcalib} presents the left and right skew scenarios. The upper plots show the left skew scenario (higher volatility for lower strikes) and the lower plots show the right skew scenario (higher volatility for higher strikes). The calibrated parameters for the left skew scenarios are $\beta_1 = 6.8783$, $\sigma_1 = 0.0892$, $R_1 = 1.31 \cdot 10^{-4}$ and for the right skew scenario are $\beta_2 = 0.3550$, $\sigma_2 = 0.0877$, $R_2 = -1.29 \cdot 10^{-4}$. See that the $\beta_1$ that appears in equation (\ref{eq:Callprice}) is a lot bigger than $\beta_2$ because this parameter fits the mean and the spot levels are quite different ($981.3$ versus $1.422$). On the other hand, the volatility level is not very high (around 9\%), the skew levels are rather mild and have opposite signs (in agreement with smirks opposite to each other). The left plots of figure \ref{fig:LRcalib} show the implied volatility of the original Heston-generated (labeled ``Original") and the perturbed-copula (labelled ``Perturbed") calibrated surfaces versus the log-moneyness-to-maturity ratio ($\ln(K/F_T)/T$, where $F_T$ is the forward of the underlying at maturity $T$) on the horizontal axis. The middle plots show the perturbed marginal probability density functions obtained after calibration from equation (\ref{eq:vi}) (labeled ``Pertubed") and the zero order term from equation (\ref{eq:pi}) (labeled ``Lognormal" and centered around $\beta_i$) also displayed for comparison purposes (these functions can be replicated with the parameters mentioned at the beginning of the paragraph). The right plots show the empirical marginal density functions obtained from equation (\ref{eq:mpdf}) out of the Heston-generated volatility surface. See that marginal densities are expressed in terms of the logarighm of the underlying rather than the underlying level itself. It can be seen that the fit of the skew provided by the perturbed marginal density function is very reasonable. It is clear to interpret that the left skew scenario increases the weight of the left queue (lower underlying values) and displaces the mode of the distribution to the right (higher underlying values). The opposite happens for the right skew scenario (increase of right queue density and move of the mode of the distribution to the left). See that the perturbed density functions incorporate a slight bump in the queues to create the skew. For extreme queue values, the density function is not as fat as the that of the empirical distribution and that is why the skew provided by the perturbed copula flattens for very extreme values of the queue (this will be clearer in section \ref{sec:CaseStudy}).

\begin{figure}[htbp]
	\centering
	\includegraphics[width=0.31\textwidth]{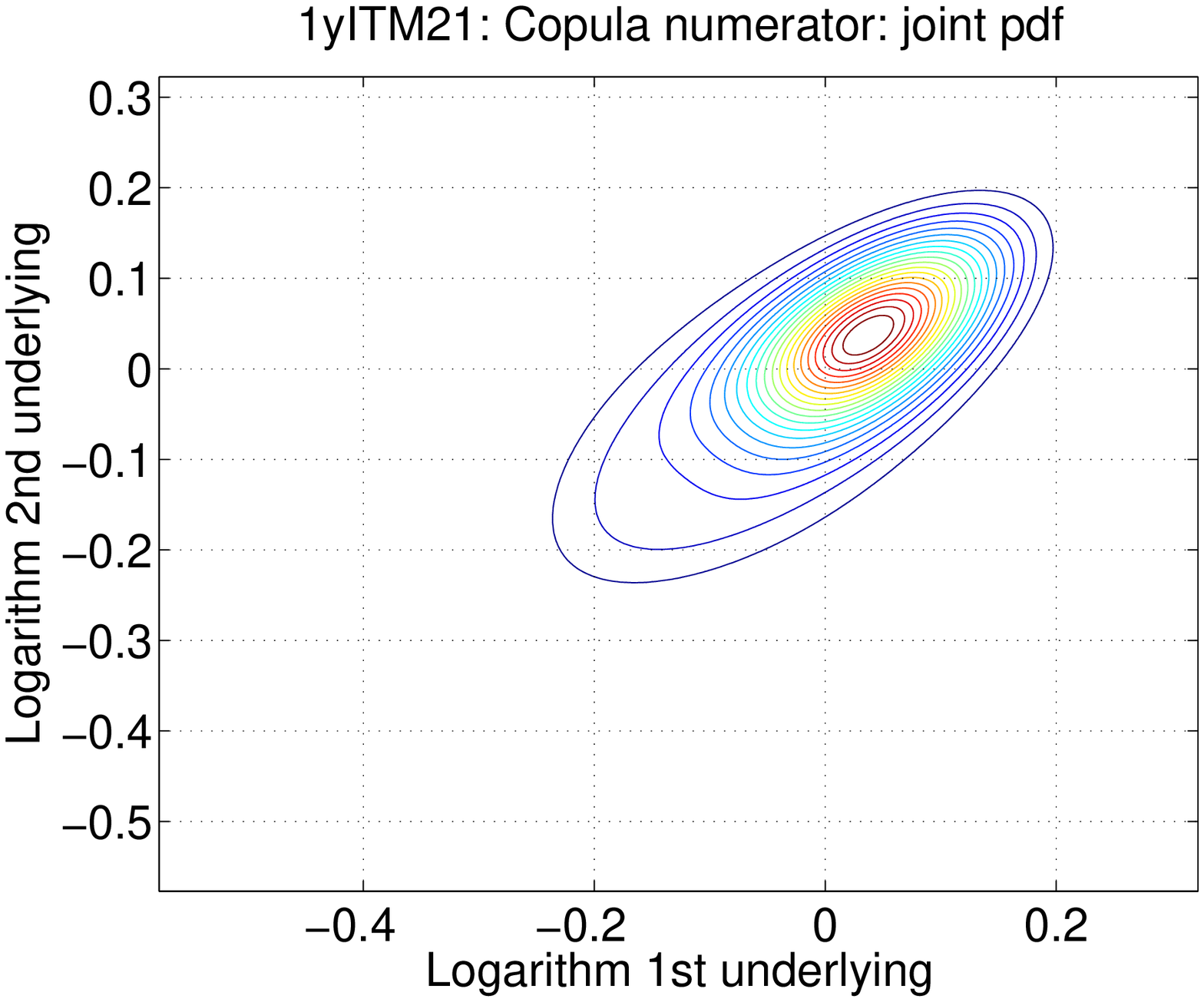}
	\includegraphics[width=0.31\textwidth]{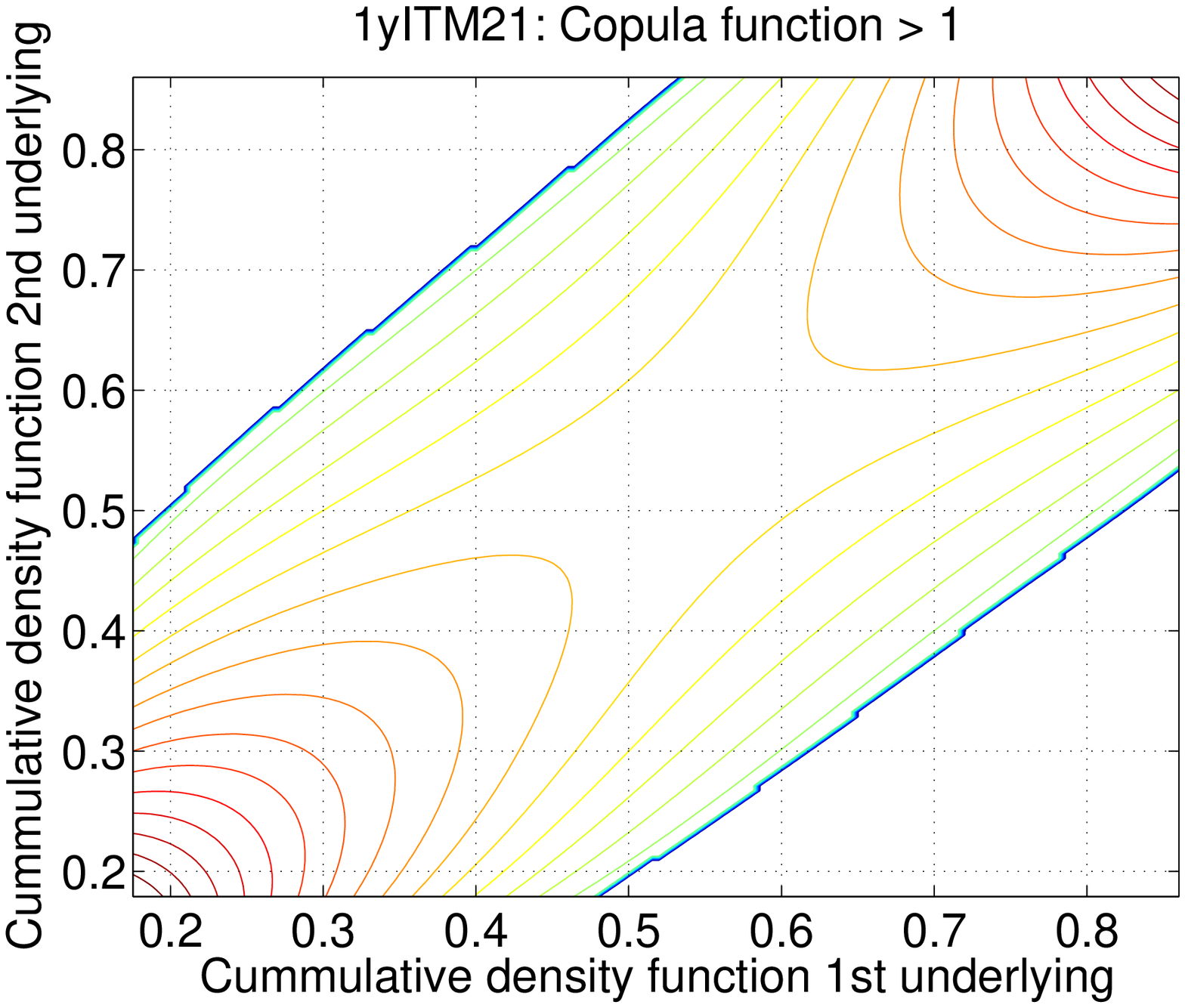}
	\includegraphics[width=0.31\textwidth]{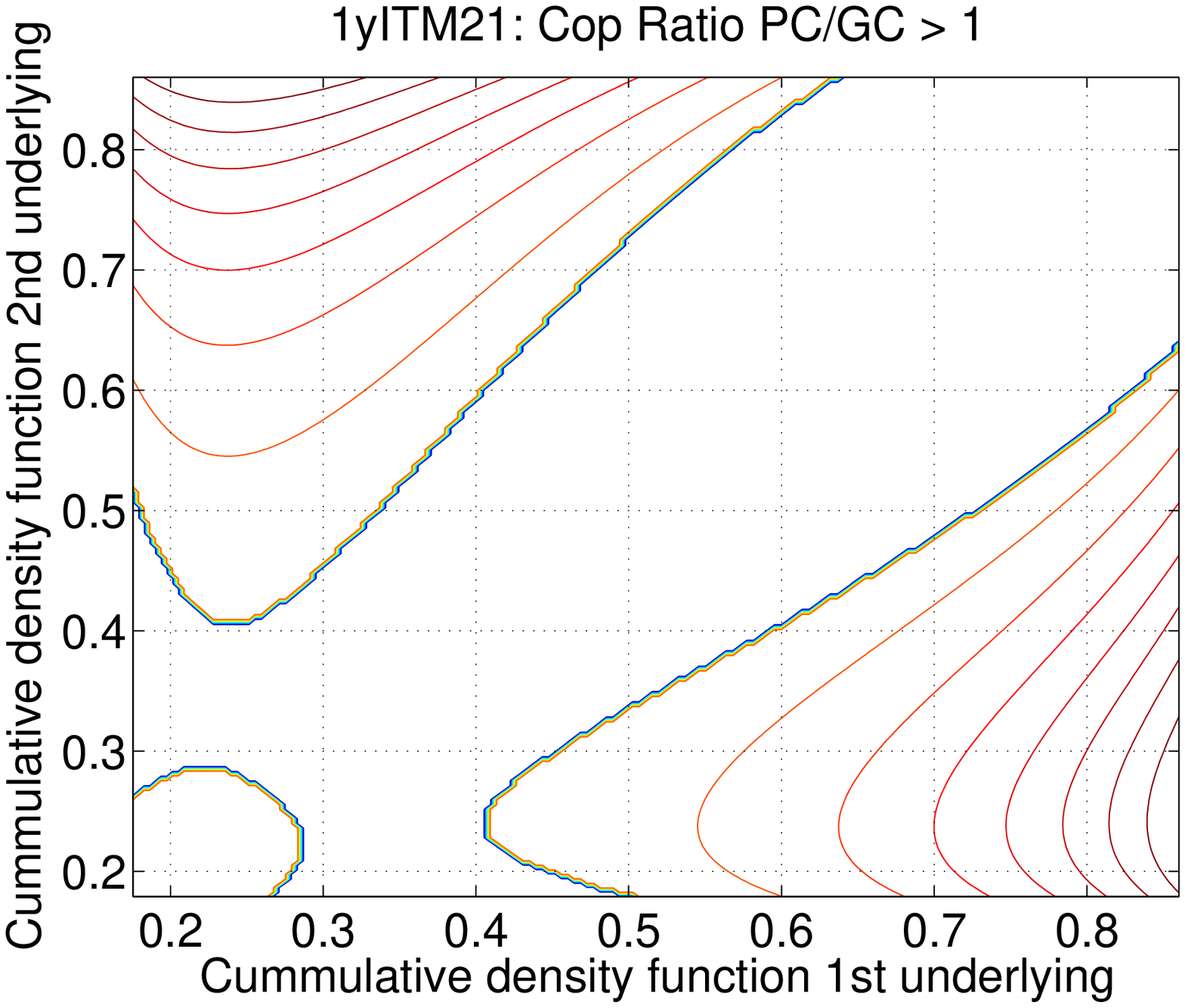}
	\includegraphics[width=0.31\textwidth]{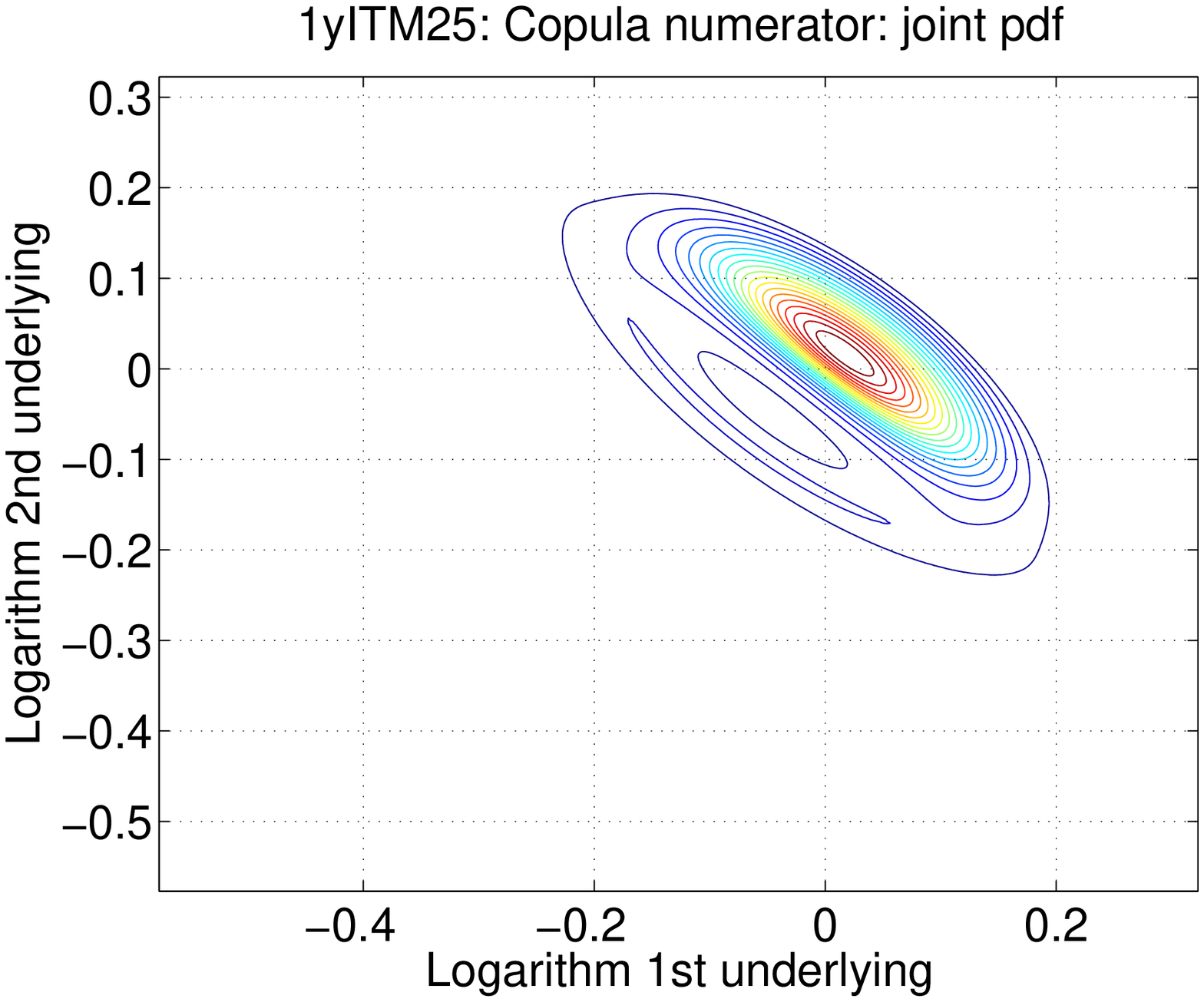}
	\includegraphics[width=0.31\textwidth]{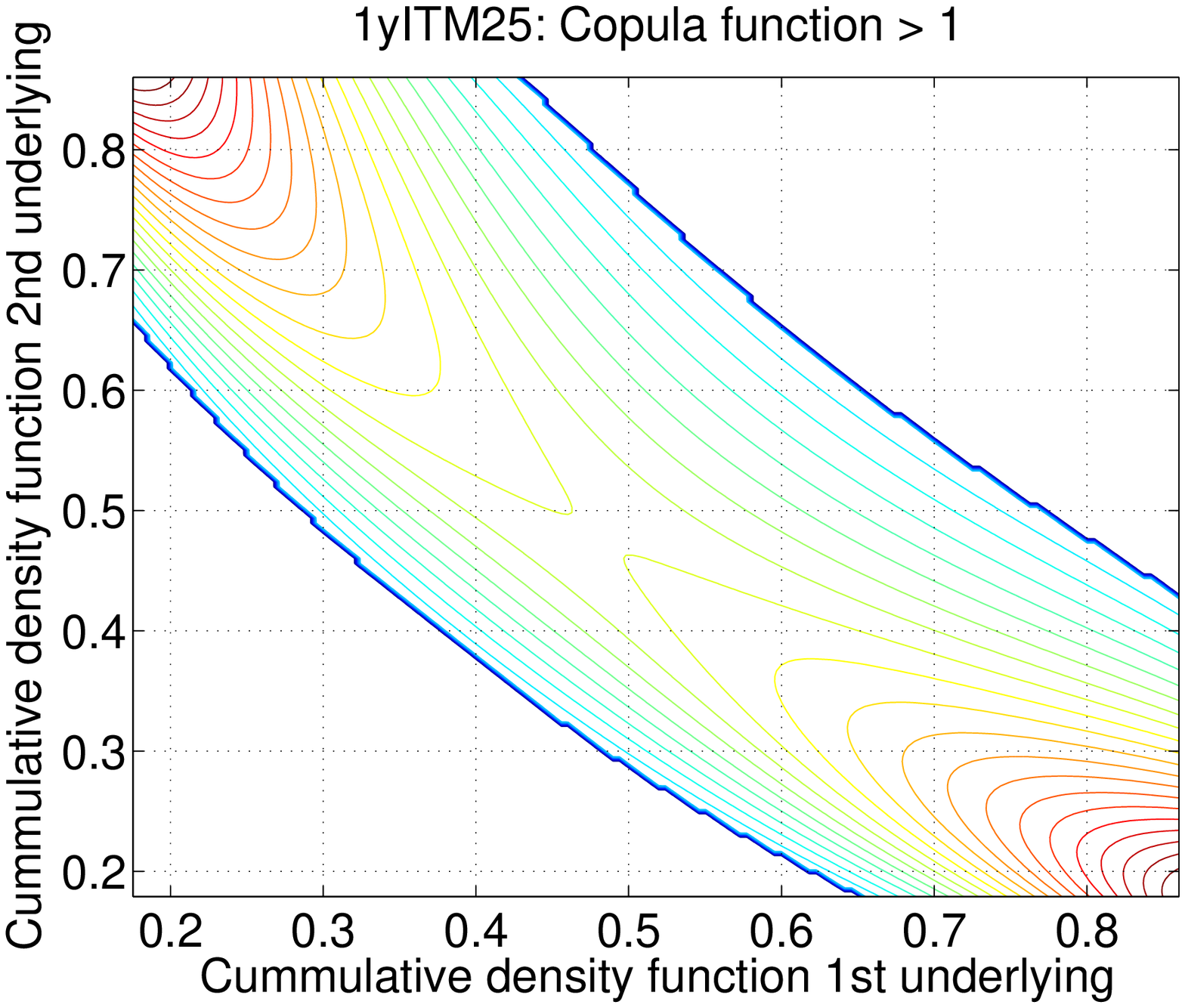}
	\includegraphics[width=0.31\textwidth]{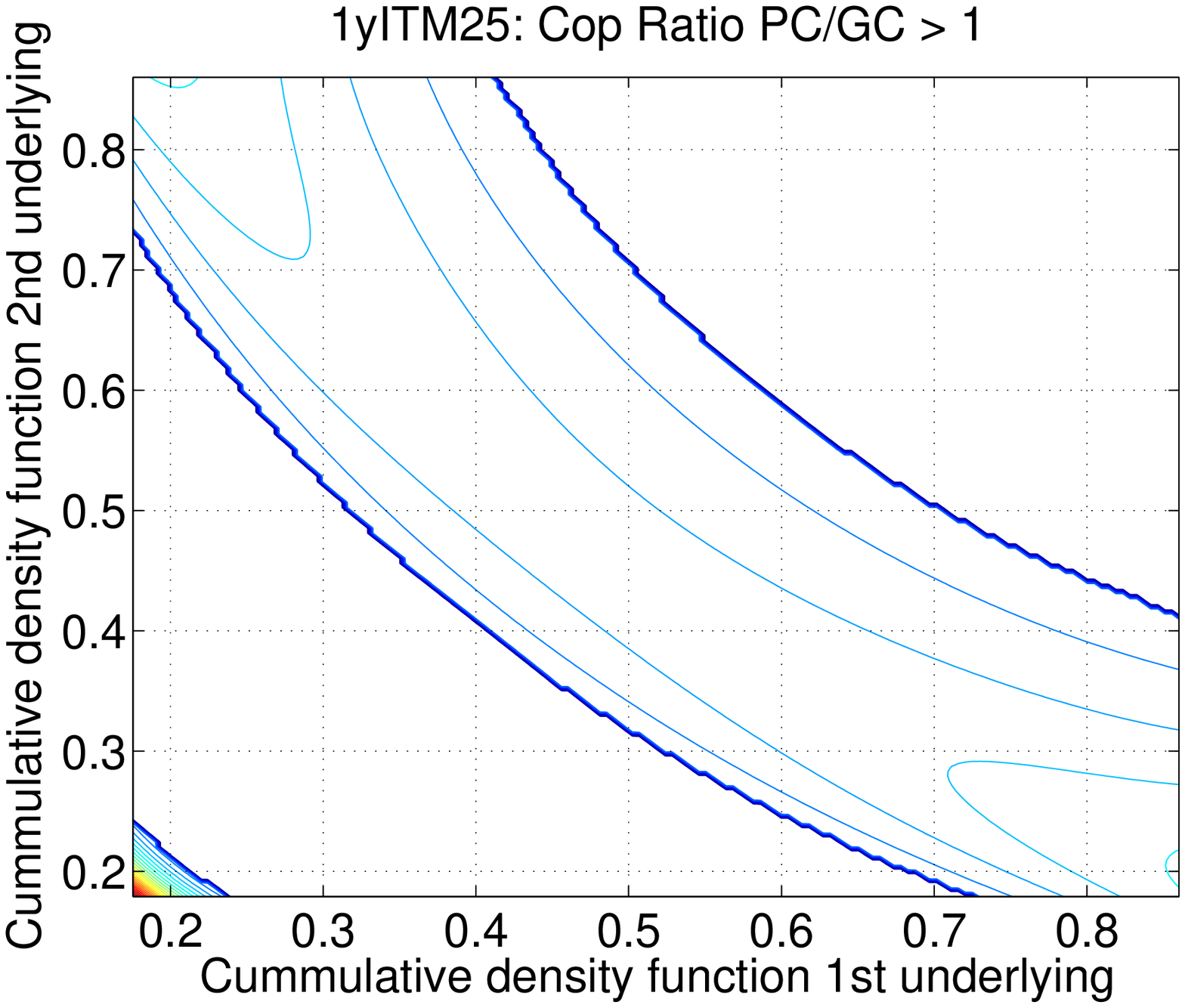}
	\caption{Copula numerator (left), copula function greater than 1 (middle) and ratio greater than 1 between perturbed and gaussian copulas (right) for a LL skew Heston scenario with positively (upper plots) and negatively (lower plots) correlated underlyings.}
	\label{fig:HesCop}
\end{figure}

\begin{figure}[htbp]
	\centering
	\includegraphics[width=0.31\textwidth]{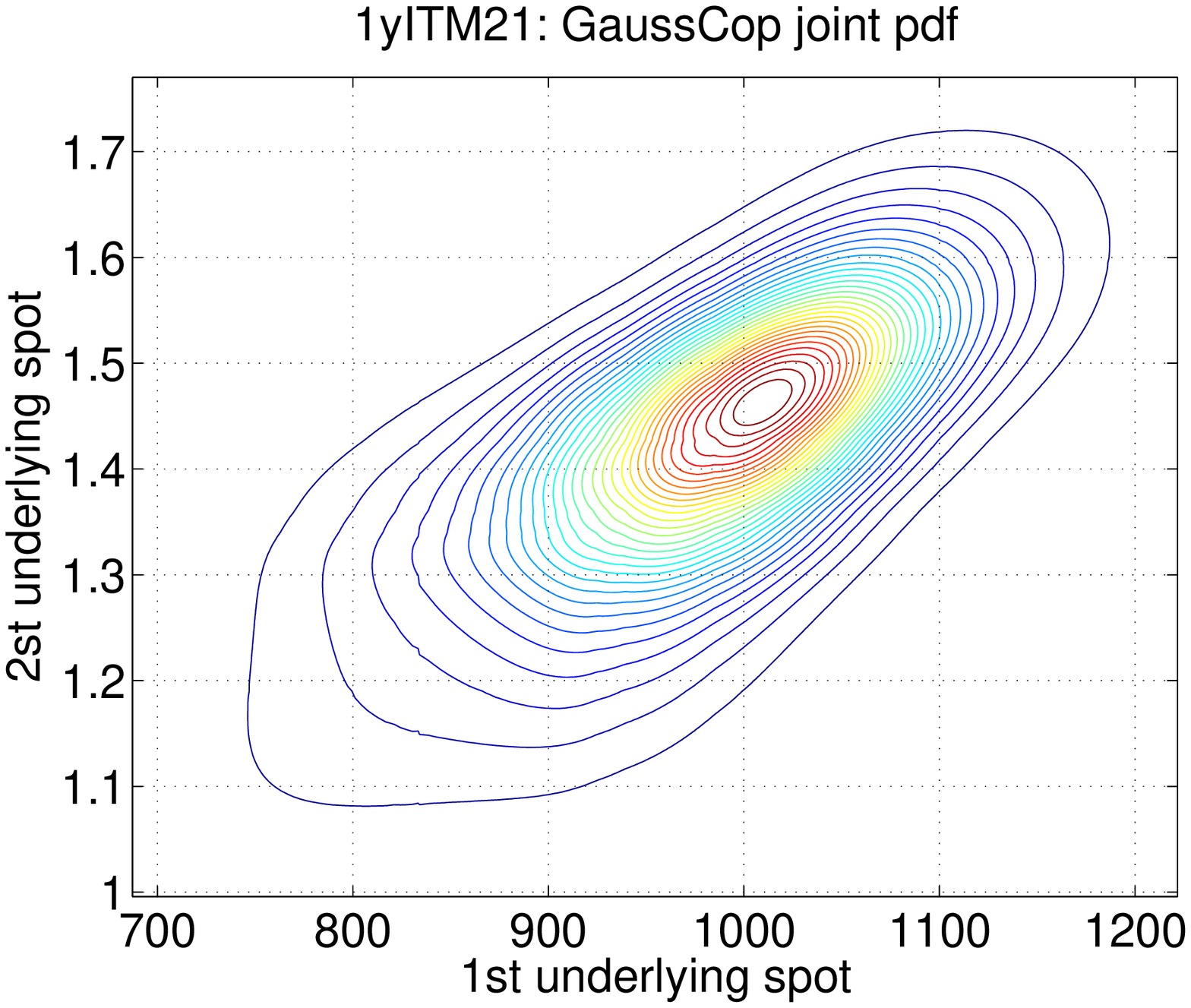}
	\includegraphics[width=0.31\textwidth]{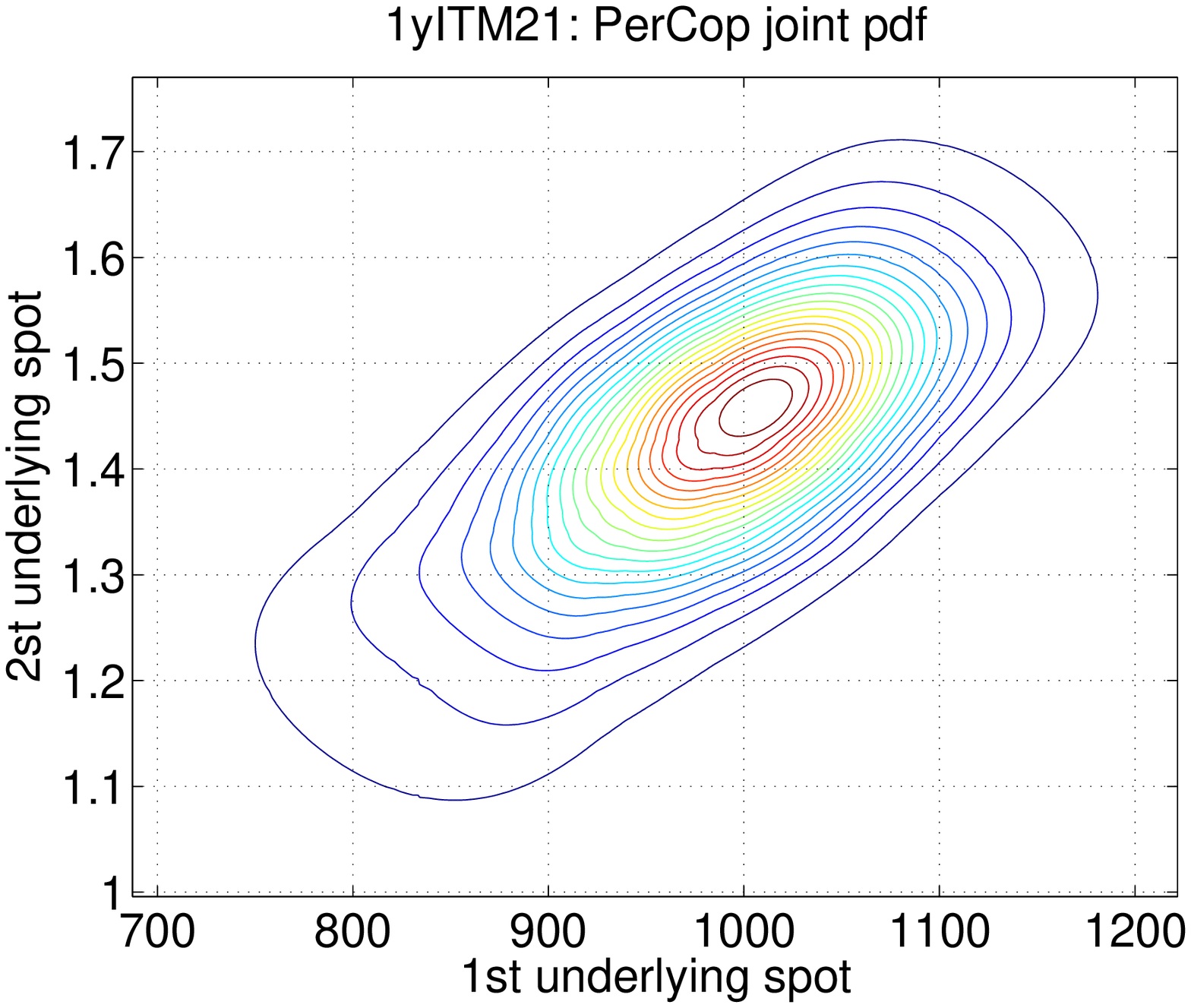}
	\includegraphics[width=0.31\textwidth]{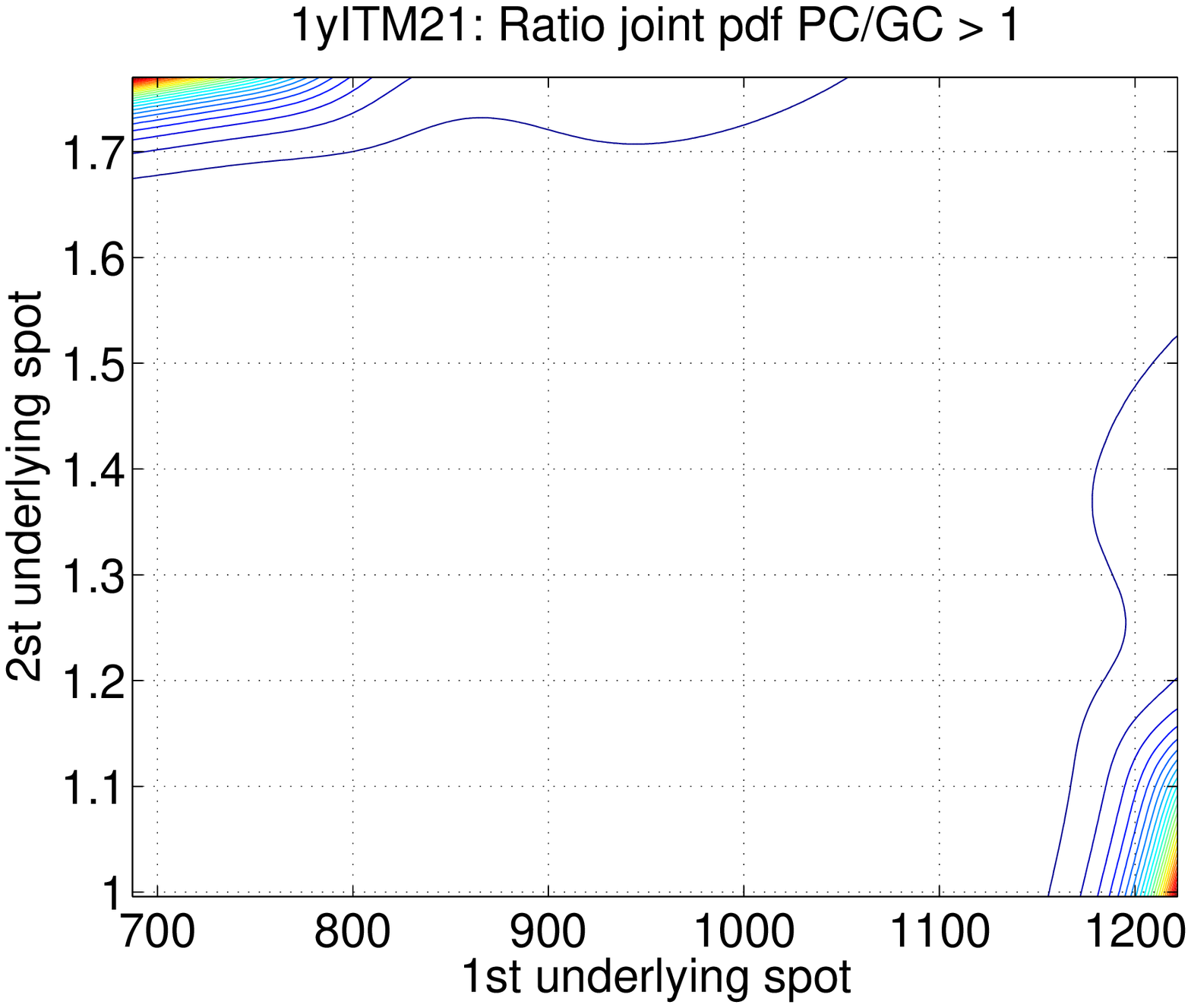}
	\includegraphics[width=0.31\textwidth]{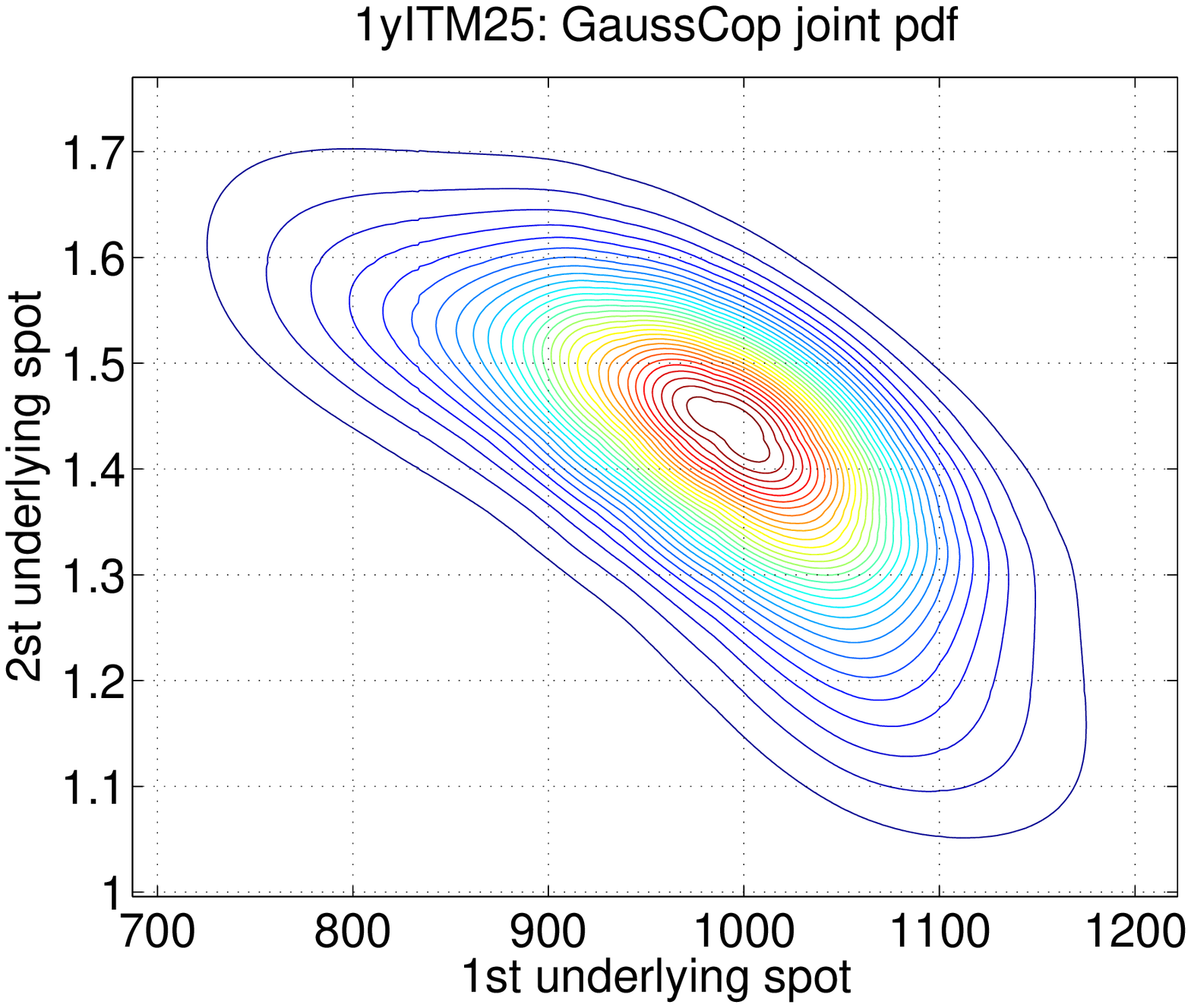}
	\includegraphics[width=0.31\textwidth]{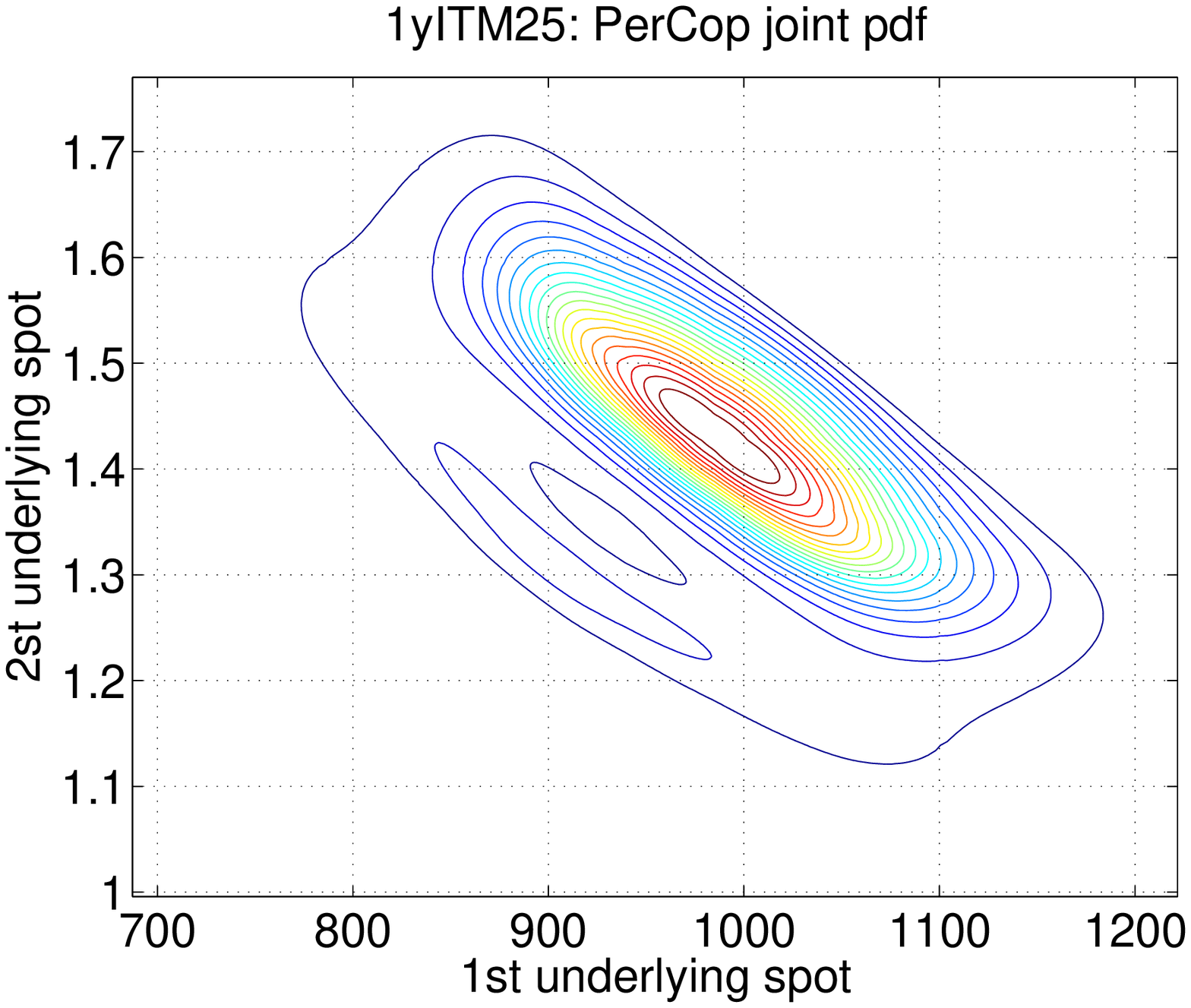}
	\includegraphics[width=0.31\textwidth]{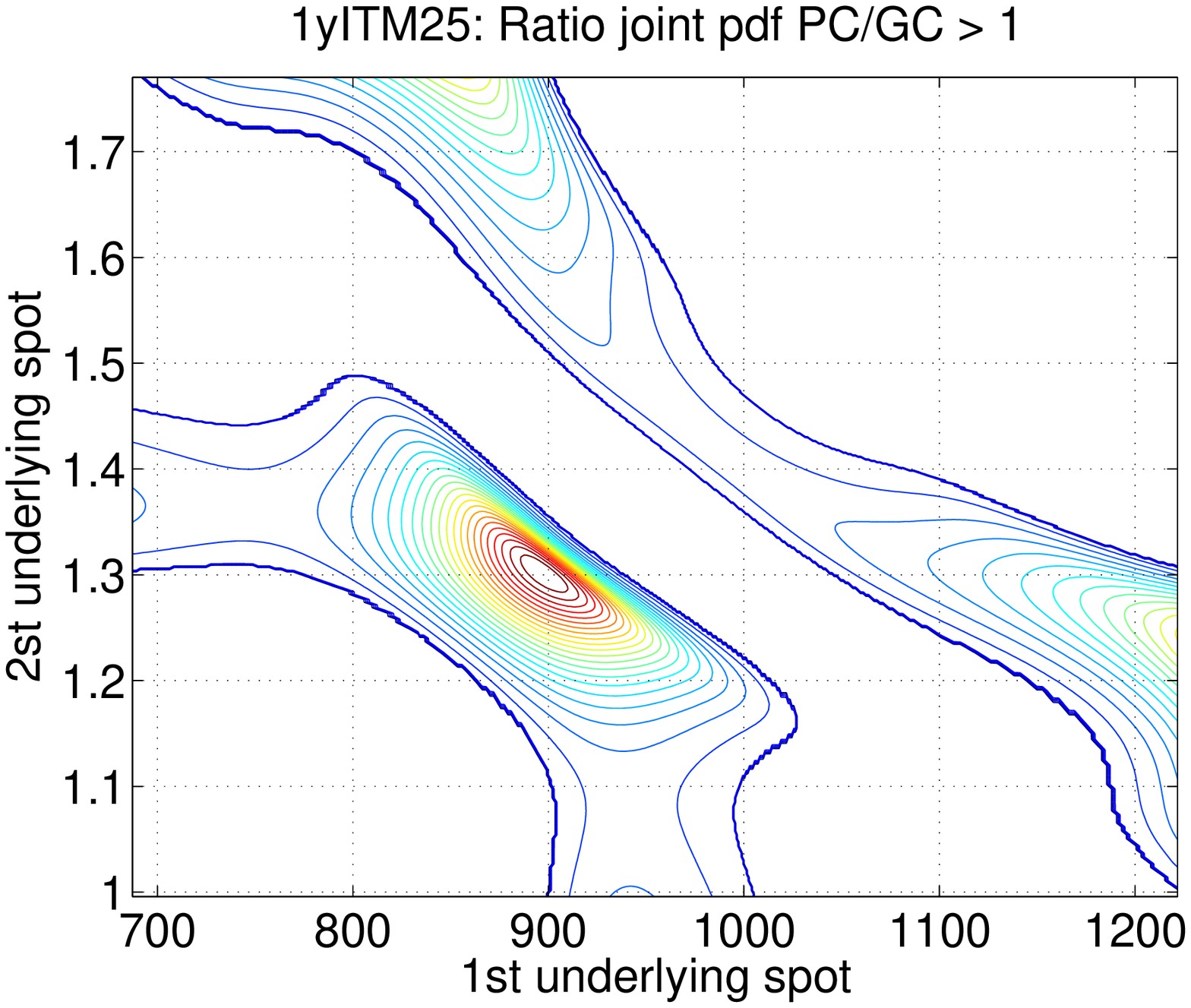}
	\caption{Joint pdf (copula function times empirical marginal densities) obtained with gaussian copula (left), perturbed copula (middle) and ratio between them greater than one (right) for a LL skew Heston scenario with positively (upper plots) and negatively (lower plots) correlated underlyings.}
	\label{fig:Hesjpdf}
\end{figure}

From the five groups of scenarios and 5 correlations, only one of them (the ``LL" scenario) for the extreme correlation values (0.6 and -0.6) is analyzed. The rest of combinations are symmetrical with respect to this reference scenario. The left plots of figure \ref{fig:HesCop} show the perturbed joint probability density function (numerator of the copula function) of both underlyings given by equation (\ref{eq:FirstOrderSolfinal}). The middle plots of figure \ref{fig:HesCop} present the copula function given by equation (\ref{eq:fcop}) for values greater than 1 (to allow for clarity). Finally, the right plots of figure \ref{fig:HesCop} display the ratio between the pertubed and gaussian copula functions also for values greater than 1. These plots might be replicated using the calibrated parameters given in figure \ref{fig:LRcalib} and calculating $R_{ij}$ and $Q_{ij}$ according to equation \ref{eq:RijRed}. The upper plots correspond to the positive (0.6) correlation scenario and the lower plots to the negative (-0.6) correlation scenario. The upper left plot of figure \ref{fig:HesCop} shows a positively sloped (from bottom left to up right) oval form which corresponds to a bi-normal distribution with positive correlation. However, the mode of the distribution is not situated in the center but it is displaced upwards to the right. A ``LL" scenario increases the left queues (lower values) and shifts the mode to the right (higher values) of both underlyings. It can be seen that the displacement produced by the skew goes to the upper right corner in the same direction as the correlation (the queues are displaced in the opposite direction towards the lower left corner). For the negative correlation scenario (bottom left plot) the displacement of the skew goes perpendicular to the direction of the correlation. The middle plots of figure \ref{fig:HesCop} show the copula function. When the copula function is equal to one, the joint probability density function is the product of the two marginals indicating independence or no co-dependence. When the copula function is greater than one, the density is increased indicating that there is more co-dependence (the opposite happens with a copula function less than one).

The middle upper plot of figure \ref{fig:HesCop} shows how the copula function increases the probability in the main diagonal as the correlation is positive (it is not clear in the plot but the lower left corner increases the density more than the oppossite corner for the effect of the skew). The middle lower plot of figure \ref{fig:HesCop} shows that the copula is greater than 1 in the anti-diagonal as the underlyings are negatively correlated. However, the anti-diagonal is displaced or biased through the the lower left corner because the effect of the ``LL" skew scenario is moving the queues in that direction.

The right plots of figure \ref{fig:HesCop} show the ratio between the perturbed and gaussian copulas and allows the comparison between them. The upper right plot corresponds to the positive correlation and shows that the effect of the perturbed copula is to increase the density of the left queues of both underlyings. For the horizontal variable, the left queue density is increased for the points with less co-dependence (the upper left points increase the most and progressively increase less until the lower left corner is reached). For the vertical variable, the left queue density (the lower side of the plot) is again increased more for the points with less co-dependence (lower right side of the plot) and progressively decreases as the points with more co-dependence are reached (lower left corner). The lower right plot of figure \ref{fig:HesCop} corresponds to the negative correlation scenario. Again, the left queue densities are increased as it can be seen with the significant increase of density in the lower left corner of the plot, precisely where the co-dependence is the smallest. For higher values of the co-dependence (the anti-diagonal), the densities are displaced towards the lower left corner.

Figure \ref{fig:Hesjpdf} compares the joint probability functions given by equation (\ref{eq:jpdf}) using the gaussian copula (left plots), the perturbed copula (middle plots) and the ratio of both (right plots) for values greater than zero (plots are clearer this way). The upper plots correspond to a scenario with positive correlation whereas the lower plots correspond to negative correlation. The upper left and middle plots of figure \ref{fig:Hesjpdf} shows the positive correlation oval shape through the diagonal. The distribution obtained with the pertubed copula shows slightly displaced or deformed towards the upper right corner (as corresponds to the movement of the mode of the joint distribution as seen in the upper left plot of figure \ref{fig:HesCop}). The upper right plot of figure \ref{fig:Hesjpdf} shows how the left queues in the region with less co-dependence have more density for the perturbed copula than the gaussian copula (this is the same as happened with the upper right plot of figure \ref{fig:HesCop}. The lower left and middle plots of figure \ref{fig:Hesjpdf} show the anti-diagonal oval shape which corresponds to negatively correlated underlyings. Again, the distribution obtained with the perturbed copula appears as a deformation of the gaussian copula towards the upper right corner. On the other hand the lower right plot of figure \ref{fig:Hesjpdf} shows that the density is increased in two regions aligned with the anti-diagonal direction. Looking at the lower right plot of figure \ref{fig:HesCop} (the ratio of the perturbed and gaussian copula functions), the mentioned regions correspond to the density increase of the perturbed copula in the anti-diagonal belt.

The corresponding plots given by figures \ref{fig:HesCop} and \ref{fig:Hesjpdf} for the ``LL" scenario which would result for the rest of scenarios can be obtained through symmetries. For a ``RR" scenario, the right queues of both underlyings would be fatter going to the upper right corner and the mode of the joint distribution would be deformed in the opposite direction (towards the lower left corner) and the plots of figure \ref{fig:HesCop} would be symmetrical with respect to the anti-diagonal line. The plots of a ``RL" scenario with positive correlation would be symmetrical to the lower plots of figure \ref{fig:HesCop} with respect to a vertical axis situated in the middle of the plots (the mode of the distribution would be deformed towards the upper left corner). The same symmetries apply to a negative correlation scenario but the plots would be symmetric to the upper plots of figure \ref{fig:HesCop} with respect to the same axis. The plots of a ``LR" scenario with positive correlation would be symmetrical to the lower plots of figure \ref{fig:HesCop} with respect to a horizontal axis situated in the middle of each plot (the joint distribution would be deformed towards the lower right corner and the queues to the upper left corner). The same symmetries apply to the negative correlation scenario but the plots would be symmetric to the upper plots of figure \ref{fig:HesCop}.

If the asymptotic approximation of the perturbed copula were carried out up to second order, the smile effect could be captured. In this situation, the perturbed copula would increase the density around the mode of the distribution as well as both queues and will reduce the density in between.

\input{Hes1yITM.tex}

Table \ref{Hes1yITM} compares the prices given by the gaussian and perturbed copulas of equation (\ref{eq:Payoff}) for the given set of 25 scenarios varying skew configurations and correlation. The strike price is $K= 834.105$ and it is in-the-money. The columns present the prices of the gaussian ("Gcop") and perturbed ("Pcop") copulas and the difference of the perturbed minus the gaussian copulas ("P-G") in per unit of notional (e.g. 0.0001 is 0.01\% of notional or a basis point). The ``SS" scenarios (11 to 15) have been omitted because they give no correction (the pertubed copula can only capture the skew effect but not the smile). The interpretation in terms of pricing is in general not as clear. The perturbed copula provides positive corrections for the ``LL" scenarios (21 to 25). This is not surprising as the joint distribution of figure \ref{fig:Hesjpdf} is deformed towards the upper right corner (the direction in which the payment increases) and the queues don't pay out as it is a call option. For the positive correlation case (upper right plot of figure \ref{fig:Hesjpdf}), the queues increase the density in areas with higher payoff (the left queue of $S_T$ does not pay out as a call option is being priced). By looking at the negative correlation case (lower right plot of figure \ref{fig:Hesjpdf}) it is not as clear from a qualitative point of view that the correction should be positive. However, table \ref{Hes1yITM} shows that indeed it is positive and even greater than the positive correlation case. The corrections for scenario ``RR" (the symmetrical of ``LL") are of opposite sign and about the same magnitude as the correponding ``LL" scenarios. The ``LR" scenarios involve deforming the joint distribution towards the right lower corner of the distribution (the fatter queues move in the opposite direction towards the upper left corner). The left tail of $S_T$ does not pay out (it is a call option) and therefore the mode of the distribution moving to the right increases the payoff and the left tail getting fatter does not reduce the price as the payoff is zero. On the other hand, the payoff for $X_T$ gets lower on one side as the density is greater for lower values but also gets higher on the other as the density is increased for very high values (the right tail gets fatter). These effects might compensate with each other or even the effect of right tail (higher values of $X_T$) be more significant. This means that it is reasonable to think that the perturbed copula gives a positive correction for a ``LR" scenario (see scenarios 1 to 5 in table \ref{Hes1yITM}). Following a similar reasoning, the symmetric scenario ``RL" should give opposite corrections (see negative corrections of scenarios 6 to 10 in table \ref{Hes1yITM}).

This set of 25 scenarios was extended to cover three moneyness levels with strikes $K = 834.10, 981.30, 1128.50$ (in-the-money, at-the-money and out-of-the-money) and two maturities (1 and 2 years). This new set of scenarios was compared with a standard local volatility model (see for example \cite{Dupire1994} and \cite{Gatheral2006}) with constant correlation set equal to the correlation used for the copulas. This means that the noises of the two underlyings were correlated using constant correlation and the distribution of the underlyings was obtained simulating each of them with their corresponding local volatility until expiration. The comparison showed that the prices obtained with the gaussian copula were not any further than 10 basis points out of the whole set of scenarios from those of the Monte Carlo method with local volatility. For this set of scenarios, the maximum differences between the gaussian and perturbed copulas were obtained for the 2 year maturity and the in-the-money cases and it was 92 basis points. The conclusion of this study is that a standard local volatility model is rather equivalent to the gaussian copula and the corrections provided by the perturbed copula can reach up to almost 100 basis points (1\% of the notional).

\section{Case study: FX quanto options}
\label{sec:CaseStudy}

This section compares the perturbed copula, the gaussian copula and the Monte Carlo method with local volatility and constant correlation for a set of scenarios build out of a real market scenario for the same FX call option on XAU/USD quantoed to EUR considered in section \ref{sec:Interpretation}. This scenario corresponds to a ``LS" scenario with the XAU/USD highly left skewed and the EUR/USD very mildly right skewed (almost a smile). Five correlations (0.6, 0.3, 0, -0.3, -0.6), two maturities (1 and 2 years) and 5 moneyness levels (0.7, 0.85, ATM, 1.15 and 1.2) with strikes $K$ = (656.46, 797.13 , 937.79, 1078.47, 1125.36) were considered. The spot price of the XAU/USD and the EUR/USD are $S_0 = 937.79$ and $X_0 = 1.4029$. 
 
\begin{figure}[htbp]
	\centering
	\includegraphics[width=0.31\textwidth]{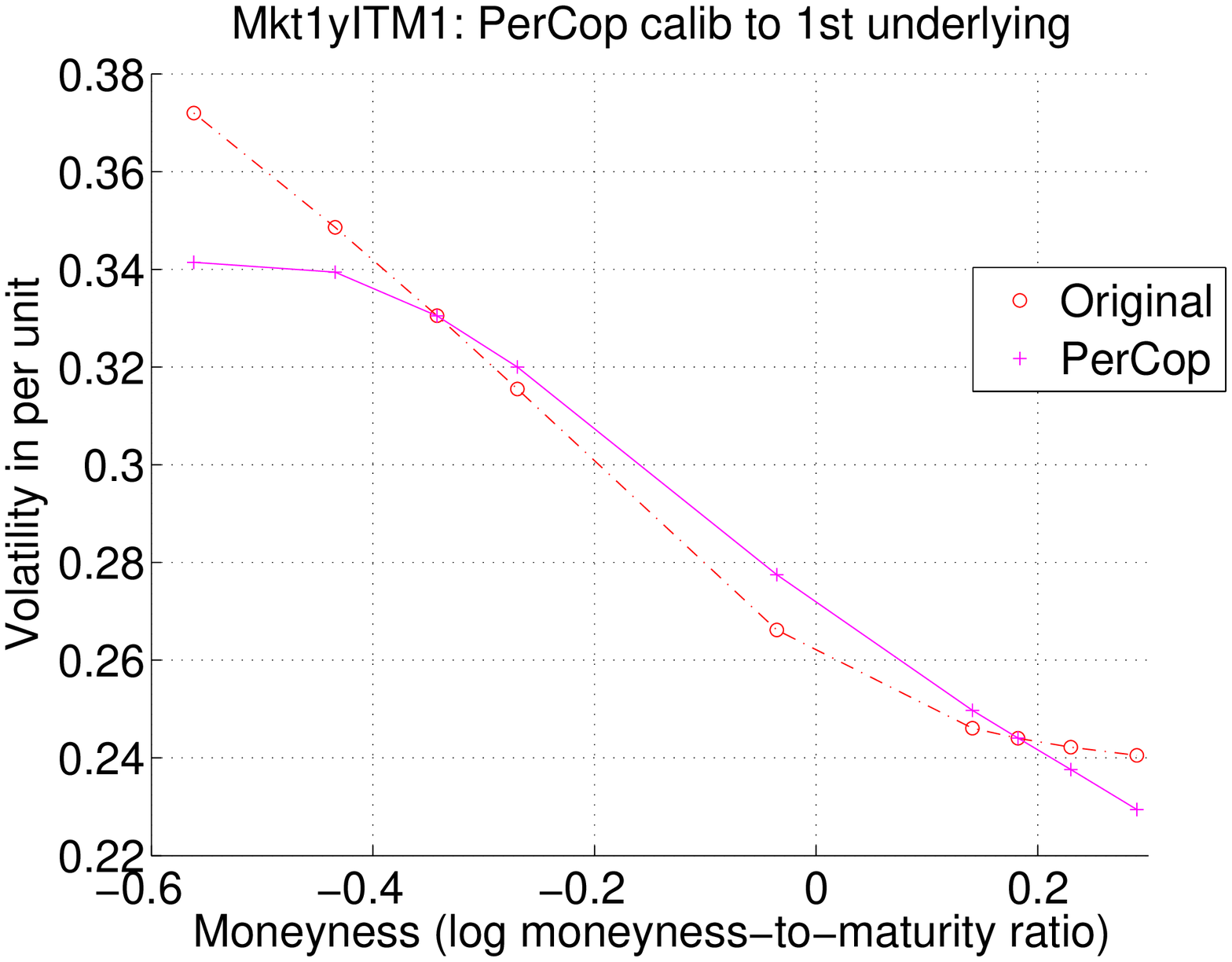}
	\includegraphics[width=0.31\textwidth]{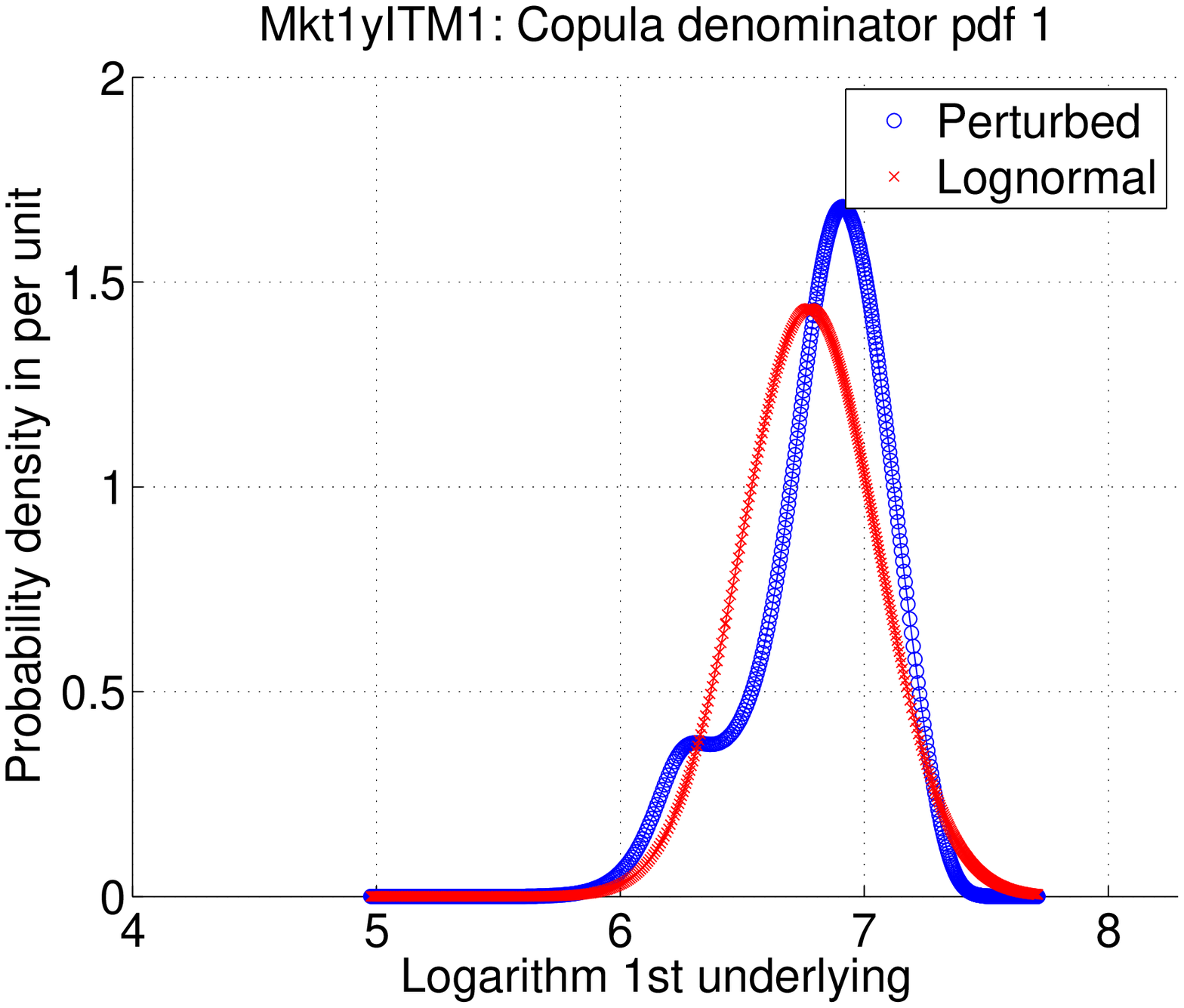}
	\includegraphics[width=0.31\textwidth]{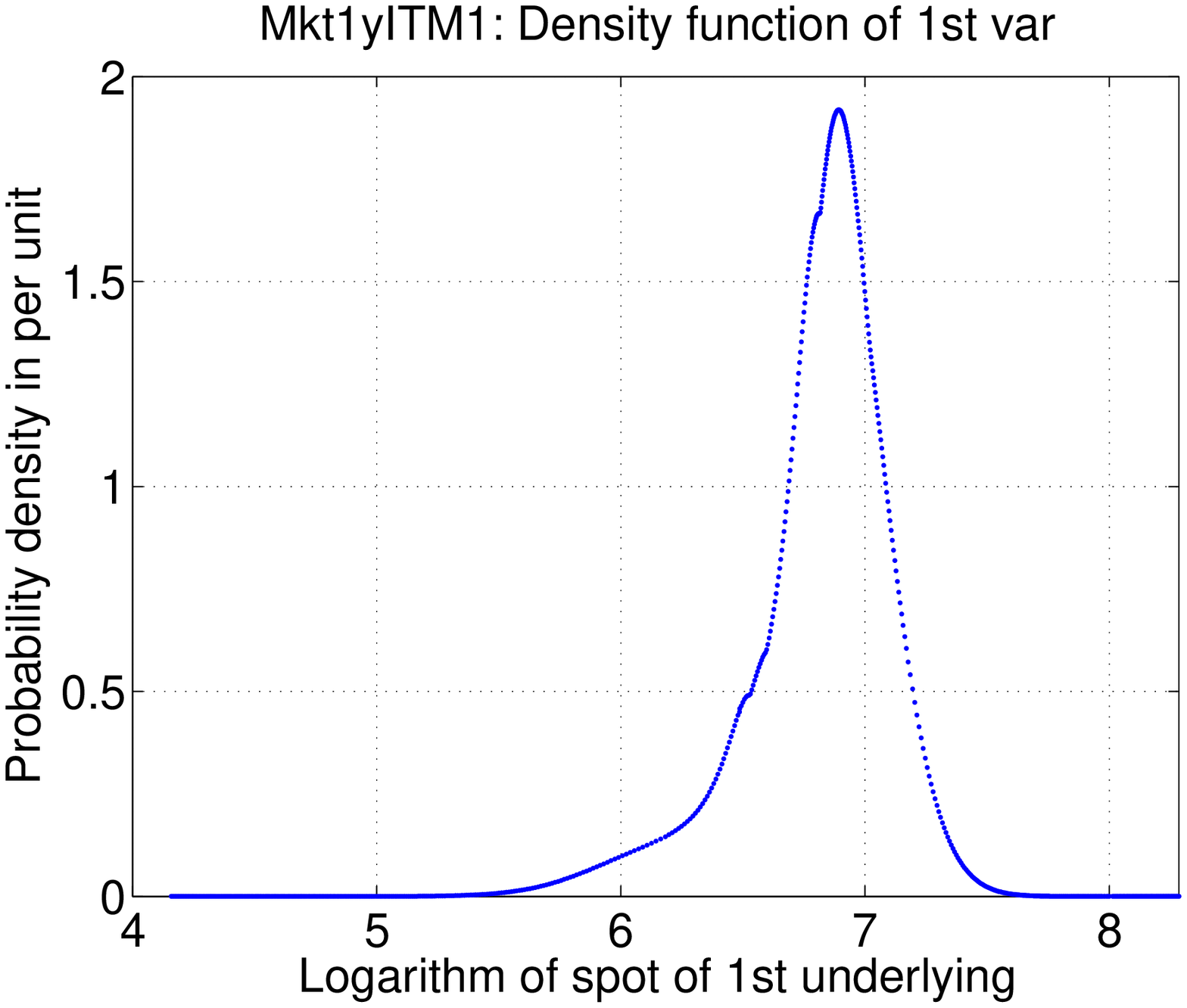}
	\includegraphics[width=0.31\textwidth]{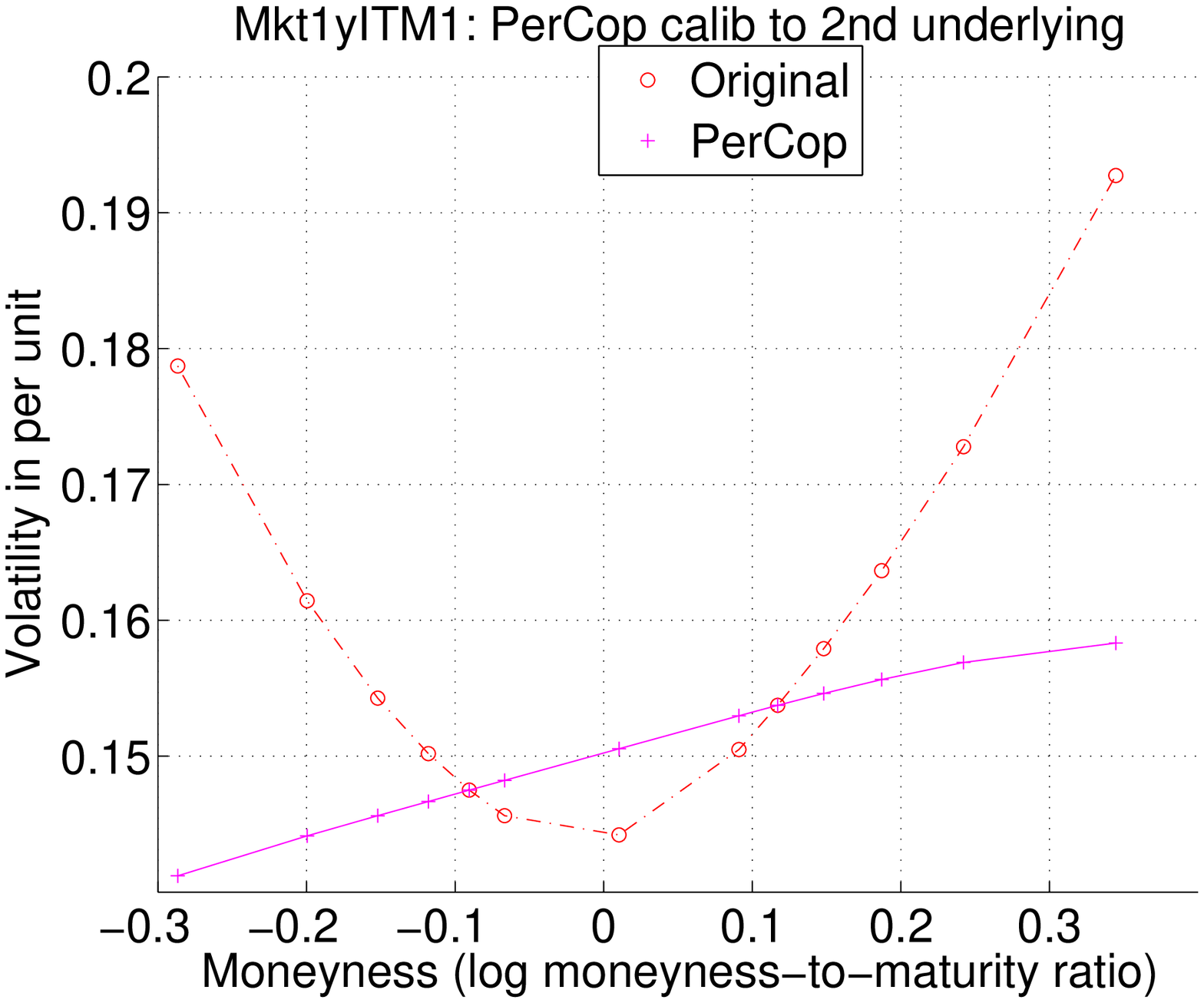}
	\includegraphics[width=0.31\textwidth]{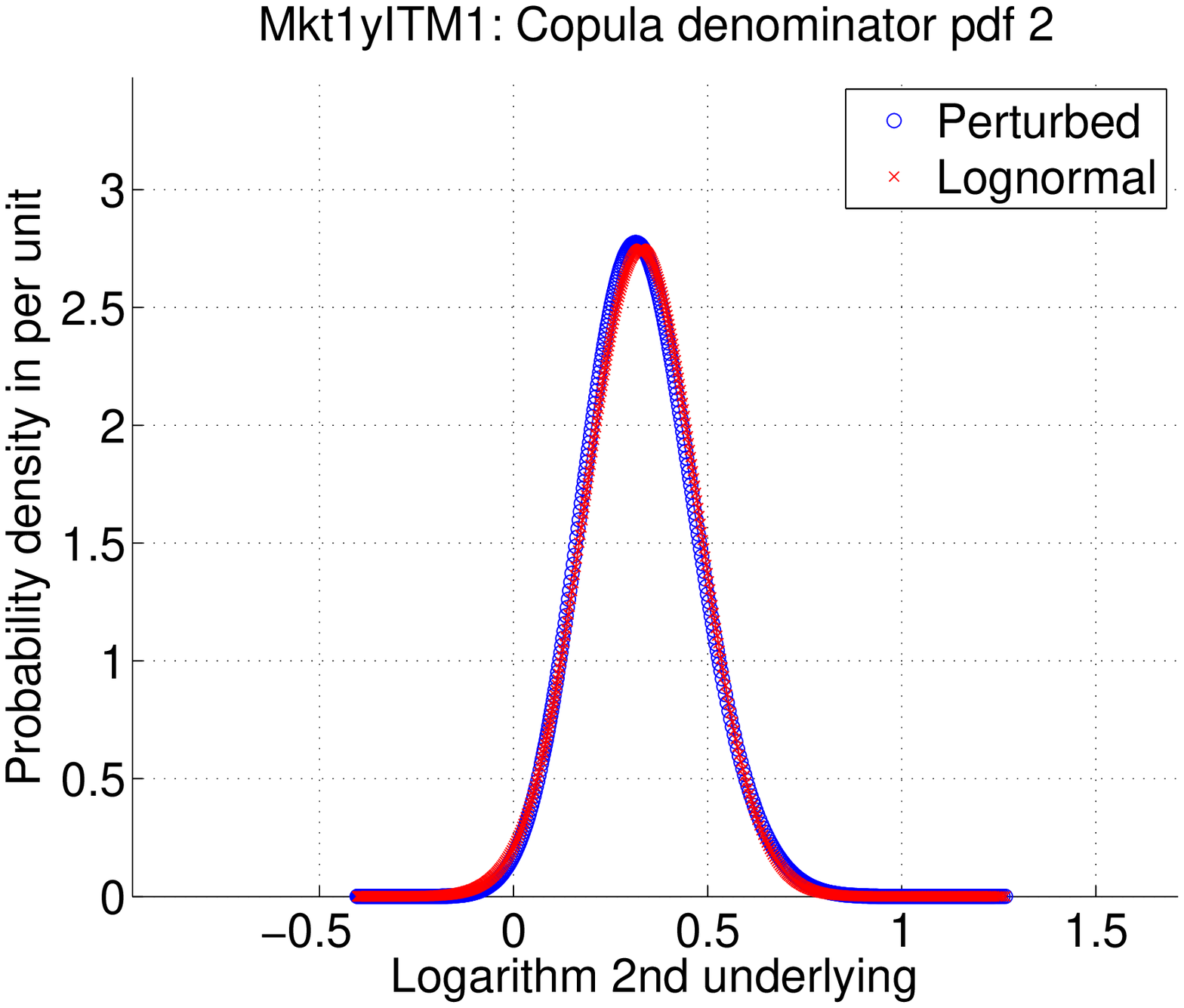}
	\includegraphics[width=0.31\textwidth]{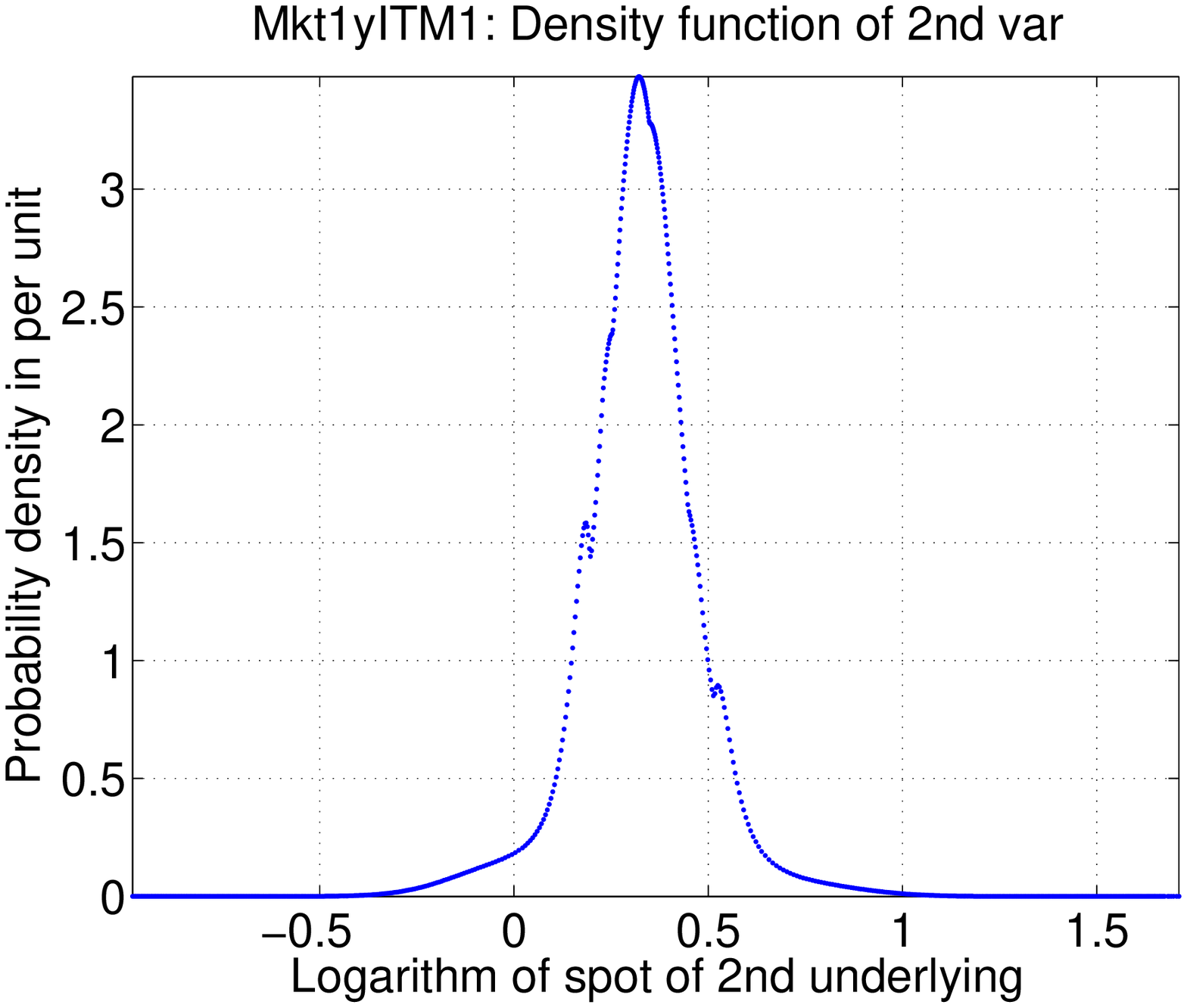}
	\caption{1 year implied volatility of left skew and smile market scenarios compared to their calibrations (left plots), corresponding perturbed pdf (middle plots) and empirical market pdf (right plots). Calibrated parameters: $\beta_1 = 6.7723$, $\sigma_1 = 0.2989$ and $R_1 = 73.43 \cdot 10^{-4}$ (XAU/USD up) and $\beta_2 = 0.3287$, $\sigma_2 = 0.1492$ and $R_1 = -1.08 \cdot 10^{-4}$ (EUR/USD down).}
	\label{fig:LScalib}
\end{figure}

Figure \ref{fig:LScalib} shows the 1 year implied volatilities (left plots), the calibrated perturbed marginal density (middle plots) and the empirical market marginal density (right plots). The upper plots correspond to the XAU/USD pair and the lower plots to the EUR/USD. The calibration parameters for the XAU/USD are $\beta_1 = 6.7723$, $\sigma_1 = 0.2989$ and $R_1 = 73.43 \cdot 10^{-4}$. For the EUR/USD they are $\beta_2 = 0.3287$, $\sigma_2 = 0.1492$ and $R_1 = -1.08 \cdot 10^{-4}$. See that the volatility levels are now more realistic than those of section \ref{sec:Interpretation} (instead of around 9\% they are now around 30\% and 15\%). The skew level of the XAU/USD is very high (around 50 times the skew levels of section \ref{sec:Interpretation}). The left plots of figure \ref{fig:LScalib} show the calibrated (labelled ``PerCop") and the original market (labelled ``Original") implied volatilities and the middle and right plots show the pertubed and empirical density functions (this figure has the same structure than figure \ref{fig:LScalib}). It can be seen that the calibration of the perturbed marginal density (see upper left plot of figure \ref{fig:LScalib}) of the XAU/USD to market is reasonable (very left skewed). If the perturbed and empirical densities are compared (upper middle and right plots), it can be seen that the pertubed copula skew is rather extreme and shows almost a bimodal distribution with a remarcable bump in the left queue (as already mentioned, the skew of the XAU/USD is around 50 times the skew considered in section \ref{sec:Interpretation}). This bimodal distribution fits the skew up to a level of moneyness of around -0.4. For lower moneyness levels, the skew flattens as the queue is not as fat as the empirical distribution for very low values of the underlying. Looking at the lower plots of figure \ref{fig:LScalib}, the calibration of the EUR/USD is rather poor but still mildly right skewed. This is the consequence of the fact that the perturbed marginal density can only capture skew but not smile. The empirical market density (lower right plot) has fatter queues and higher mode point than the perturbed density. However, the perturbed density has higher density in between the queues and the mode.

\begin{figure}[htbp]
	\centering
	\includegraphics[width=0.31\textwidth]{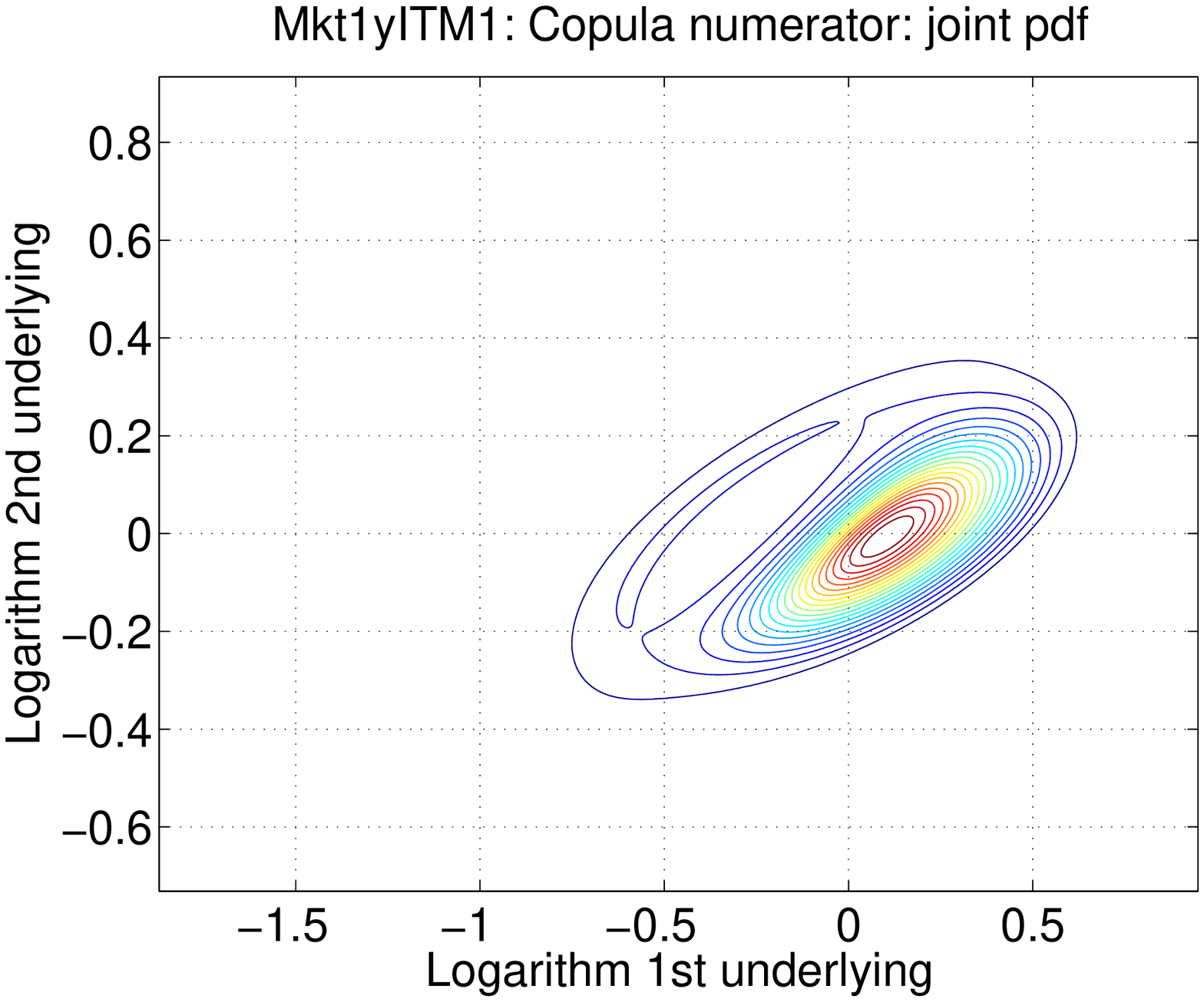}
	\includegraphics[width=0.31\textwidth]{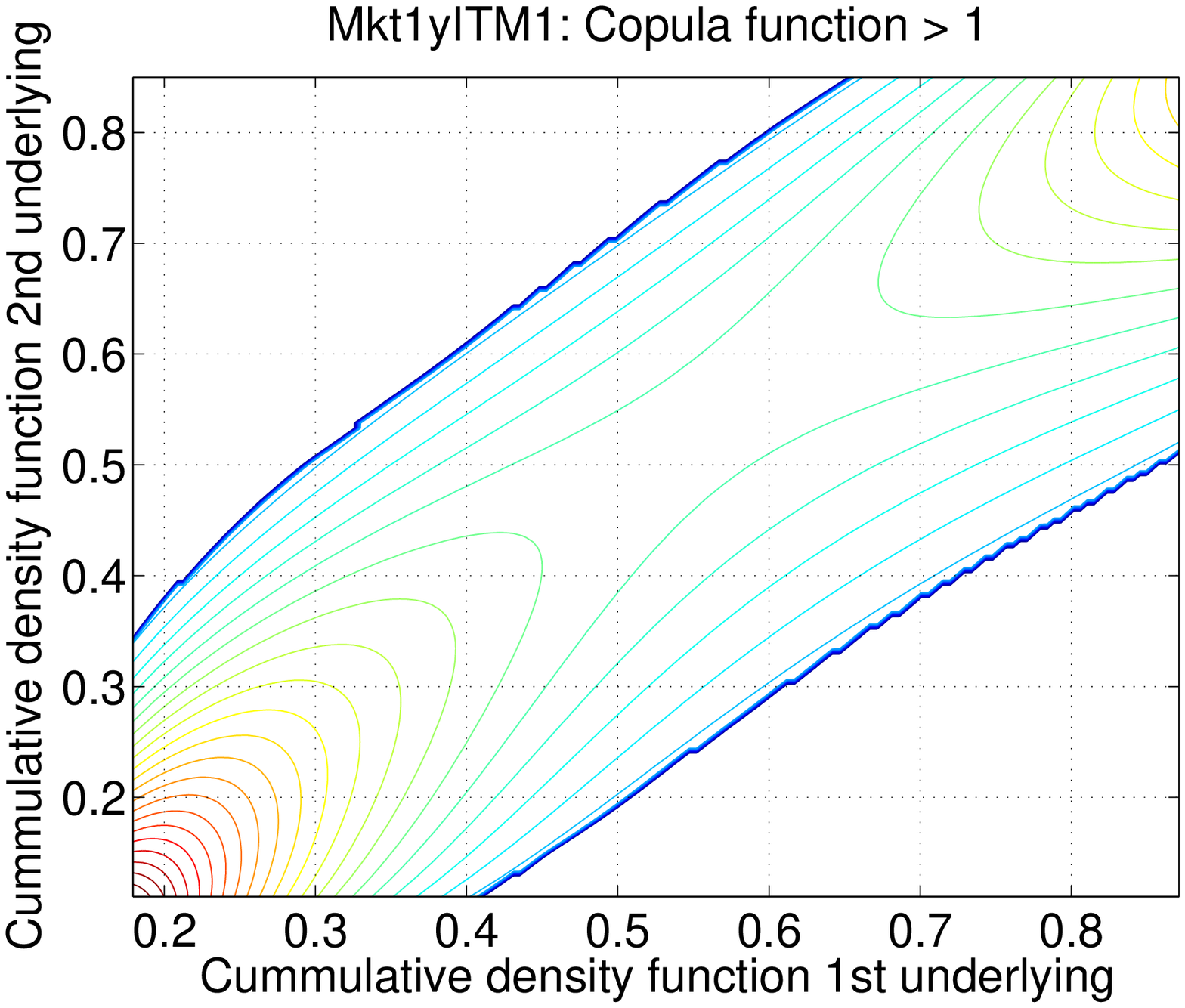}
	\includegraphics[width=0.31\textwidth]{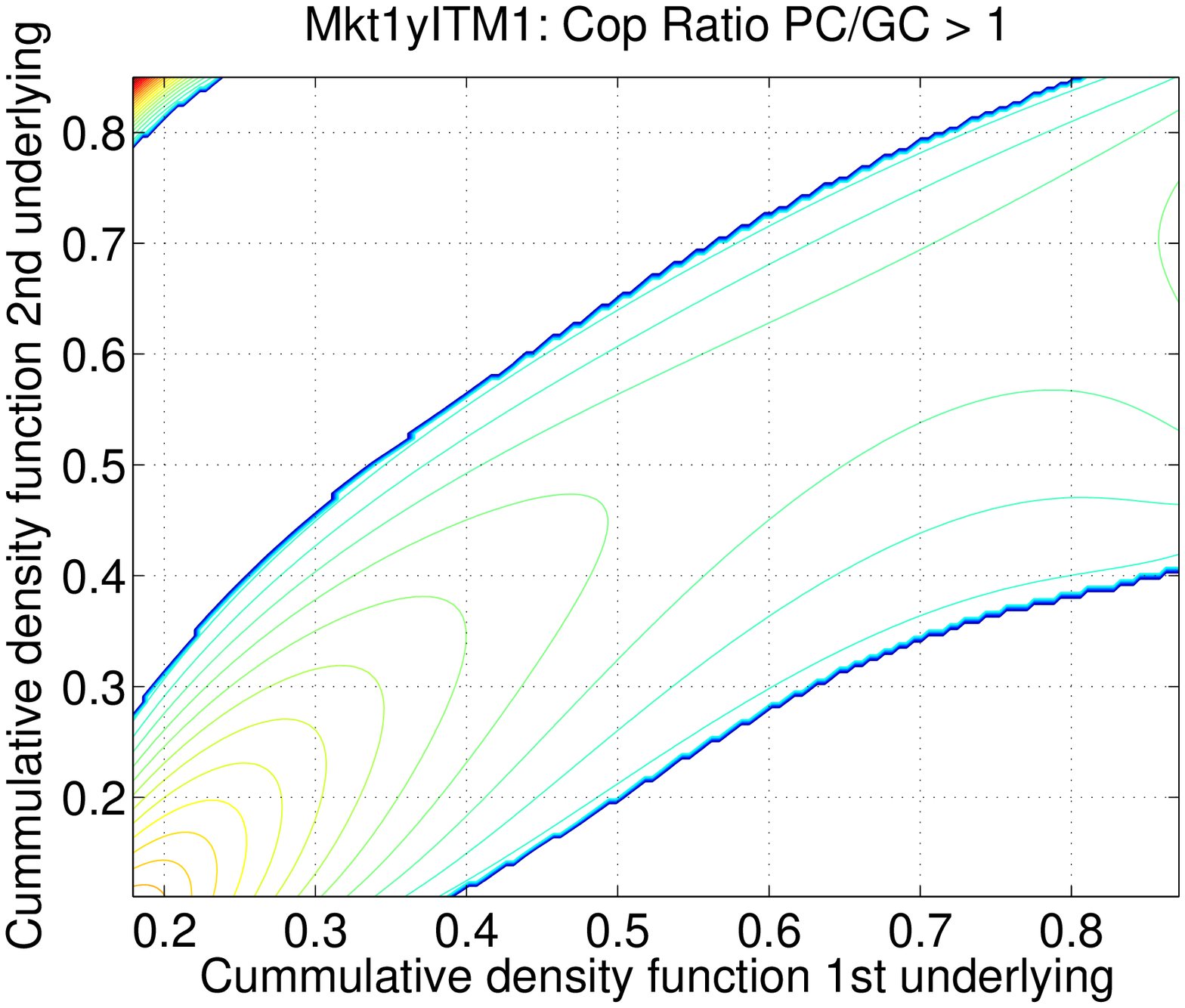}
	\includegraphics[width=0.31\textwidth]{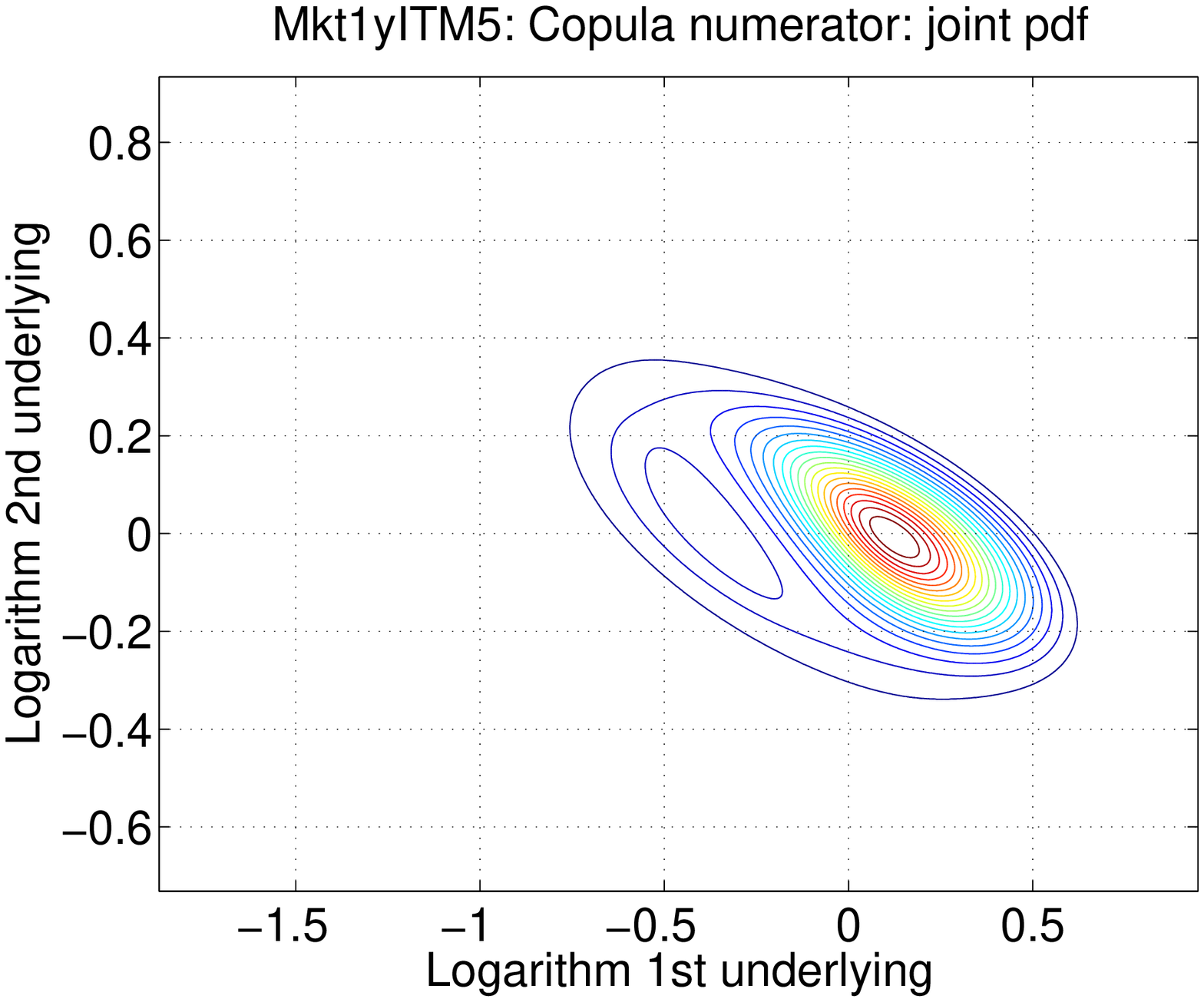}
	\includegraphics[width=0.31\textwidth]{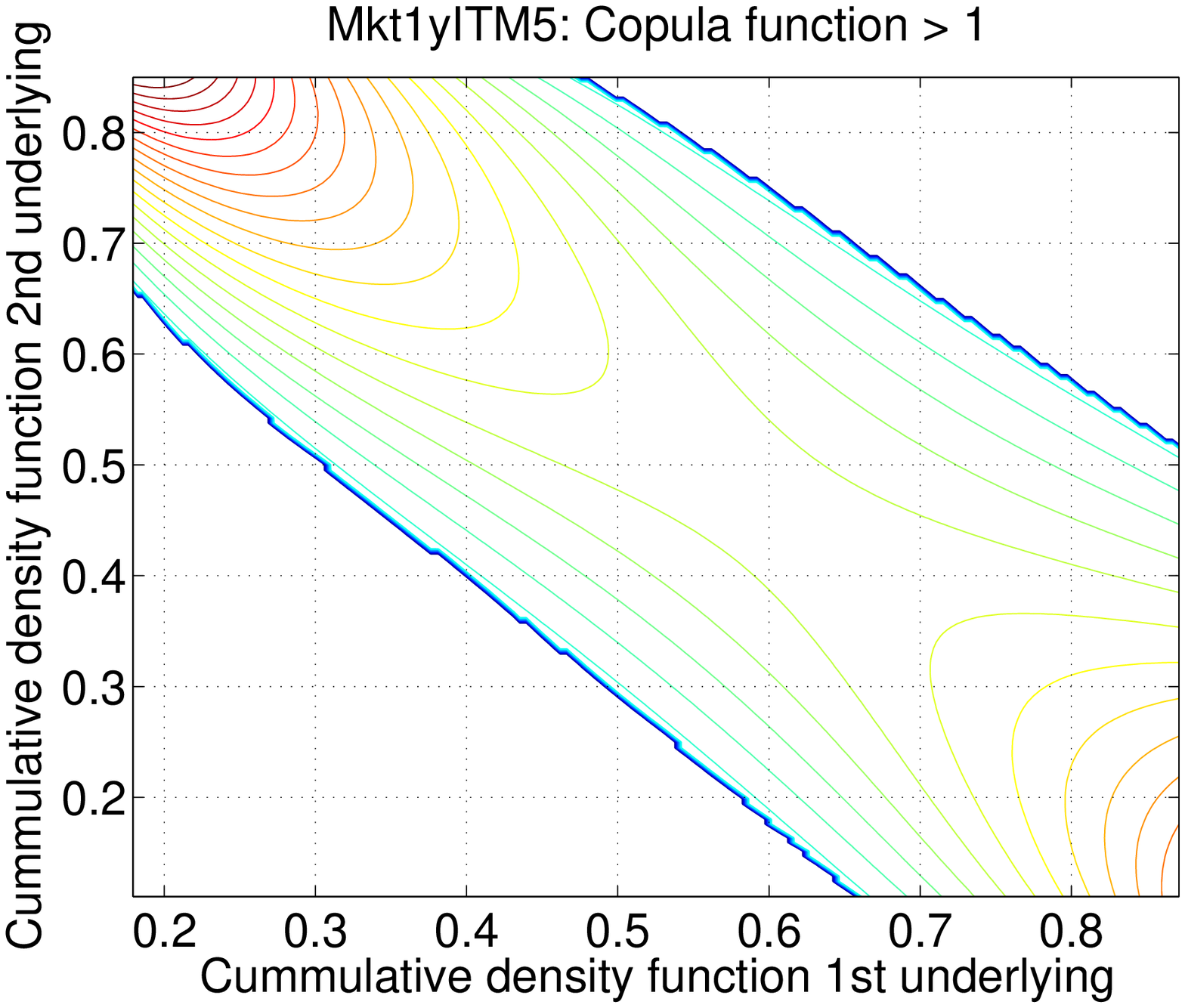}
	\includegraphics[width=0.31\textwidth]{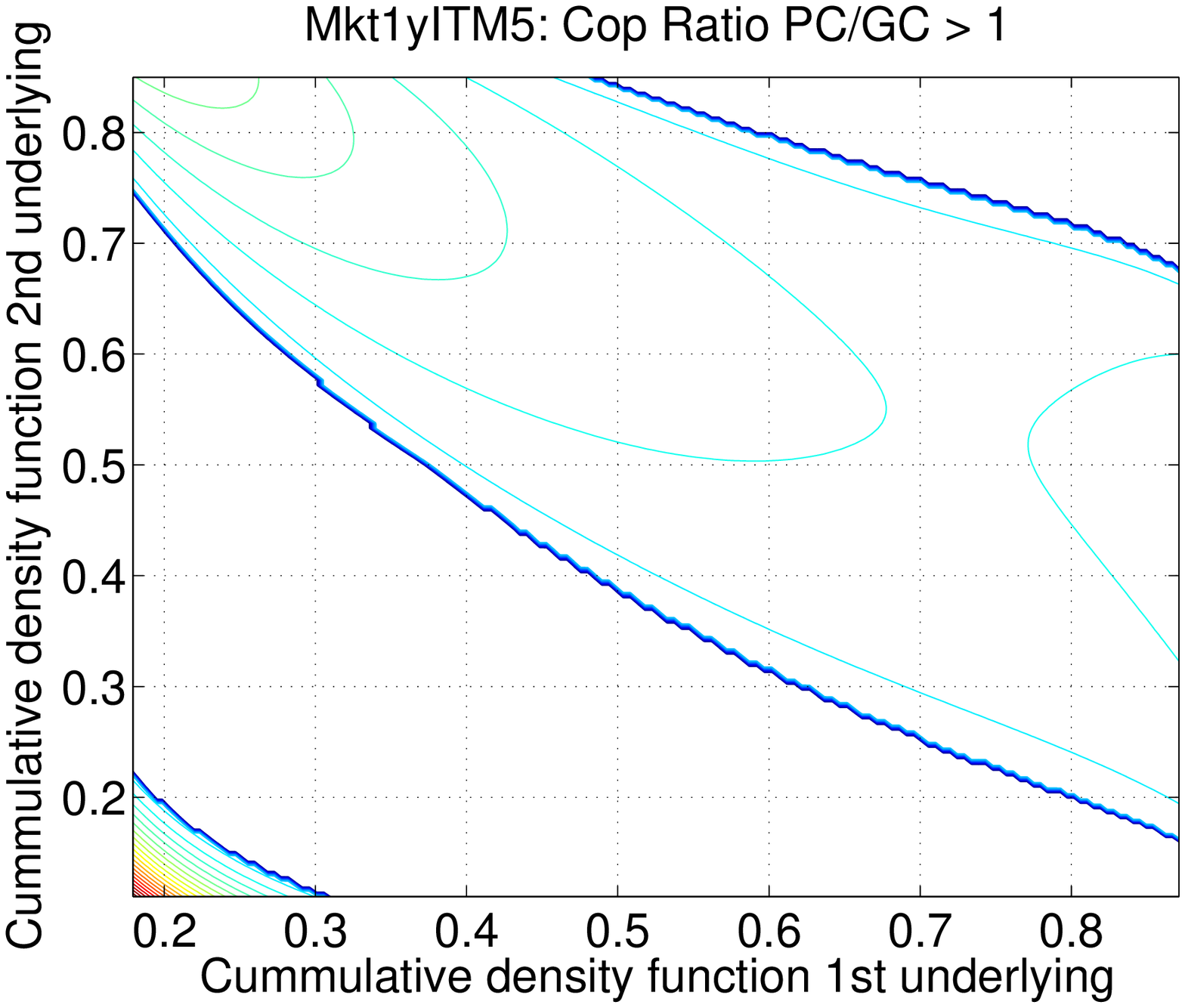}
	\caption{1 year copula numerator (left), copula function greater than 1 (middle) and ratio between perturbed and gaussian copulas (right) for values greater than one considering a LS skew market scenario with positively (upper plots) and negatively (lower plots) correlated underlyings.}
	\label{fig:MktCop}
\end{figure}

\begin{figure}[htbp]
	\centering
	\includegraphics[width=0.31\textwidth]{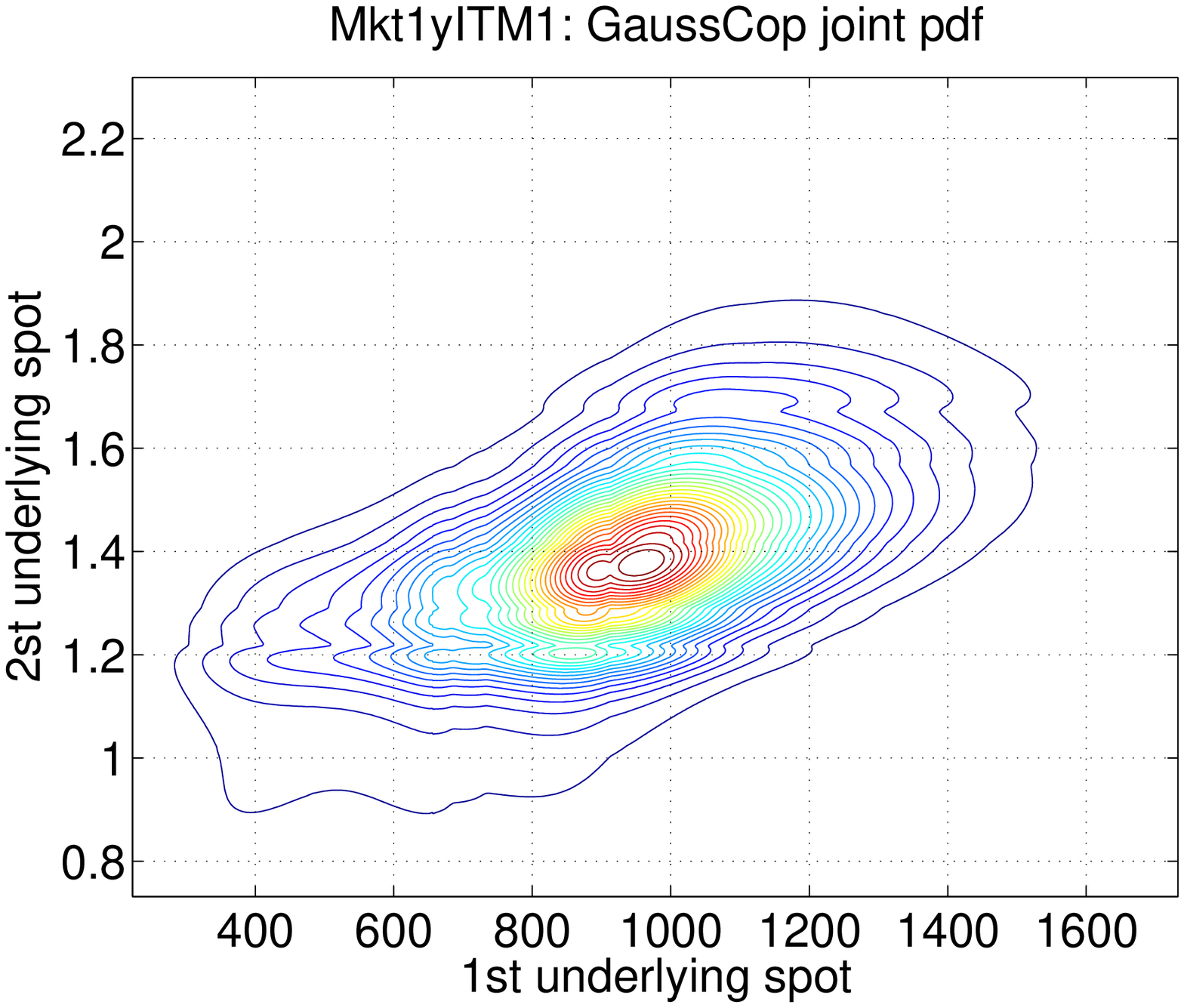}
	\includegraphics[width=0.31\textwidth]{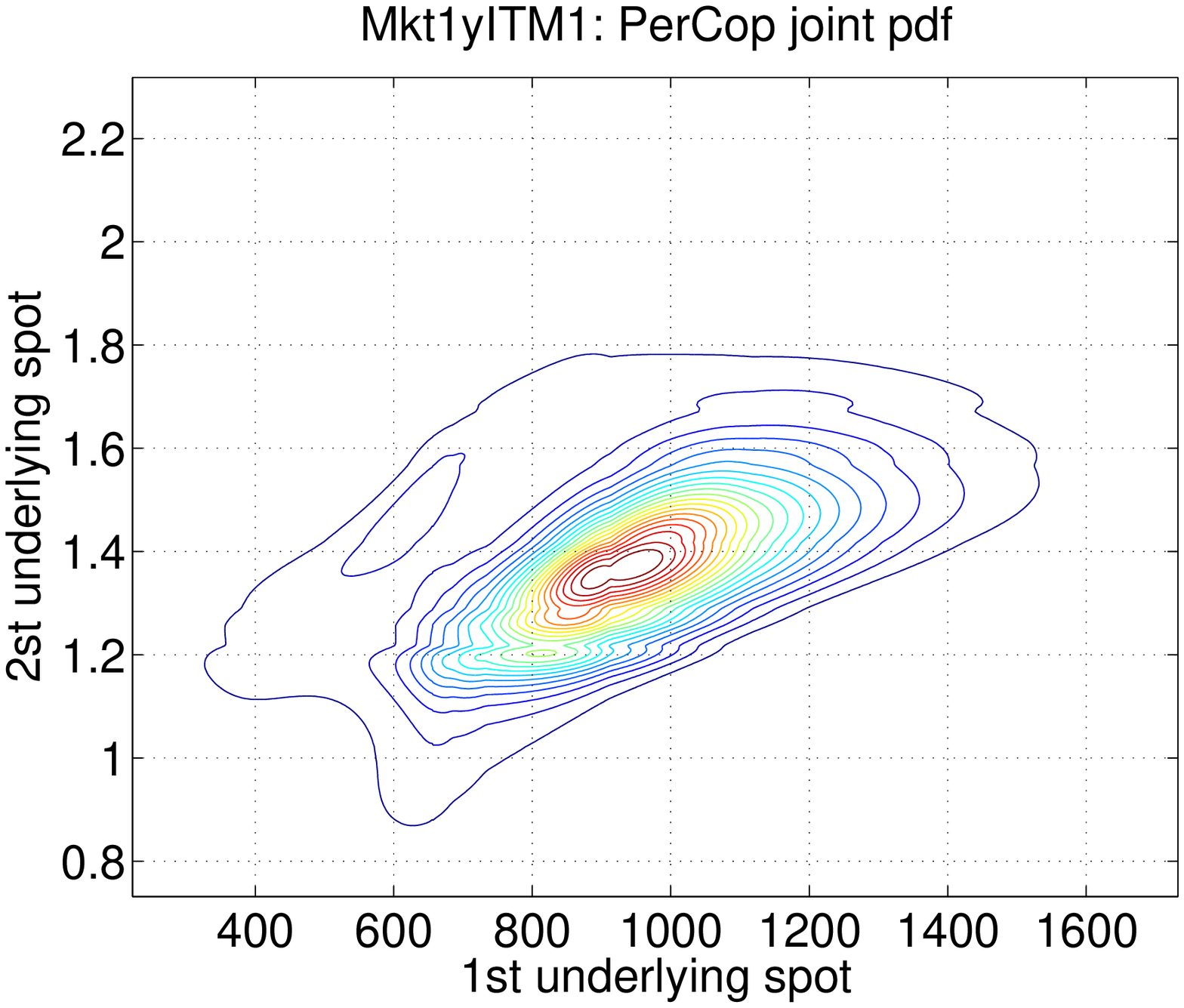}
	\includegraphics[width=0.31\textwidth]{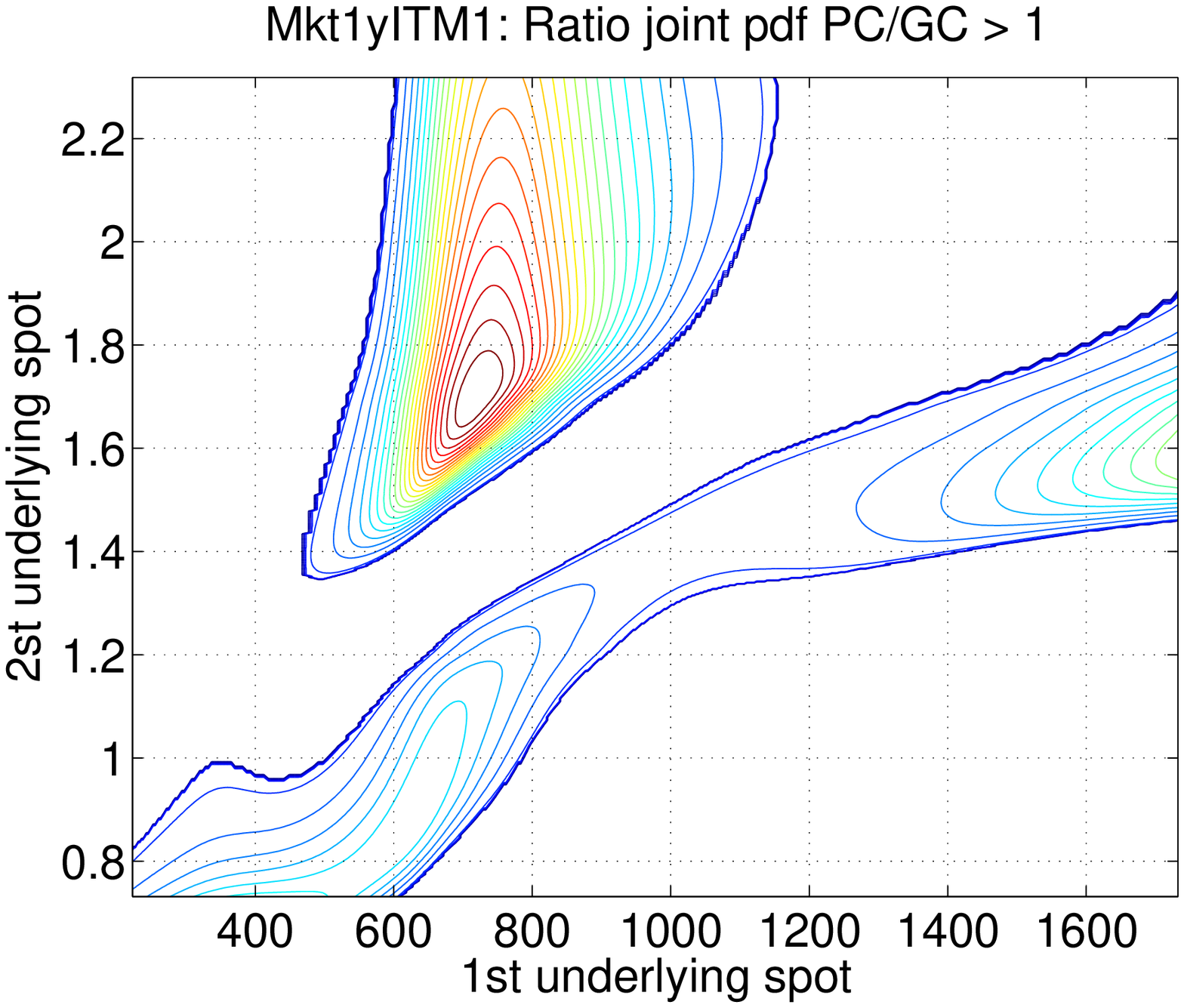}
	\includegraphics[width=0.31\textwidth]{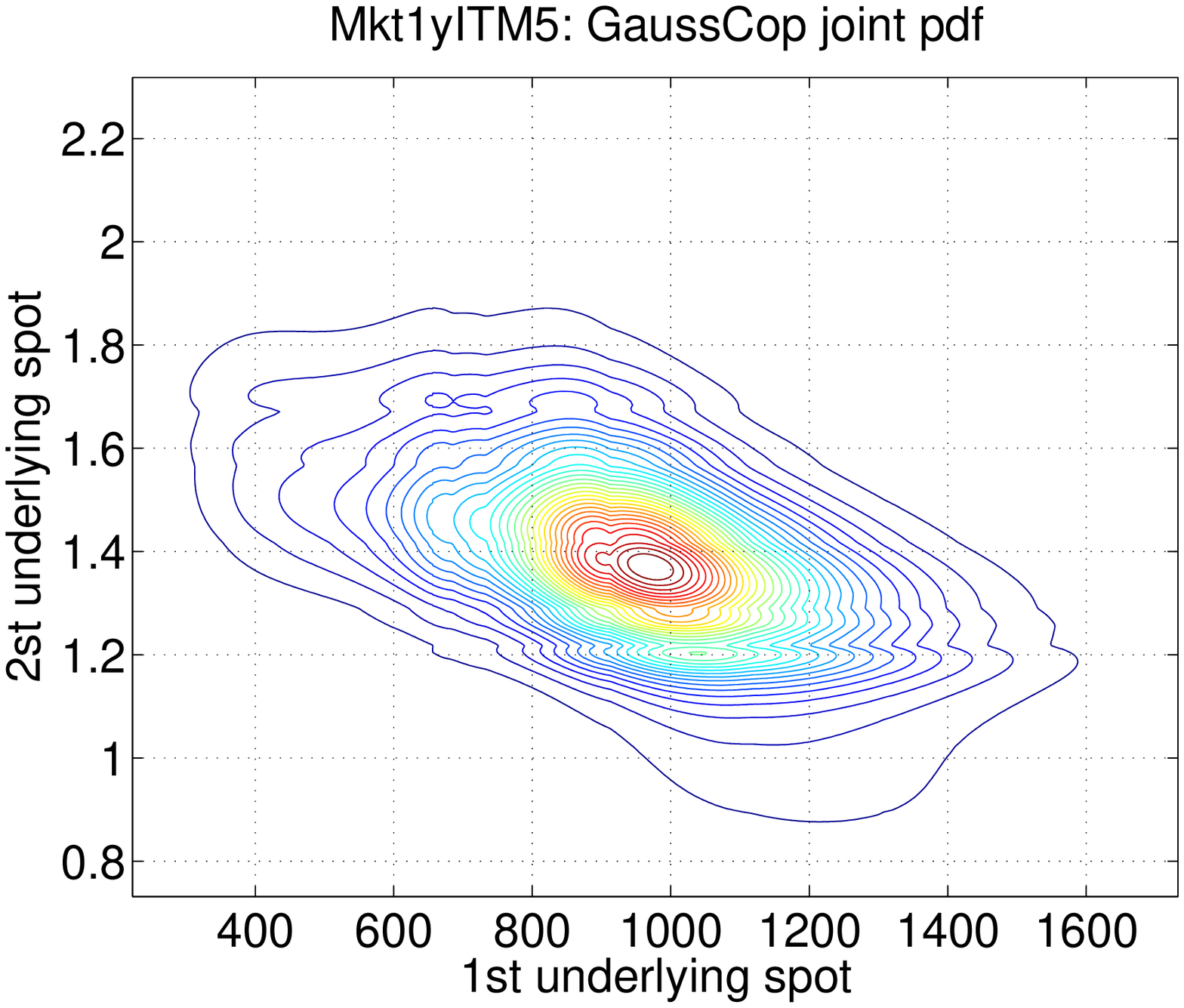}
	\includegraphics[width=0.31\textwidth]{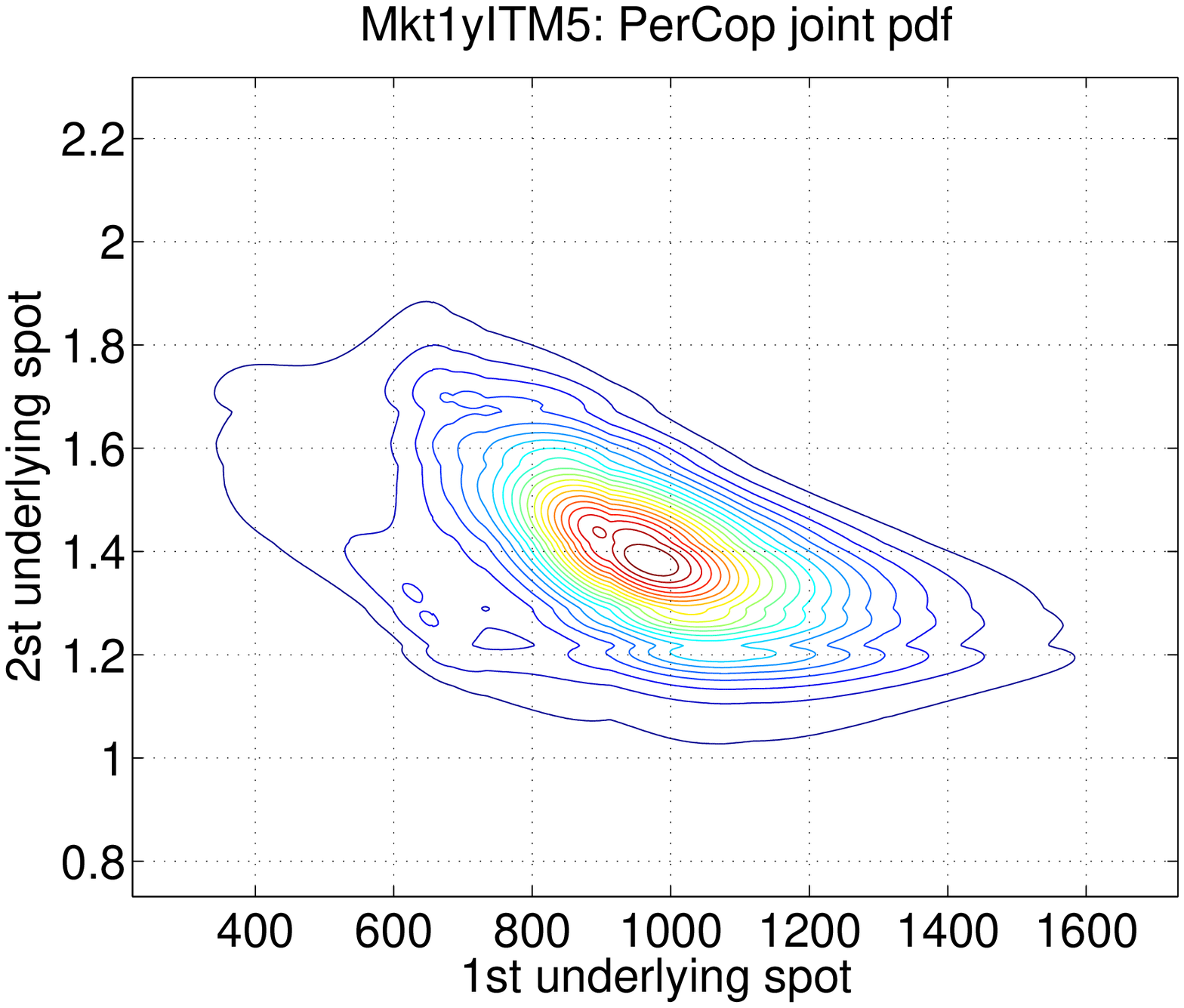}
	\includegraphics[width=0.31\textwidth]{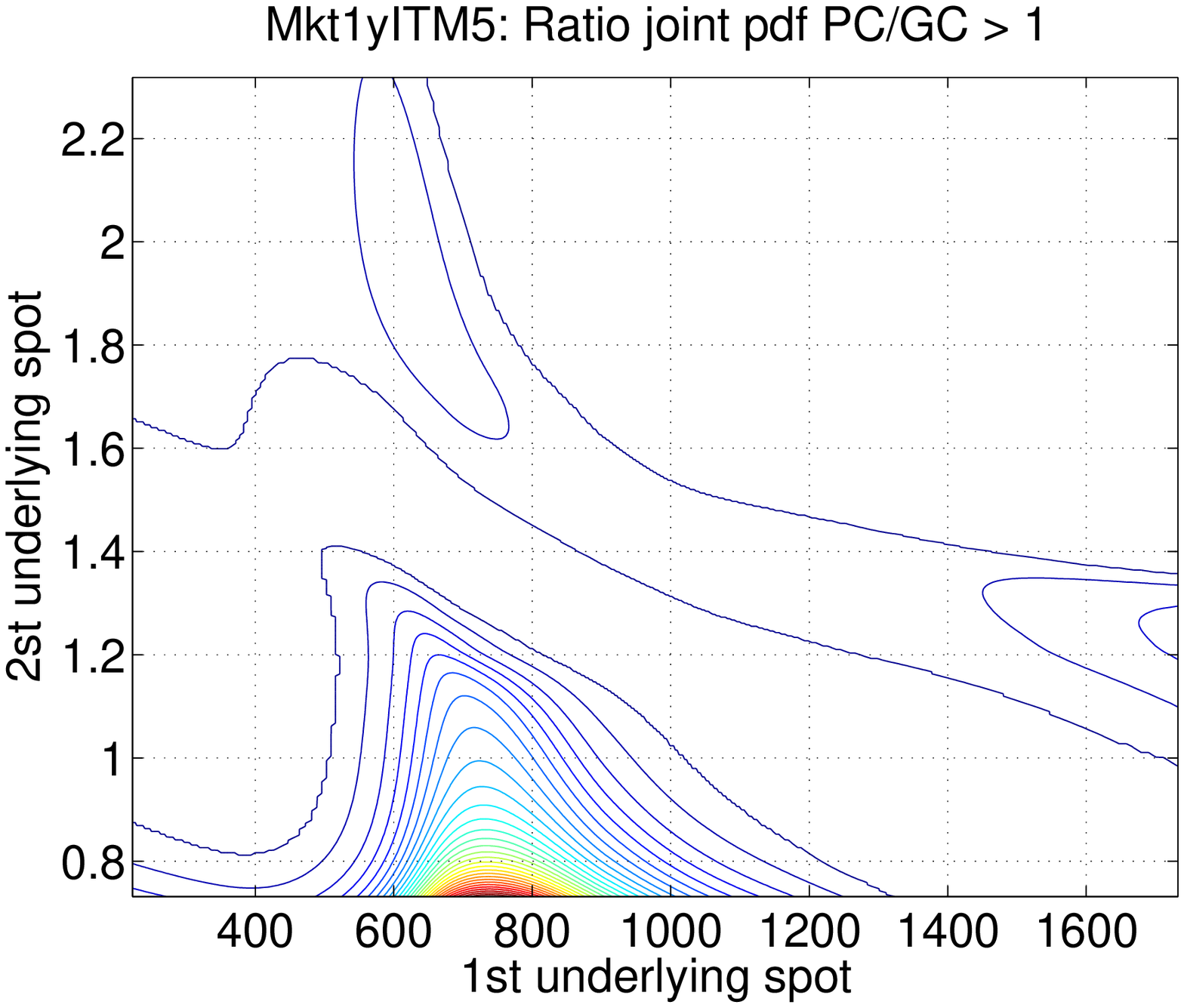}
	\caption{Joint pdf (copula function times empirical marginal densities) obtained with gaussian copula (left), perturbed copula (middle) and ratio between them (right) for values greater than one considering a LS market scenario with positively (upper plots) and negatively (lower plots) correlated underlyings.}
	\label{fig:Mktjpdf}
\end{figure}

\input{20111012_MktMatMon_P-G.tex}

\input{20090908_MktMatMon_LVATMcorr-G.tex}

Figure \ref{fig:MktCop} has the same structure than figure \ref{fig:HesCop} and presents the 1 year perturbed joint density functions given by equation (\ref{eq:jpdf}) (left plots), the perturbed copula functions given by equation (\ref{eq:fcop}) (middle plots) and the ratio of the perturbed over the gaussian copula for values greater than one (right plots). Again, these plots might be replicated using the calibrated parameters given in figure \ref{fig:LScalib} and calculating $R_{ij}$ and $Q_{ij}$ according to equation (\ref{eq:RijRed}). The upper and lower plots correspond respectively to positive and negative correlation between the XAU/USD (horizontal axis) and the EUR/USD (vertical axis). The upper and lower plots are approximately symmetrical to each other with respect to a horizonal axis in the middle of the plots. This is because the skew with respect to the EUR/USD (the vertical axis) is very mild (almost a smile). The lower plots of figure \ref{fig:MktCop} are very similar to the lower plots of figure \ref{fig:HesCop}. Now the joint distribution (lower left plot of figure \ref{fig:MktCop}) gets deformed only to the right (and not upwards to the right). In addition, only the left tail of the horizontal axis is fatter (the lower middle plot has higher copula values for the upper left corner and not as big for the opposite corner). Again, the lower right plot of figure \ref{fig:MktCop} shows how the left tail density increases more for the points with less co-dependence (lower left corner has the biggest increase in density with respect to the gaussian copula) and progressively increases less as the co-dependence rises.

Figure \ref{fig:Mktjpdf} presents the final joint density function from equation (\ref{eq:jpdf}) for 1 year maturity obtained with the gaussian (left plots) and perturbed (middle plots) copulas and the ratio of the perturbed over the gaussian density for values greater than one (right plots). The upper plots correspond to positive correlation and the lower to negative correlation between the XAU/USD and the EUR/USD. The effect of the left skew can be seen comparing the left and middle plots of figure \ref{fig:Mktjpdf}: the distribution gets deformed towards the right for the perturbed copula case compared to the gaussian copula (the mode of the distribution is moved towards the right and the tail gets fatter towards the left). The left tails of the XAU/USD rise the density starting from the points with less co-dependence and progressively increasing less when approching the points with greater co-dependence (these points also increase their density but as they already have big density, the rise by the effect of the skew is less than for the points with almost no co-dependence). The upper right plot of figure \ref{fig:Mktjpdf} shows a big area of increase of density in the upper middle side of the plot. This is the effect of the bump which appeared in the marginal density function (see the left tail of the pertubed marginal density of figure \ref{fig:LScalib}).  For the negative correlation scenario (lower plots of figure \ref{fig:Mktjpdf}), a symmetric behaviour  shows up with respect to a horizontal axis through the middle of the plots.

Table \ref{20111012_MktMatMon_P-G} shows the difference in basis points between the perturbed and gaussian copulas for the ``LS" scenario varying moneyness, maturity and correlation. In order to properly compare perturbed and Gaussian copulas, the correlations which appear in the columns of table \ref{20111012_MktMatMon_P-G} refer to those input in the Gaussian copula. The correlations used for the pertubed copula are implied so that the quanto forward, ${\bf E}\left( {S_T X_T } \right)$, is equal for both the Gaussian and perturbed copulas. The upper group of rows show different moneyness levels for 1 year maturity and the lower group of rows for 2 year maturity. Differences increase with maturity and they can get beyond 40 basis points (0.4\% of the notional). As the overall effect of the skew of the XAU/USD is to deform the joint distribution towards higher values of the horizontal axis (in this direction the payoff is higher), the corrections given by the perturbed copula are positive for almost every case. The fact that the left tail gets fatter does not have a big impact as the payoff for the left tail is zero (a call option is considered). It is not easy to interpret the premium corrections in terms of the correlation (whether positive correlation should correct more or less than negative correlation). In fact, the same qualitative plot given by figure \ref{fig:Mktjpdf} appears for scenarios which invert the size of corrections between positive and negative correlations (all corrections were however positive for these cases).

Table \ref{20090908_MktMatMon_LVATMcorr-G} is similar to table \ref{20111012_MktMatMon_P-G} but shows the difference of the Monte Carlo method with local volatility and constant correlation minus the gaussian copula. See that the biggest difference does not go beyond 17 basis points. This means that the Monte Carlo method is rather equivalent to the gaussian copula (as it was concluded in section \ref{sec:Interpretation} varying moneyness, maturity and all combinations of skew).

\section{Conclusions}
\label{sec:Conclusions}

The perturbed copula approach of \cite{Fouque2006} has been successfully applied for valuation of derivatives which depend on two underlyings for which prior information about their co-dependence is unknown. The application of this perturbed copula allows to introduce the skew information of the underlyings in the co-dependence. The perturbed copula formulation has been generalized to deal with widely used lognormal-inspired underlyings where the skew is introduced through the dependence of the volatility with respect to a common stochastic volatility factor. The analytic formula of the perturbed copula is obtained through an asymptotic expansion under the assumption that this common factor is fast mean reverting.

The original formulation of the peturbed copula has been simplified through additional hypothesis which allow to reduce the copula to five intuitive and easy to interpret parameters: two control the volatility levels, two the skew levels and one the correlation between both underlyings.

An exact fit calibration procedure of this five parameters is proposed giving initial values for them which are close enough to the solution so that a regular Newton-Raphson search algorithm finds the solution quickly.

The effect of the perturbed copula is interpreted in comparison with the gaussian copula in terms of the direction of the skew and the correlation of each underlying. Intuitive criteria are provided to qualitatively predict the effects of the perturbed copula in the pricing of a derivative. These qualitative ideas are applied to the particular case of FX options quantoed to a currency different from the currencies of the underlying pair.

A real market case study is also analyzed for FX quanto options to a third currency. It has been seen that the price impact of considering the skew in the co-dependence is not negligible (it can go beyond 40 basis points or 0.4\% of the notional amount in some of the cases analyzed). This rises a warning concern about the risk of quanto options in the presence of skew.

The cases analyzed have also been compared with a Monte Carlo local volatility model with constant correlation equal to the correlation used in the copulas. It has been concluded that this method is almost equivalent to the gaussian copula. This means that the regular widely used local volatility models do not incorporate the skew effect in the co-dependence and some potential model risk might be unidentified.

\appendix
\section{Initial parameters for calibration}
\label{app:InitialParams}

This appendix derives the initial guess of the calibration parameters. This derivation has already been carried out in \cite{Fouque2000} and therefore it will be omitted here. However, there are a few subtle differences in the derivation process which are addressed in this section.

Consider the system $S_t^{(i)}$ and $Y_t$ of stochastic equations (\ref{eq:ProcProbNeutral}). The same procedure used for the derivation of the perturbed copula is followed here to obtain the asymptotic expansion of the vanilla option price. The operators $\mathcal{L}_1$ and $\mathcal{L}_2$ given by equations (\ref{eq:L1}) and (\ref{eq:L2}) are replaced by $\tilde \mathcal{L}_1$ and $\tilde \mathcal{L}_2$ given by equations (\ref{eq:L1Und}) and (\ref{eq:L2Und}). The operator $\mathcal{L}_0$, which corresponds to the Ornstein-Uhlenbeck process, does not change. See that the operator $\tilde \mathcal{L}_2$ is the operator $\mathcal{L}_{BS}$ resulting out of the Black Scholes process with volatility equal to $f(y)$. If the underlying process is denoted by $S_t^{(i)}$ and the underlying variable by $S_i$, $S_i$ multiplies the partial derivative $\frac{\partial }{\partial S_i \partial y}$ and $S_i^2$ multiplies $\frac{\partial ^2}{\partial S_i^2}$ because the lognormal-inspired process is considered instead of the normal process of equation (\ref{eq:PerCopModel}) used for the derivation of the copula. The term $(r-q) S_i \frac{\partial}{ \partial S_i}$ corresponds to the drift of the process $S_t^{(i)}$. The dot notation at the end of (\ref{eq:L2Und}) means the function after the operator and this last term corresponds to the discounting.

\begin{equation}
\tilde \mathcal{L}_1  = \nu \sqrt 2 \rho _{iY} f_i \left( y \right)S_i \frac{{\partial ^2 }}{{\partial S_i \partial y}} - \sqrt 2 v\Lambda \left( y \right)\frac{\partial }{{\partial y}}
  \label{eq:L1Und}
\end{equation}

\begin{equation}
\tilde \mathcal{L}_2  = \frac{\partial }{{\partial t}} + \frac{1}{2}f_i^2 \left( y \right)S_i^2 \frac{{\partial ^2 }}{{\partial S_i^2 }} + r_t\left( {\frac{{\left( {r_t - q_t^{(i)}} \right)}}{r_t}S_i \frac{\partial }{{\partial S_i }} -  \cdot } \right) = \mathcal{L}_{BS} \left( {f(y)} \right)
  \label{eq:L2Und}
\end{equation}

If the same procedure of section \ref{sec:Formulation} is followed, equations (\ref{eq:PDE_P0Und}) and (\ref{eq:PDE_P1Und}) are obtained. These equations give the zero and first order terms of the solution, which are now denoted by $P_0$ and $P_1$. Equation (\ref{eq:PDE_P0Und}) is equivalent to equation (\ref{eq:PDEu0}) and gives the zero order term $P_0$ whose final condition corresponds to the vanilla call ($\varphi = 1$) or put ($\varphi = -1$) option with strike $K$. When the expectation of $\mathcal{L}_2$ with respect to the invariant distribution is calculated, the Black Scholes operator $\mathcal{L}_{BS}(\bar \sigma_i)$ is obtained. This means that the zero order solution is the vanilla Black Scholes price with volatility equal to $\bar \sigma _i$. To obtain the first order term $P_1$, consider equation (\ref{eq:L0m1_L2TildeUnd}) which is equivalent to equation (\ref{eq:invL0}). Now, there is only one term which corresponds to underlying $i$ (the other underlying and the correlation term are not present). The function $\tilde \phi_i$ is the solution of a Poisson equation similar to the first equation of (\ref{eq:PoissonEq}).

\begin{equation}
\begin{array}{l}
 \left\langle {\mathcal{L}_2 } \right\rangle P_0  = \mathcal{L}_{BS} \left( {\bar \sigma _i } \right)P_0  = 0 \\ 
 P_0 \left( {T,S_i } \right) = \left[ {\varphi (S_i  - K)} \right]^ +   \\ 
 \end{array}
  \label{eq:PDE_P0Und}
\end{equation}

\begin{equation}
\left\langle {\tilde \mathcal{L}_2 } \right\rangle P_1  = \left\langle {\tilde \mathcal{L}_1 \mathcal{L}_0^{ - 1} \left( {\tilde \mathcal{L}_2  - \left\langle {\tilde \mathcal{L}_2 } \right\rangle } \right)} \right\rangle P_0 
  \label{eq:PDE_P1Und}
\end{equation}

\begin{equation}
\mathcal{L}_0^{ - 1} \left( {\tilde \mathcal{L}_2  - \left\langle {\tilde \mathcal{L}_2 } \right\rangle } \right)P_0  = \frac{1}{2}\tilde \phi _i \left( y \right)S_i^2 \frac{{\partial ^2 P_0 }}{{\partial S_i^2 }}
  \label{eq:L0m1_L2TildeUnd}
\end{equation}

Applying the operator $\tilde \mathcal{L}_1$ to equation (\ref{eq:L0m1_L2TildeUnd}), taking expectations with respect to the limit distribution of $\mathcal{L}_0$ and multiplying by $\sqrt{\epsilon}$ yields equation (\ref{eq:V2V3new}), where $\tilde R_i$ and $U_i$ are given by equation (\ref{eq:RiUiDef}). See that when the operator $\tilde \mathcal{L}_1$ is applied, the only function which depends on $y$ is $\tilde \phi_i(y)$.

\begin{equation}
\begin{array}{l}
 \left\langle {\tilde \mathcal{L}_1 \mathcal{L}_0^{ - 1} \left( {\mathcal{L}_2  - \left\langle {\mathcal{L}_2 } \right\rangle } \right)} \right\rangle \sqrt \varepsilon  P_0  = \tilde R_i S_i \frac{\partial }{{\partial S_i }}\left( {S_i^2 \frac{{\partial ^2 P_0 }}{{\partial S_i^2 }}} \right) + U_i S_i^2 \frac{{\partial ^2 P_0 }}{{\partial S_i^2 }} \\ 
 \begin{array}{*{20}c}
   {} & {} & {}  \\
\end{array} = \left( {2\tilde R_i  + U_i } \right)S_i^2 \frac{{\partial ^2 P_0 }}{{\partial S_i^2 }} + \tilde R_i S_i^2 \frac{{\partial ^3 P_0 }}{{\partial S_i^3 }} \\ 
 \end{array}
  \label{eq:V2V3new}
\end{equation}

\begin{equation}
\tilde R_i  = \frac{{\nu \rho _{1Y} \sqrt \varepsilon  }}{{\sqrt 2 }}\left\langle {f_i \tilde \phi _i '} \right\rangle \begin{array}{*{20}c}
   {} & {} & {}  \\
\end{array}U_i  = - \frac{{ \sqrt 2 }}{2}v\left\langle {\Lambda \tilde \phi _i '} \right\rangle 
  \label{eq:RiUiDef}
\end{equation}

If equation (\ref{eq:PDE_P1Und}) is multiplied by $\sqrt{\epsilon}$, the expectation of $\tilde \mathcal{L}_2$ with respect to the invatiant distribution is replaced in the left hand side of equation (\ref{eq:PDE_P1Und}) and equation (\ref{eq:V2V3new}) is replaced in the right hand side of equation (\ref{eq:PDE_P1Und}), equation (\ref{eq:LBS_P1}) is obtained. This equation is the same as (5.36) of \cite{Fouque2000}. Identifying terms yields $V_2$ and $V_3$ as defined in \cite{Fouque2000} in terms of $\tilde R_i$ and $U_i$ according to equation (\ref{eq:V2V3}).

\begin{equation}
\mathcal{L}_{BS} \left( {\bar \sigma _i } \right)\left( {\sqrt \varepsilon  P_1 } \right) = \left( {2\tilde R_i  + U_i } \right)S_i^2 \frac{{\partial ^2 P_0 }}{{\partial S_i^2 }} + \tilde R_i S_i^2 \frac{{\partial ^3 P_0 }}{{\partial S_i^3 }}
  \label{eq:LBS_P1}
\end{equation}

\begin{equation}
V_2  = 2\tilde R_i  + U_i \begin{array}{*{20}c}
   {} & {} & {}  \\
\end{array}V_3  = \tilde R_i 
  \label{eq:V2V3}
\end{equation}

It is easy to check that the solution of equation (\ref{eq:LBS_P1}) (the vanilla option price) is given by equation (\ref{eq:P1}) (considering that applying $\mathcal{L}_{BS}$ on any of the derivatives with respect to $S_i$ gives zero because $\mathcal{L}_{BS}P_0 = 0$). See that for pricing purposes, the market price of risk $\Lambda(Y_t)$ is set to zero and therefore, $U_i = 0$.

\begin{equation}
\sqrt \varepsilon  P_1 \left( {t,S_i } \right) =  - \left( {T - t} \right)\left( {\left( {2\tilde R_i  + U_i } \right)S_i^2 \frac{{\partial ^2 P_0 }}{{\partial S_i^2 }} + \tilde R_i S_i^2 \frac{{\partial ^3 P_0 }}{{\partial S_i^3 }}} \right)
  \label{eq:P1}
\end{equation}

\begin{equation}
P\left( {t,S_i } \right) = P_0 \left( {t,S_i } \right) + \sqrt \varepsilon  P_1 \left( {t,S_i } \right)
  \label{eq:P}
\end{equation}

Equation (\ref{eq:SigImplFouque}) gives an approximation of the implied volatility in terms of $V_2$ and $V_3$ where $r=\frac{1}{(T-t)} \int_t^T r_sds$ and $q_i=\frac{1}{(T-t)} \int_t^T q_s^{(i)}ds$. This equation is taken from equation (5.55) of section 5.3 of \cite{Fouque2000} (see that $(r-q_i)$ appears now instead of $r$ because the process here includes dividends). Setting $V_2 = 2\tilde R_i$ ($U_i$ is equal to zero as the market price of risk is zero) and replacing $V_3 = \tilde R_i$ yields the final calibration equation (\ref{eq:SigImplFinal}), where $F_{iT}=S_i(t)e^{(r-q_i)(T-t)}$ is the forward price of underying $i$ at expiration.

\begin{equation}
\sigma _i^{impl}  =  - \frac{{V_3 }}{{\bar \sigma _i^3 }}\left( {\ln \left( {\frac{K}{{S_i (t)}}} \right)\frac{1}{{T - t}}} \right) + \bar \sigma _i  + \frac{{V_3 }}{{\bar \sigma _i^3 }}\left( {(r - q_i) + \frac{3}{2}\bar \sigma _i^2 } \right) - \frac{{V_2 }}{{\bar \sigma }} + O(\epsilon)
  \label{eq:SigImplFouque}
\end{equation}

\begin{equation}
\sigma _i^{impl}  =  - \frac{{2\tilde R_i }}{{\bar \sigma _i^3 }}\left( {\ln \left( {\frac{K}{{F_{iT} }}} \right)\frac{1}{{T - t}}} \right) + \bar \sigma _i  + \frac{{\tilde R_i }}{{2 \bar \sigma _i }} + O(\epsilon)
  \label{eq:SigImplFinal}
\end{equation}

\end{document}

%% file: Hes1yITM.tex
\begin{tiny}\begin{table}[htbp]\centering	\begin{tabular}{|r|c|c|c|}
	\hline
				\textbf{Sce}&$\,\,$\textbf{\Large\phantom{I}\normalsize Gcop}$\,\,$ & $\,\,$\textbf{Pcop}$\,\,$ & $\,\,$\textbf{P-G}$\,\,$\\
		\hline
		$\,\,$\textbf{\Large\phantom{I}\normalsize  1-LR $\rho=+0.6$}$\,\,$ & \,\,~0.2640\,\, & \,\,~0.2671\,\, & ~0.0031\,\,\\
		\hline
		$\,\,$\textbf{\Large\phantom{I}\normalsize  2-LR $\rho=+0.3$}$\,\,$ & \,\,~0.2593\,\, & \,\,~0.2623\,\, & ~0.0030\,\,\\
		\hline
		$\,\,$\textbf{\Large\phantom{I}\normalsize  3-LR $\rho=+0.0$}$\,\,$ & \,\,~0.2558\,\, & \,\,~0.2583\,\, & ~0.0025\,\,\\
		\hline
		$\,\,$\textbf{\Large\phantom{I}\normalsize  4-LR $\rho=-0.3$}$\,\,$ & \,\,~0.2523\,\, & \,\,~0.2544\,\, & ~0.0021\,\,\\
		\hline
		$\,\,$\textbf{\Large\phantom{I}\normalsize  5-LR $\rho=-0.6$}$\,\,$ & \,\,~0.2476\,\, & \,\,~0.2491\,\, & ~0.0015\,\,\\
		\hline\hline
		$\,\,$\textbf{\Large\phantom{I}\normalsize  6-RL $\rho=+0.6$}$\,\,$ & \,\,~0.2601\,\, & \,\,~0.2585\,\, & -0.0016\,\,\\
		\hline
		$\,\,$\textbf{\Large\phantom{I}\normalsize  7-RL $\rho=+0.3$}$\,\,$ & \,\,~0.2552\,\, & \,\,~0.2530\,\, & -0.0022\,\,\\
		\hline
		$\,\,$\textbf{\Large\phantom{I}\normalsize  8-RL $\rho=+0.0$}$\,\,$ & \,\,~0.2515\,\, & \,\,~0.2495\,\, & -0.0021\,\,\\
		\hline
		$\,\,$\textbf{\Large\phantom{I}\normalsize  9-RL $\rho=-0.3$}$\,\,$ & \,\,~0.2479\,\, & \,\,~0.2461\,\, & -0.0018\,\,\\
		\hline
		$\,\,$\textbf{\Large\phantom{I}\normalsize 10-RL $\rho=-0.6$}$\,\,$ & \,\,~0.2430\,\, & \,\,~0.2416\,\, & -0.0015\,\,\\
		\hline\hline
		$\,\,$\textbf{\Large\phantom{I}\normalsize 16-RR $\rho=+0.6$}$\,\,$ & \,\,~0.2614\,\, & \,\,~0.2595\,\, & -0.0020\,\,\\
		\hline
		$\,\,$\textbf{\Large\phantom{I}\normalsize 17-RR $\rho=+0.3$}$\,\,$ & \,\,~0.2559\,\, & \,\,~0.2533\,\, & -0.0026\,\,\\
		\hline
		$\,\,$\textbf{\Large\phantom{I}\normalsize 18-RR $\rho=+0.0$}$\,\,$ & \,\,~0.2520\,\, & \,\,~0.2489\,\, & -0.0032\,\,\\
		\hline
		$\,\,$\textbf{\Large\phantom{I}\normalsize 19-RR $\rho=-0.3$}$\,\,$ & \,\,~0.2484\,\, & \,\,~0.2447\,\, & -0.0037\,\,\\
		\hline
		$\,\,$\textbf{\Large\phantom{I}\normalsize 20-RR $\rho=-0.6$}$\,\,$ & \,\,~0.2438\,\, & \,\,~0.2402\,\, & -0.0037\,\,\\
		\hline\hline
		$\,\,$\textbf{\Large\phantom{I}\normalsize 21-LL $\rho=+0.6$}$\,\,$ & \,\,~0.2642\,\, & \,\,~0.2662\,\, & ~0.0020\,\,\\
		\hline
		$\,\,$\textbf{\Large\phantom{I}\normalsize 22-LL $\rho=+0.3$}$\,\,$ & \,\,~0.2594\,\, & \,\,~0.2620\,\, & ~0.0026\,\,\\
		\hline
		$\,\,$\textbf{\Large\phantom{I}\normalsize 23-LL $\rho=+0.0$}$\,\,$ & \,\,~0.2559\,\, & \,\,~0.2590\,\, & ~0.0030\,\,\\
		\hline
		$\,\,$\textbf{\Large\phantom{I}\normalsize 24-LL $\rho=-0.3$}$\,\,$ & \,\,~0.2525\,\, & \,\,~0.2559\,\, & ~0.0034\,\,\\
		\hline
		$\,\,$\textbf{\Large\phantom{I}\normalsize 25-LL $\rho=-0.6$}$\,\,$ & \,\,~0.2479\,\, & \,\,~0.2509\,\, & ~0.0030\,\,\\
		\hline
	\end{tabular}
	\caption{Comparison of Gaussian and perturbed copula methods.}
\label{Hes1yITM}
\end{table}\end{tiny}

%% file: 20111012_MktMatMon_P-G.tex
\begin{tiny}\begin{table}[htbp]\centering	\begin{tabular}{|r|c|c|c|c|c|}
	\hline
				\textbf{Sce | corr}&$\,\,$\textbf{\Large\phantom{I}\normalsize +0.6}$\,\,$ & $\,\,$\textbf{+0.3}$\,\,$ & $\,\,$\textbf{+0.0}$\,\,$ & $\,\,$\textbf{-0.3}$\,\,$ & $\,\,$\textbf{-0.6}$\,\,$\\
		\hline
		$\,\,$\textbf{\Large\phantom{I}\normalsize 0.7 1y}$\,\,$ & \,\,~17.35\,\, & \,\,~25.74\,\, & \,\,~29.65\,\, & \,\,~31.55\,\, & ~27.49\,\,\\
		\hline
		$\,\,$\textbf{\Large\phantom{I}\normalsize 0.85 1y}$\,\,$ & \,\,~9.11\,\, & \,\,~20.06\,\, & \,\,~27.36\,\, & \,\,~29.45\,\, & ~15.19\,\,\\
		\hline
		$\,\,$\textbf{\Large\phantom{I}\normalsize ATM 1y}$\,\,$ & \,\,~24.39\,\, & \,\,~23.69\,\, & \,\,~29.08\,\, & \,\,~31.05\,\, & ~23.84\,\,\\
		\hline
		$\,\,$\textbf{\Large\phantom{I}\normalsize 1.15 1y}$\,\,$ & \,\,~32.70\,\, & \,\,~25.30\,\, & \,\,~27.92\,\, & \,\,~29.07\,\, & ~27.27\,\,\\
		\hline
		$\,\,$\textbf{\Large\phantom{I}\normalsize 1.2 1y}$\,\,$ & \,\,~32.04\,\, & \,\,~24.55\,\, & \,\,~26.51\,\, & \,\,~27.23\,\, & ~25.75\,\,\\
		\hline\hline
		$\,\,$\textbf{\Large\phantom{I}\normalsize 0.7 2y}$\,\,$ & \,\,-7.16\,\, & \,\,~11.00\,\, & \,\,~24.76\,\, & \,\,~33.31\,\, & ~8.85\,\,\\
		\hline
		$\,\,$\textbf{\Large\phantom{I}\normalsize 0.85 2y}$\,\,$ & \,\,-0.93\,\, & \,\,~13.83\,\, & \,\,~29.87\,\, & \,\,~36.32\,\, & ~7.12\,\,\\
		\hline
		$\,\,$\textbf{\Large\phantom{I}\normalsize ATM 2y}$\,\,$ & \,\,~21.28\,\, & \,\,~21.66\,\, & \,\,~34.77\,\, & \,\,~40.00\,\, & ~22.56\,\,\\
		\hline
		$\,\,$\textbf{\Large\phantom{I}\normalsize 1.15 2y}$\,\,$ & \,\,~37.22\,\, & \,\,~27.95\,\, & \,\,~37.33\,\, & \,\,~41.31\,\, & ~33.18\,\,\\
		\hline
		$\,\,$\textbf{\Large\phantom{I}\normalsize 1.2 2y}$\,\,$ & \,\,~39.76\,\, & \,\,~29.17\,\, & \,\,~37.46\,\, & \,\,~40.87\,\, & ~34.32\,\,\\
		\hline
	\end{tabular}
	\caption{Perturbed minus gaussian copula difference in basis points varying moneyness, correlation and maturity  for market scenarios.}
\label{20111012_MktMatMon_P-G}
\end{table}\end{tiny}

%% file: 20090908_MktMatMon_LVATMcorr-G.tex
\begin{tiny}\begin{table}[htbp]\centering	\begin{tabular}{|r|c|c|c|c|c|}
	\hline
				\textbf{Sce | corr}&$\,\,$\textbf{\Large\phantom{I}\normalsize +0.6}$\,\,$ & $\,\,$\textbf{+0.3}$\,\,$ & $\,\,$\textbf{+0.0}$\,\,$ & $\,\,$\textbf{-0.3}$\,\,$ & $\,\,$\textbf{-0.6}$\,\,$\\
		\hline
		$\,\,$\textbf{\Large\phantom{I}\normalsize 0.7 1y}$\,\,$ & \,\,-15.4566\,\, & \,\,-12.2356\,\, & \,\,-9.3169\,\, & \,\,-6.5506\,\, & -1.7595\,\,\\
		\hline
		$\,\,$\textbf{\Large\phantom{I}\normalsize 0.85 1y}$\,\,$ & \,\,-16.7825\,\, & \,\,-15.0450\,\, & \,\,-12.7075\,\, & \,\,-10.1333\,\, & -7.0619\,\,\\
		\hline
		$\,\,$\textbf{\Large\phantom{I}\normalsize ATM 1y}$\,\,$ & \,\,-14.1529\,\, & \,\,-13.7226\,\, & \,\,-11.6279\,\, & \,\,-9.1954\,\, & -8.2319\,\,\\
		\hline
		$\,\,$\textbf{\Large\phantom{I}\normalsize 1.15 1y}$\,\,$ & \,\,-10.2258\,\, & \,\,-9.2708\,\, & \,\,-7.3272\,\, & \,\,-6.1700\,\, & -5.2791\,\,\\
		\hline
		$\,\,$\textbf{\Large\phantom{I}\normalsize 1.2 1y}$\,\,$ & \,\,-9.6079\,\, & \,\,-8.5011\,\, & \,\,-6.9462\,\, & \,\,-6.0872\,\, & -5.1886\,\,\\
		\hline\hline
		$\,\,$\textbf{\Large\phantom{I}\normalsize 0.7 2y}$\,\,$ & \,\,~4.6959\,\, & \,\,-5.1750\,\, & \,\,-7.6028\,\, & \,\,-8.4318\,\, & -5.9855\,\,\\
		\hline
		$\,\,$\textbf{\Large\phantom{I}\normalsize 0.85 2y}$\,\,$ & \,\,~6.2041\,\, & \,\,-2.6924\,\, & \,\,-5.9902\,\, & \,\,-7.1165\,\, & -3.6275\,\,\\
		\hline
		$\,\,$\textbf{\Large\phantom{I}\normalsize ATM 2y}$\,\,$ & \,\,-1.5824\,\, & \,\,-8.5340\,\, & \,\,-13.2798\,\, & \,\,-13.9504\,\, & -10.0221\,\,\\
		\hline
		$\,\,$\textbf{\Large\phantom{I}\normalsize 1.15 2y}$\,\,$ & \,\,~0.3596\,\, & \,\,-6.4414\,\, & \,\,-9.4217\,\, & \,\,-9.3834\,\, & -6.7082\,\,\\
		\hline
		$\,\,$\textbf{\Large\phantom{I}\normalsize 1.2 2y}$\,\,$ & \,\,~0.2626\,\, & \,\,-6.5320\,\, & \,\,-8.8819\,\, & \,\,-8.6984\,\, & -6.6643\,\,\\
		\hline
	\end{tabular}
	\caption{Local volatility with ATM correlation minus gaussian copula difference in basis points varying moneyness, correlation and maturity for market scenarios.}
\label{20090908_MktMatMon_LVATMcorr-G}
\end{table}\end{tiny}